\documentclass[12 pt,oneside]{book}
\usepackage{doublespace}
\usepackage{graphicx}
\usepackage{amsmath}
\begin{document}
\pagestyle{myheadings}
\pagenumbering{roman}

\include{cover}
\tableofcontents
\listoffigures
\newpage
\title{Inclusive electron scattering from nuclei at $x>1$ and high
  $Q^2$ with a 5.75 GeV beam}
\author{Nadia Fomin}
\maketitle
\newpage
\bibliographystyle{thesisprsty}


\pagenumbering{arabic}

\chapter*{Abstract}

Experiment E02-019, performed in Hall C at the Thomas Jefferson
National Accelerator Facility (TJNAF), was a measurement of inclusive
electron cross sections for several nuclei ($^{2}$H,$^{3}$He,
$^{4}$He, $^{9}$Be,$^{12}$C, $^{63}$Cu, and $^{197}$Au) in the quasielastic region
at high momentum transfer.  

In the region of low energy transfer, the cross sections were analyzed
in terms of the reduced response, F(y), by examining its $y$-scaling
behavior.  The data were also examined in terms of the nuclear
structure function $\nu W_2^A$ and its behavior in $x$ and the
Nachtmann variable $\xi$.  The
data show approximate scaling of $\nu W_2^A$ in $\xi$ for all targets at all
kinematics, unlike scaling in $x$, which is confined to the DIS regime.  However, $y$-scaling observations are limited to the
kinematic region dominated by the quasielastic response
(\mbox{$y<$0}), where some scaling violations arising from FSIs are observed.
\chapter{Introduction}
The first scattering experiments were performed by Hans Geiger (1882-1945) and
Ernest Marsden (1889-1970), under the guidance of Ernest Rutherford
(1871-1937) and serve as the foundation for the nuclear model of the atom~\cite{Finn:1992a}.    These experiments verified that the
nuclear charge is a multiple of $e$, the electron charge,
specifically, $Ze$ and that it is concentrated near the center of the atom.  This was determined by
observing the angular distribution of the scattered particles.

Experimental technology has advanced greatly since Rutherford's time,
but the principles behind the experiments are still the same:
particles are scattered off a target, detected after the interaction
and their energy and scattering angle are measured.

\section{Background}
Quantum Electrodynamics (QED) is the theory that describes the interaction between charged
particles via the exchange of photons.  It has been very successful in
making accurate predictions of many quantities such as the anomalous magnetic moment of
the electron.  The interaction of the electron with
the charge of the nucleus is well understood in QED, which along with
the fact that it's a weak probe and does not disturb the target, makes the
electron ideal for studying nuclear structure.  This was first understood more than
50 years ago when electron scattering experiments from
nuclear targets were performed by Lyman~\cite{PhysRev.84.634}.

The nucleus receives energy $\nu$ and momentum $\textbf{q}$ from the
electron through the exchange of a single virtual photon.
The resolution of the electron probe is determined by the wavelength
of the virtual photon, $\lambda =\frac{h}{\sqrt{Q^2}}$, which is in turn of
a function of the 4-momentum transfer, Q$^2$.  The larger 
wavelengths, given by the lower values of Q$^2$, probe the bulk
properties of the nucleus and give us information about its
structure.  Increasing Q$^2$ and decreasing the wavelength of the
probe allows for study of the nucleons inside the nucleus.  
 As the Q$^2$ is increased further, the
resolution becomes finer, and we are able to probe the fundamental
building blocks of matter, i.e. quarks.  

\section{Scaling}
Scaling refers to the regime where the measured cross-section becomes
independent of Q$^2$ and $\nu$, the traditional variables, and can
instead be expressed as a function of just one
kinematic variable, which is itself a function of Q$^2$ and $\nu$.  The appeal of the scaling approach lies in the
fact that when we observe scaling, it usually means that a simple
reaction mechanism dominates the process and structural information
can be extracted in a model-independent way~\cite{Day:1990mf}.
In different kinematic regimes (deep inelastic, quasi-elastic),
scaling in different variables is observed and reflects the different
dominant processes.

\subsection{Elastic Scattering}
In Rutherford's experiment alpha particles were scattered elastically 
 by the Coulomb field of a gold nucleus.  For
 this case, a quantum mechanical formulation of the cross section
 yields the same result as a classical one.  Treating this cross
section in the Born approximation (ingoing and outgoing particle is each
described by a plane wave), Rutherford's scattering formula is given
by:
\begin{equation}
\label{rutherford}
\left(\frac{d\sigma}{d\Omega}\right)_R=\frac{4m^2(Z_1Ze^2)^2}{q^4}
\end{equation}
where $Z_1e$ and $m$ are the charge and mass of the incoming particle,  $Ze$ is that
of the nucleus, and $q^4$ is the 4th power of the 3-momentum transfer, \textbf{q}.  In this formula, the incoming and target particles
are both taken to be structureless objects of spin 0.  Additionally,
the target is treated is being infinitely massive and does not recoil.

 However, in the case of electron
scattering, its spin has to be taken into account and the cross
section expression is modified to give:
\begin{equation}
\label{mott}
\left(\frac{d\sigma}{d\Omega}\right)_{Mott}=4(Ze^2)^2\frac{E^2}{q^4}\left(1-\beta ^2
\sin^2{(\theta/2})\right) ,
\end{equation}
where $E, \beta$ are the energy and velocity (fraction of $c$) of the
incident electron. This expression is known as the Mott cross
section.  So far, the recoil motion of the target has been ignored
with the assumption that the target is very massive and therefore the
recoil will be negligible.  To properly account for this recoil motion, the Mott cross section is corrected by
$1/(1+2E/M\sin ^2 (\theta/2))$.

The Mott cross section needs to be further modified when the target is
an extended object (has internal structure).
This is done through the use of \textit{form factors}, which depend on the square of
the momentum transfer, and modify the cross section:
\begin{equation}
\frac{d\sigma}{d\Omega}=\left(\frac{d\sigma}{d\Omega}\right)_{Mott}^R |F(\textbf{q}^2)|^2,
\label{cs_ff}
\end{equation}
where the superscript $R$ refers to the recoil factor in the Mott
cross section.  Eq.~\ref{cs_ff} shows that the form factor can be determined
experimentally by measuring the cross section at any scattering angle
and dividing by the Mott cross section at the same kinematics.  To
relate it to theoretical calculations, the
nuclear form factor can be written down as the Fourier transform of
the probability density.
For a spin 0 target, the form factor in Eq.~\ref{cs_ff} describes the
distribution of electric charge. Elastic electron scattering has
been used to accurately map out the nuclear charge distribution of
many nuclei~\cite{Frois87}.  If the target has spin 1/2 (for example, a nucleon), then an additional factor is needed to describe
the distribution of the magnetization over the target volume.

In the case of a nucleon, incorporating both distributions into the cross section, yields
the following expression, referred to as the \textit{Rosenbluth} cross
section:
\begin{equation}
\frac{d\sigma ^{(p,n)}}{d\Omega}=\left(\frac{d\sigma}{d\Omega}\right)_{Mott}\left[
  \frac{G^{2}_{E (p,n)}+\tau G^{2}_{M (p,n)}}{1+\tau}\: +\:2\tau
  G^{2}_{M (p,n)} \tan ^2 (\theta/2)\right]
\label{rosen_cs}
\end{equation}
with $\tau=Q^2/4m^2$ and $G_E$ and $G_M$ being the electric and magnetic
form factors, which depend only on Q$^2$.  Note that we have adopted the
relativistic definition of 4-momentum transfer, where
Q$^2=4EE'\sin^2(\theta/2)$.  The electron and magnetic form factors
are named as such because in the limit of $Q^2=0$,
they take on the values of $q_N/e$ (charge) and $\mu/\mu _N$ (magnetic
moment) for  $G_E$ and
$G_M$, respectively, with $e$ being the positive unit of charge  and
$\mu _N$ the nuclear magneton.  

One can see from Eq.~\ref{rosen_cs} that the form factors can be
separated by performing scattering experiments at fixed Q$^2$ and
different scattering angles.  This
is known as Rosenbluth separation.  

The so-called ``dipole'' fit, given by 
\begin{equation}
\label{dipole}
G_D(Q^2)=\frac{1}{1+Q^2/0.71}
\end{equation}
describes the separated form factors well and arises from an exponential charge
distribution.  The other form factors, $G^n_E$,  $G^p_M$,  $G^n_M$, can be expressed in
terms of it for a reasonable approximation:
\begin{eqnarray}
\label{ff_approx}
G^p_E(q^2)&\approx& \frac{G^p_M(q^2)}{(\mu_p/\mu_N)} \approx
  \frac{G^n_M(q^2)}{(\mu_n/\mu_N)}=G_D(Q^2)  \\ \nonumber
G^n_E&=&-\frac{\mu_N}{1+5.6\tau}G^p_E
\end{eqnarray}
The electric form factor of the neutron, $G^n_E$, given in
Eq.~\ref{ff_approx} follows a modified dipole behavior, due to
Galster~\cite{Galster:1971kv}. Of the 4 form factors, $G^n_E$ has been
the most difficult to measure as no free neutron
targets exist and the form factor is very small.

There are two experimental methods used to separate the elastic form
factors:  polarization transfer and Rosenbluth separation.  The latter
method involves measuring the cross section at several values of beam
energy and scattering angles for a constant Q$^2$. The information about form factors can
then be extracted from the reduced cross section via:
\begin{equation}
\left( \frac{d\sigma}{d
  \Omega}\right)_{reduced}=\frac{\epsilon(1+\tau)}{\tau} \left(
  \frac{d\sigma}{d\Omega}\right)_{exp} / \left(
  \frac{d\sigma}{d\Omega}\right)_{Mott}=G_M^2+\frac{\epsilon}{\tau}G^2_E,
\label{rosen_sep}
\end{equation}
where $\epsilon =(1+2(1+\tau ) \tan ^2 \theta /2)^{-1}$ is the polarization of the virtual photon.  $G_E$ and $G_M$ are obtained from the slope and intercept when the
Eq.~\ref{rosen_sep} is fit for a range of $\epsilon$ values for a
given Q$^2$.  The linear dependence of the reduced cross section on
$\epsilon$ is a direct consequence of the one-photon approximation.

The polarization transfer method involves scattering longitudinally
polarized electrons and measuring the polarization
of the recoil proton, which will have two non-zero polarization
components, $P_z$ (from $G_M$) and $P_x$ (from the $G_M$ and $G_E$
interference term).  The ratio of the form factors then
gives~\cite{Perdrisat:2006hj}:
\begin{equation}
\frac{G_E}{G_M}=-\frac{P_x}{P_z}\frac{E+E'}{2M} \tan(\frac{\theta}{2})
\label{pol_transfer}
\end{equation}
where $E$, $E'$, and $\theta$ refer to the energy of the beam and the
  energy and angle of the scattered electron.

\begin{figure*}[h!]
\center
\includegraphics[width=0.65\textwidth] {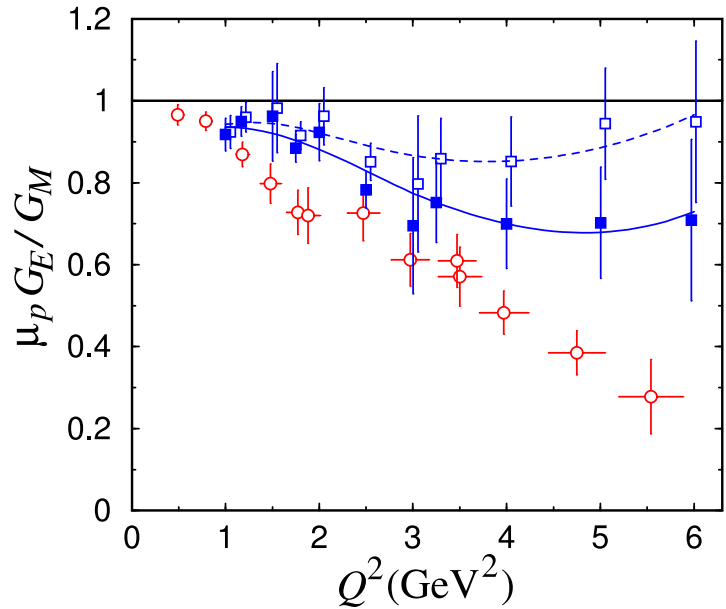}
\caption[$\mu _p G_E/G_M$ vs Q$^2$, world data.]{$\mu _p G_E/G_M$ vs
    Q$^2$ from~\cite{Blunden:2003sp}.  Results from polarization
    transfer experiments~\cite{Jones:1999rz,Gayou:2001qd}  are given
    by the hollow circles.  Rosenbluth separation
    results~\cite{Walker:1993vj,Andivahis:1994rq} are given by
    hollow squares.  The filled squares show the effect on the
    Rosenbluth data when TPE corrections are included.  The dashed and
    solid lines represent a global fit ~\cite{Arrington:2003df}, and
    the size of the effect on it due to TPE.}
\label{ff_disagreement}
\end{figure*}

Results from experiments using the two techniques are not in agreement
with each other, as can be seen in Fig.~\ref{ff_disagreement}.  After
the possibility of experimental errors had been ruled out, the
proposed explanation of the discrepancy between the two measurements
is a two-photon exchange (TPE) ~\cite{Guichon:2003qm,Blunden:2003sp}
contribution to the scattering, which is a more significant effect in
the Rosenbluth results.  No existing model
can completely reconcile the two measurements and more data are
needed.
%
%
%
\section{Inelastic Scattering \label{inelastic_intro}}
While the initial and final state of the
target are the same for elastic scattering, this is not true for inelastic scattering.
In this case, the final hadronic state has no limitations:  the target
can be excited, or broken up, with multiple particles in the final state.  
It is convenient to define the different kinematic regions in terms of
$W$ and Q$^2$, where $W^2$ is the invariant mass of the final hadronic
state and Q$^2$ is the 4-momentum
transfer squared.  These are given by:
\begin{eqnarray}
Q^2  &=& -q^2=\vec{q^2}-\nu ^2=4EE' \sin ^2(\theta/2) \\ \nonumber
W^2  &=& 2M\nu+M^2-Q^2
\label{kin_quants}
\end{eqnarray}
where $E$ and $E'$ are the energies of the incoming and scattered
electron, $M$ is the nucleon mass, and $\theta$ is the scattering
angle.

Experiments done in the \textit{deep inelastic} regime, defined by
kinematics where \mbox{$W>$2GeV} and
\mbox{$Q^2>$1GeV$^2$}, found that inelastic cross sections were only weakly Q$^2$ dependent, signaling that electrons were scattering
from the point-like components of the nucleon.  These point-like
objects were later identified with quarks~\cite{Feynman:1969ej}, but were at first
referred to simply as \textit{partons}.

Follow-up experiments to study nucleon structure have been carried
out, establishing that quarks are spin 1/2 objects.  In fact, the differential cross section for deep inelastic scattering is
written in a form similar to Eq.~\ref{rosen_cs}, in terms of two form
factors, $W_1$ and $W_2$, referred to as the inelastic \textit{structure functions}:
\begin{equation}
\label{dis_cs}
\frac{d\sigma}{d\Omega dE'}=\frac{\alpha^2 \cos ^2 (\theta /2)}{4E^2\sin^4 (\theta/2)}\left[ W_2(Q^2,\nu)+2W_1(Q^2,\nu) \tan ^2(\theta/2) \right]
\end{equation}
Not that the expression in Eq.~\ref{dis_cs} that multiplies the
structure functions is the Mott cross section.
\begin{figure*}
\center
\includegraphics[width=.75\textwidth,clip]{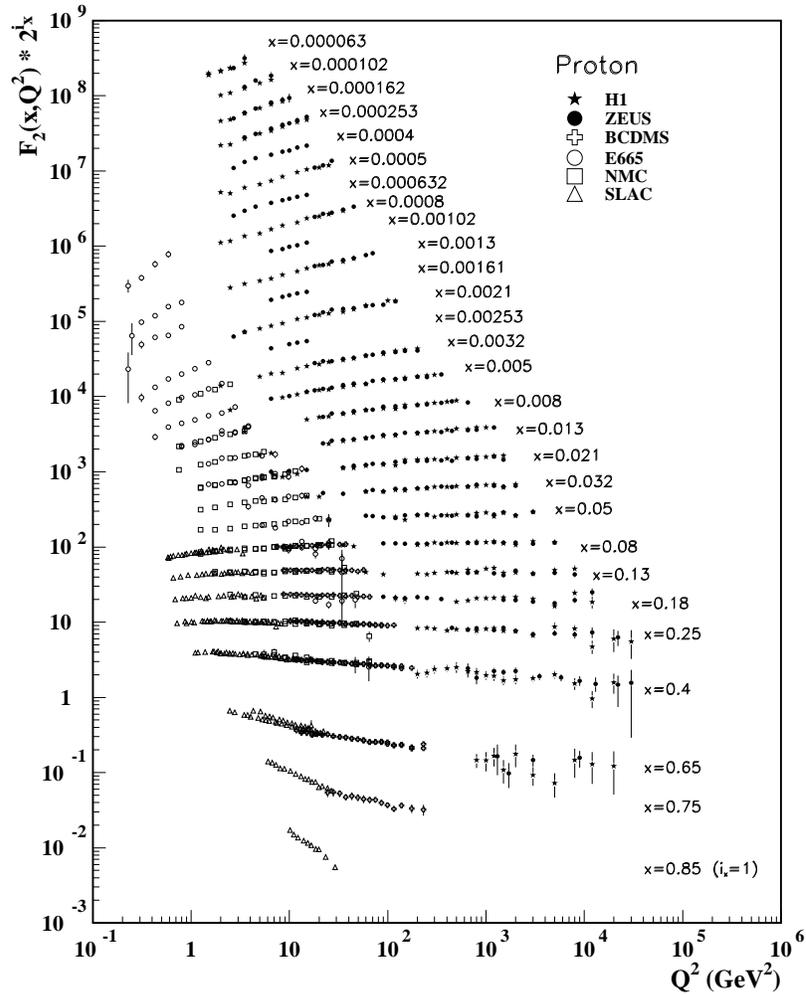}
\caption[$F_2^p$ as a function of Q$^2$]{A representative, but not
    complete sample of the measurements of the $F^p_2$ structure
    function for a range of $x$ values taken from~\cite{PDBook}.  The points have been scaled
    by $2^{i_x}$, where $i_x$ indicates the number of the $x$ bin
    (1-28).  The structure function appears to be Q$^2$-independent
    for $0.1<x<0.4$.  The scale breaking seen at lower values of $x$
    is known to be the result of a large contribution from gluons,
    which dominate at low $x$ and high Q$^2$.}

\label{f2_p_pdg}
\end{figure*}
For inelastic scattering, Bjorken~\cite{Bjorken:1969ja} put forth the idea
that in the limit of infinite momentum and energy transfers, these
structure functions should become dependent on only one variable, $x$,
rather than both Q$^2$ and $\nu$.  This turned out to be the case~\cite{Friedman:1971zi}, and
the structure functions have been found to \textit{scale} in the limit
of high energy and momentum transfers and are usually redefined as:
\begin{equation}
\label{sf_dis_limit}
F_1(x)=mW_1(x)\:\:\:\:\:\textrm{and} \:\:\:\:\:\ F_2(x)=\nu W_2(x)
\end{equation}
In this limit, the two structure functions are related to each other through the Callan-Gross relation \cite{PhysRevLett.22.156},
$F_2(x)=2xF_1(x)$, based on the interpretation of structure functions
as linear combinations of quark distribution functions.  Note that the new variable that the structure
functions depend on is $x$, or \textit{Bjorken}  $x$, given by
$x=\frac{Q^2}{2m\nu}$.  A representative plot of $F_2^p$ measurements
is seen in Fig.~\ref{f2_p_pdg}.  Scaling behavior can be seen for
$0.1<x<0.4$ and $ln$ Q$^2$ behavior is seen for low values of $x$.
\begin{figure*}[htbp]
\center
\includegraphics[height=3in] {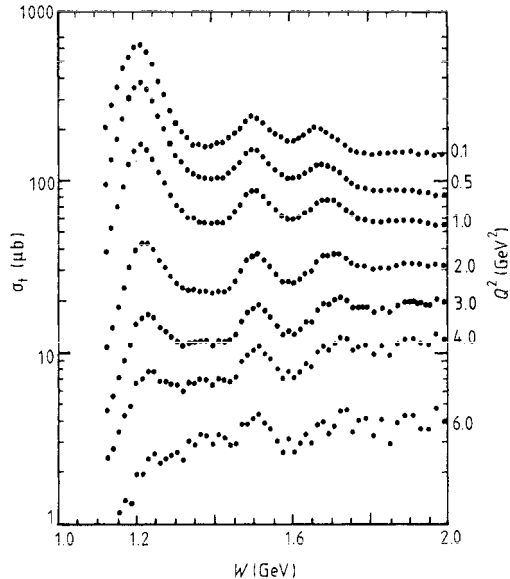}
\caption[Nucleon Resonances]{Electron-Proton cross sections measured
  at several sets of kinematics, plotted as a function of the invariant mass
  $W$ ~\cite{Foster:1983kn}.  The $\Delta$ resonance is well defined
  at $W=1232$ MeV, and the
  second and third resonance regions are also seen. }
\label{res_plot}
\end{figure*}

Outside of the scaling regime, structure functions are most often
presented in terms of the
cross sections for absorption of transverse and longitudinal virtual
photons. The inelastic cross section is related to these through:
\begin{equation}
\frac{d\sigma}{d\Omega dE'}=\Gamma \left[ \sigma _T(W^2,Q^2)+\epsilon
  \sigma _L(W^2, Q^2) \right]
\label{hel_cs}
\end{equation}
The ratio of the two contributions defines $R=\sigma
_L/\sigma _T$, which is
related to the structure functions through:
\begin{equation}
R\equiv \frac{\sigma _L}{\sigma _T}=\frac{W_2}{W_1}\left (
1+\frac{\nu ^2}{Q^2}\right) -1
\end{equation}
The individual structure functions can be determined if the cross
section is measured at several angles for a fixed $\nu$ and Q$^2$, in
the same way that the form factors are separated.
In the deep-inelastic region, $R$ is known to be
small~\cite{Whitlow:1990gk} and zero or small constant values have been
used in some analyses. 
\subsection{Resonances}
Inclusive measurements of electron scattering have revealed the existence of several resonances in the inelastic
region. These resonances correspond to excited nucleon states.  In
Fig.~\ref{res_plot}, the three resonance regions are clearly seen with
\mbox{1 GeV $<W<$ 2 GeV}.
The resonance transition form factors differ from elastic ones and
the ability to measure them depends on the knowledge of the relative
contributions of longitudinal and transverse
cross sections, discussed above. 
Many early studies of the resonance form factors relied on assumptions
about the behavior of $R=\sigma _L/\sigma _T$ in this region.  Prior
to Jefferson Lab experiment E94-110 ~\cite{Liang:2004tj}, no high
precision Rosenbluth-separated data was available to allow for a detailed study
of the resonance structure. The longitudinal contribution was found to be
significant, which was contrary to the existing assumptions.

Comparisons of the Q$^2$-dependence of the resonance form factors to
elastic ones require them to be defined in terms of $\sigma _T$.
 The transition form factors~\cite{Carlson:1988gt,Stoler:1991vi} can then be
 written as 
\begin{equation}
F^2(Q^2)=\frac{1}{4\pi \alpha}\frac{\Gamma _R W_R}{Q^2}(W_R^2-M^2)\sigma _T
\end{equation}
with $W_R$ being the energy of the resonance and $\Gamma _R$, its
 width. 
\begin{figure*}[htpt]
\center
\includegraphics[height=4in,clip]{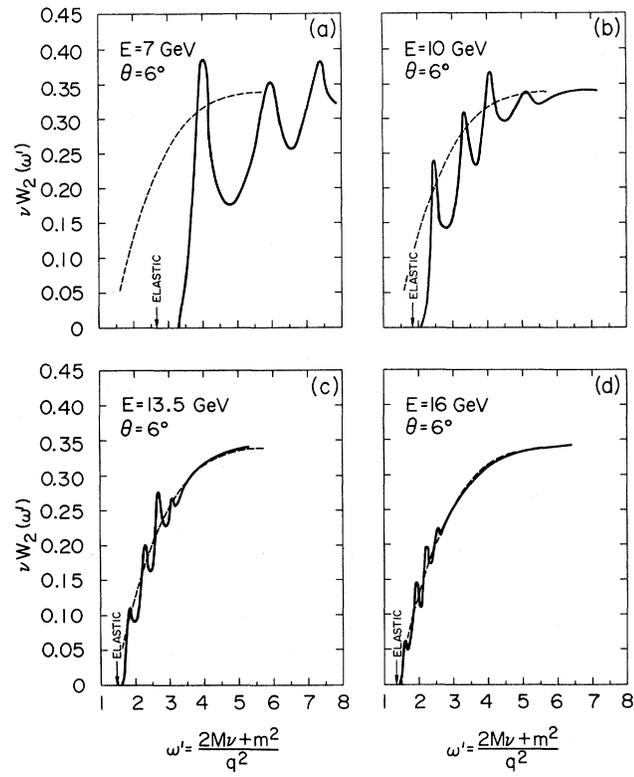}
\caption[ $\nu W^A_2$ data compared to the scaling curve.]{The
  structure function $\nu W^A_2$ plotted as a function of
  $w'=1/x+M^2/Q^2$ for increasing values of Q$^2$.  Dashed line
  corresponds to the scaling curve. Figure from~\cite{Bloom:1970xb}
  with data from~\cite{Bloom:1969kc}.}
\label{duality}
\end{figure*}
In a famous observation~\cite{Bloom:1970xb}, it was
noted that the structure function $\nu W^A_2$, when 
examined in terms of the dimensionless variable $w'=1/x+M^2/Q^2$, has the same Q$^2$ behavior as the resonance form
factors.  In fact, averaging over the resonance peaks produces the
scaling curve for the structure functions, which is seen in Fig.~\ref{duality}.  The resonance peaks do not
disappear with increasing values of Q$^2$, but decrease in strength
and move to lower values of $w'$, following the magnitude of the
scaling limit curve.  This is known as the Bloom-Gilman duality and
continues to be a subject of great interest.
%
%
%
%
%
\subsection{EMC Effect}
\begin{figure*}
\center
\includegraphics[width=.75\textwidth,clip]{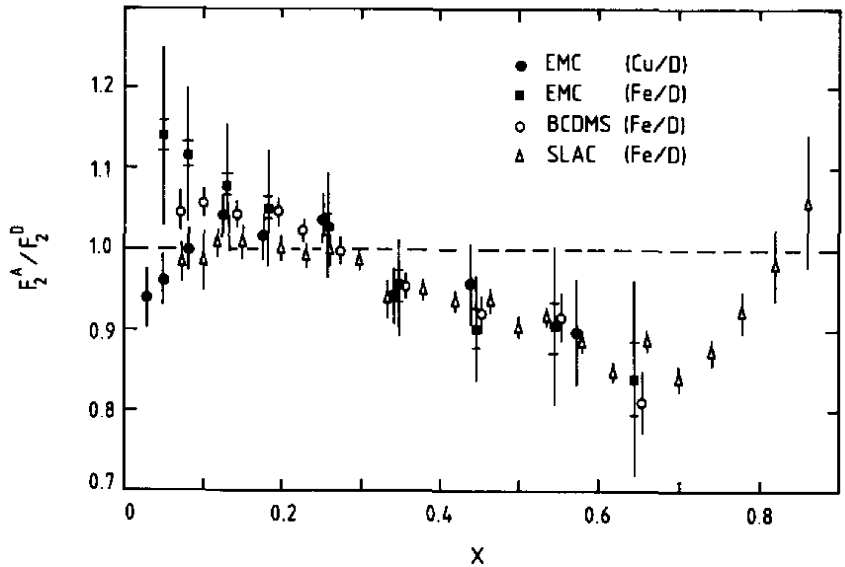}
\caption[The EMC Effect in Fe, Cu.]{Ratio of the $F_2$ structure
  functions for copper and iron to that for deuterium.  Data collected
  by the EMC collaboration as well as earlier data from SLAC is shown~\cite{Ashman:1988bf}.}
\label{emc_orig}
\end{figure*}

An experiment carried out at CERN by the European Muon
Collaboration (EMC) \cite{Ashman:1988bf}, made the unexpected and
surprising discovery that the deep inelastic
structure function, $F_2$, for heavy targets (e.g. iron, copper) was not the
same as that for deuterium.  This indicated modification by the
nuclear medium - a bound nucleon behaved differently from a free
one. Given the fact that the binding energies (a few MeV) are
extremely small when compared to the experimental energy transfer (on
the order of GeV), no noticeable effect was expected.  Fig.~\ref{emc_orig} shows the ratio of the $F_2$ structure
functions from heavy nuclei to those of deuterium.  The shape at low
$x$ ($x<$0.1) is thought to be the result of nuclear ``shadowing''\cite{Close:1988xw}, where the
virtual photon interacts with the neighbors of the struck nucleon.
The shape of the curve at higher $x$ (up to $x\approx 0.7$) is considered to be due to the
binding of the struck nucleon in the nucleus and for $x$ higher than
0.7,  the result of the motion of
the bound nucleons \cite{Berger:1987er}. Measurements of the EMC
effect for different nuclei and over a wide kinematic range
have been performed since the original discovery.
Jefferson Lab experiment E03-103 ~\cite{Arrington:2003emc} was one such effort and
it was complementary to the experiment described here.  E03-103 ran
concurrently and examined the cross sections at the low end of the $x$-coverage.
\subsection{Quasielastic Scattering \label{qe_intro}}
The quasielastic kinematic region is defined by energy and
momentum transfers where single nucleon knockout is the dominant
process.  In quasielastic scattering, the virtual photon is
absorbed on a single nucleon.  In electron-nucleus
scattering, the quasielastic contribution is dominant at
momentum transfers between 500~MeV/c and 2~GeV/c,
although inelastic contributions can be significant at large Q$^2$.
The dominant feature in the inclusive spectrum is a broad peak that
is a result of scattering from a moving bound nucleon in the nucleus.
A more detailed description of the reaction follows in Sec.~\ref{qe_theory} and a comprehensive review
can be found in~\cite{Benhar:2006wy}.

\begin{figure*}[h!]
\center
\includegraphics[width=.85\textwidth] {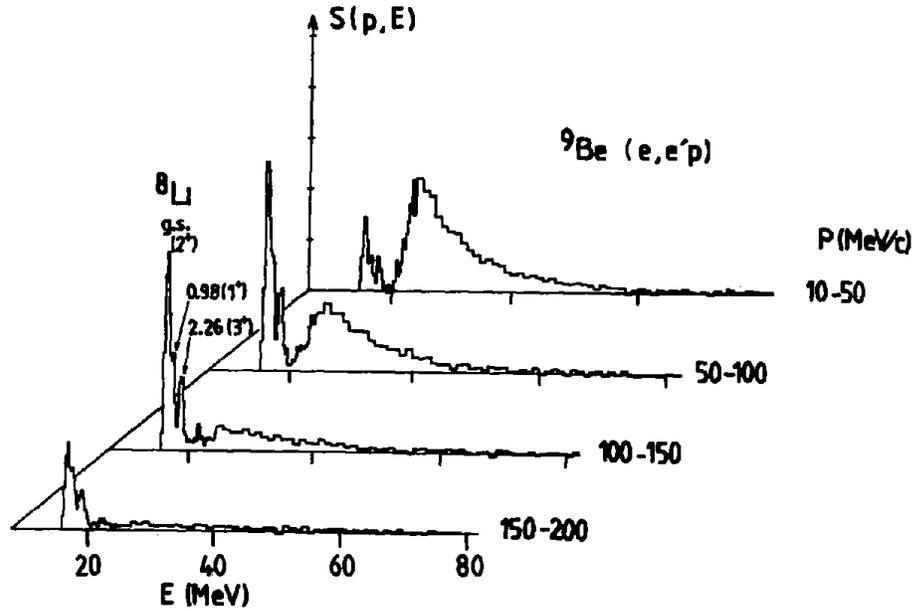}
\caption[Spectral Function for Be.]{Spectral function for Be is shown
  from~\cite{Mougey:1979xk}, where the binding energy spectra can be
  seen for several bins of recoil momentum.  The peaks in the binding
  energy spectra correspond to holes in the 1-$p$ state.  See Sec.~\ref{qe_intro} for details.}
\label{be_spect}
\end{figure*}

The information about the nuclear response in encapsulated inside
 the \textit{Spectral Function}, which is not an experimental
observable.  It describes the energy and momentum distributions of the
nucleons inside nuclei and can most easily be introduced by
 considering the ($e,e'\: p$) reaction in the quasielastic region.
 Here, an electron ejects a proton from the target nucleus and the
 final state system consists of A-1 nucleons, which can be bound or
 not.  In the Independent Particle Shell Model, the spectral function
 for A and A-1 is given by:
\begin{equation}
S(\vec{p_0}, E_0)=\sum _{\alpha} N_{\alpha} |\Psi _{\alpha}(p_0)|^2
\delta (E_0+\Sigma _{\alpha})
\label{spectral_func_intro}
\end{equation}
%
where $\Psi _{\alpha}(p_0)$ is the single proton wave function in a
  ground state $\alpha$, in the momentum-space representation, $p_0$ is
  the initial momentum of the nucleon, $\Sigma
  _{\alpha}$ is the energy eigenvalue, and $N_{\alpha}$ is the occupancy.
  An experimentally determined beryllium spectral function is shown in
Fig.~\ref{be_spect}, where several peaks in energy can be seen. 
These peaks are identified with holes
in the orbitals - i.e. the knocked out nucleons from different ground
  state orbitals.  In this model, the nucleons are assumed to occupy
single particle states and move independently of each other. The IPSM predicts that the spectroscopic factor, obtained by
integrating the momentum distribution of a given shell, is equal to the
number of nucleons that can occupy that shell.
 
However, experiments (see, for example~\cite{PhysRevC.49.955})
yielded a spectroscopic factor that was $\approx$70$\%$ of the
prediction for a wide range of nuclei.  This deficiency is thought to be largely due to Short
Range Correlations that are not included in IPSM and which produce high momentum nucleons as a result of
the strong repulsive $N-N$ force at short distances.  SRCs will be discussed in more
detail in Sec.~\ref{src_section}.

\begin{figure*}
\center
\includegraphics[width=.85\textwidth] {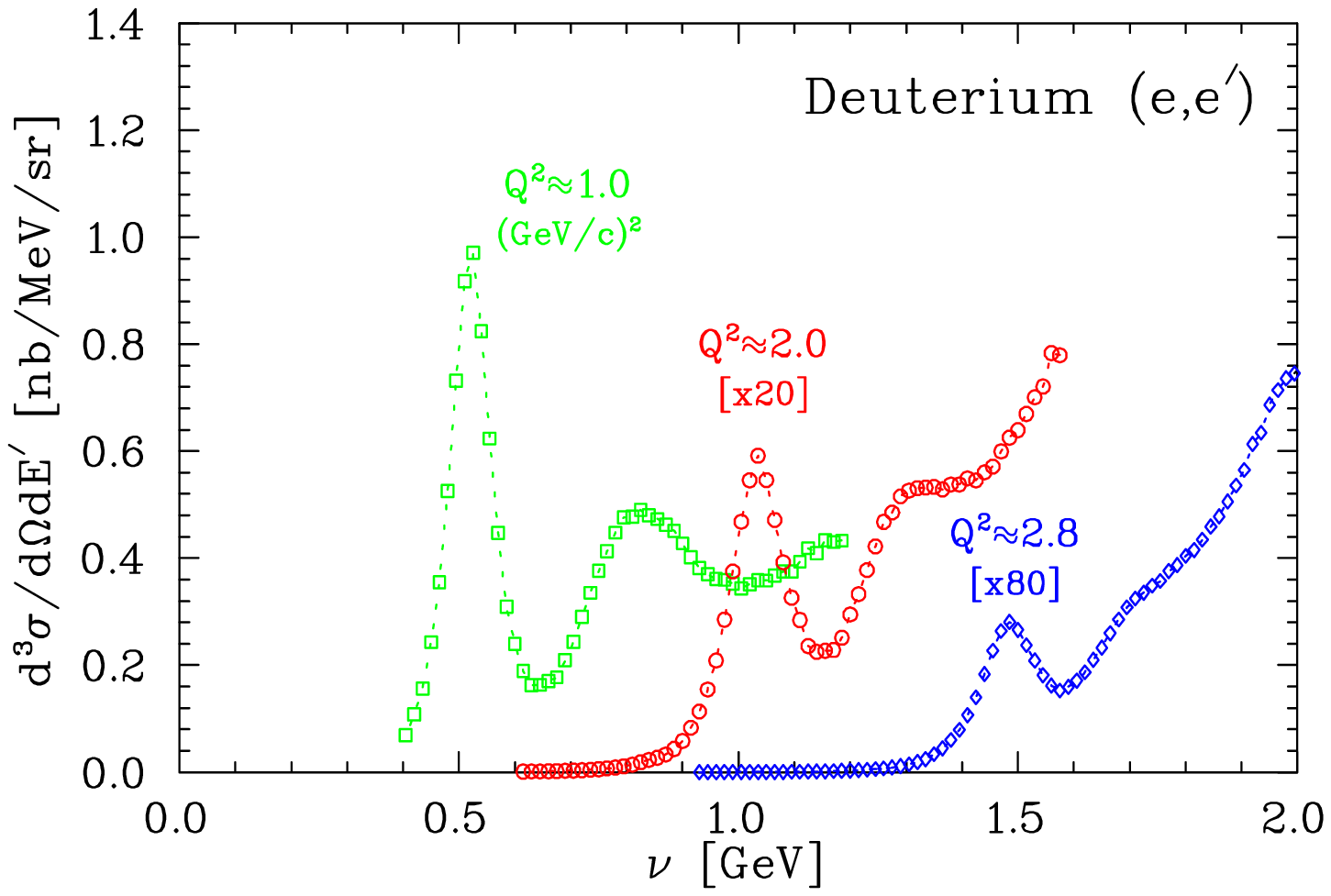}
\caption[Inclusive Deuterium Cross section]{The differential cross
  section for deuterium is shown as a function of energy loss $\nu$
  for three sets of kinematics.  The Q$^2$ values shown at those at
  the quasielastic peak. Data shown is that from JLab experiment E89-008.~\cite{john_thesis}}
\label{transition}
\end{figure*}

On the low energy loss side of the quasielastic peak (see Fig.~\ref{transition}), we will examine
the quasielastic response function F($|\textbf{q}|,y$), which is an
energy and momentum integral of the spectral function.  F($|\textbf{q}|,y$) is expected
to scale in $y$, the minimum momentum of the struck particle in the
direction of momentum transfer in the kinematic regime where inelastic
contributions and Final State Interactions are minimal.  
On the side of the quasielastic peak where the energy loss is higher,
inelastic scattering begins to dominate quickly, and it is more
appropriate to examine the data through the study of the
inelastic structure function, $F_2^A$.  The superscript $A$, in this case, signals that
this is a nuclear structure function, which differs from the nucleon
structure function discussed earlier in Sec.~\ref{inelastic_intro} due to medium modification.
The nuclear structure function is expected to
scale in the same way as those of free nucleons.  When scattering from
a nucleus, momentum is
shared between the constituents of the nucleus, and $x$ can vary
between 0 and $\approx A$, rather than 0 and 1, as for a free nucleon.

This nuclear structure function, $\nu W^A_2$, has been observed to scale in $x$ 
in the deep inelastic region and over a wider kinematic range if
examined~\cite{Filippone:1992iz} in the Nachtmann variable
$\xi=\frac{2x}{1+\sqrt{1+4M^2x^2/Q^2}}$, which represents the
fractional quark lightcone momentum, and which reduces to $x$ for very large
values of Q$^2$.

\section{Experiment E02-019}
\begin{figure*}[htbp]
\center
\includegraphics[angle=270,width=.85\textwidth] {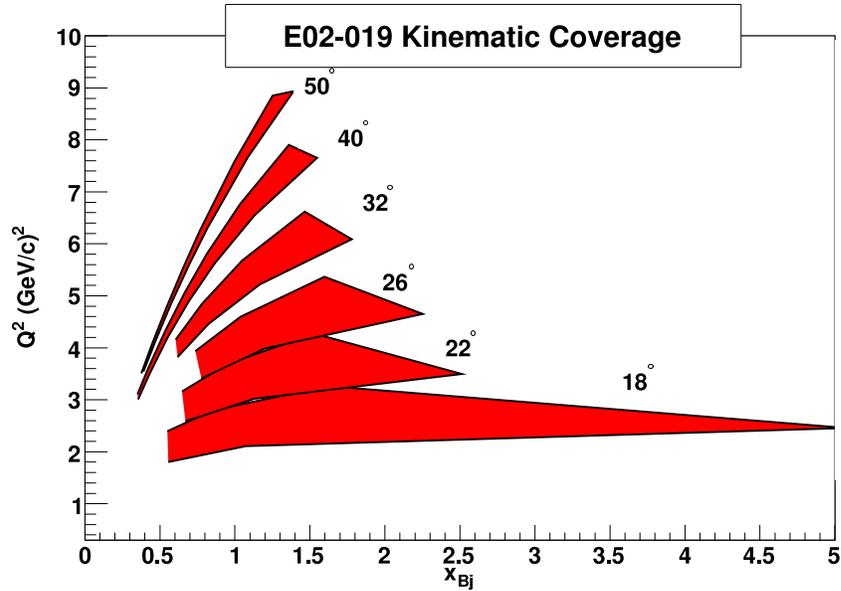}
\caption[Kinematic coverage for JLab experiment E02-019.]{Kinematic
  Coverage for JLab experiment E02-019.  The highest $x$ values are
  not accessible to the light targets.}
\label{kinem_coverage}
\end{figure*}

Experiment E02-019 ran in the fall of 2004 at the Thomas Jefferson
National Accelerator Facility in experimental Hall C.  It was an extension of a previous
experiment, E89-008, with a higher beam energy, which allows
measurements at high  Q$^2$ values with a wide range of targets from
very light ($^2$H) to very heavy ($^{197}$Au).  The kinematic coverage of E02-019
is shown in Fig.~\ref{kinem_coverage}.

In the next chapter, we will examine the physics motivation for the experiment
in much more detail.  Several different interpretations for the data will be presented.
In Chapter 3, the experimental setup will be discussed,
including beam line, target, and spectrometer.  In Chapter 4, a
detailed discussion of the analysis procedure will follow.  It will
include the steps for extracting experimental cross sections as well
as descriptions of all the corrections that were applied to the data.
Finally, in Chapter 5, results will be presented, including cross
sections, structure and scaling functions.  Comparisons with
theoretical calculations will be done, where possible.  Implications of the data
as well as prospects for further measurements will be discussed.

%

\chapter{Theoretical Overview}

The quasielastic scattering process will be presented in the Plane
Wave Impulse Approximation (PWIA)
which will allow us to develop the scaling of a reduced cross section
in terms of $y$ and $F(y,|\textbf{q}|)$, where $y$ is the longitudinal
momentum of the struck nucleon.  We will examine the approach to
scaling with increasing momentum transfer and discuss the assumptions
implicit in the presentation.

The scaling limit of the inelastic nuclear structure function $\nu
W_2^A(x,Q^2)$ will also be examined in both $x$ and $\xi$.
\section{Quasi-elastic Cross Section \label{qe_theory}}

\begin{figure*}[h!]
\center
\includegraphics[height=2.5in,clip]{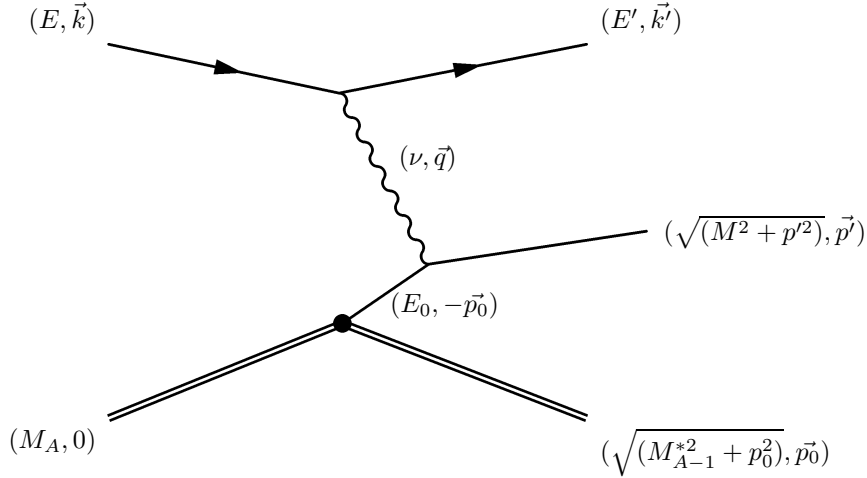}
\caption[Diagram of the quasielastic reaction]{The quasi-elastic reaction. E,$\vec{k}$ and  E,$\vec{k'}$ are
the energies and momenta of the incoming and scattered electron.  The
virtual photon of energy and momentum ($\nu$, $\vec{q}$) interacts with 
the nucleus M$_A$, at rest in the laboratory.  The result is a knocked out nucleon with energy and momentum $\sqrt{M^2+p'^2}$,
$\vec{p'}$ and the recoil nucleus with energy and momentum $\sqrt{M_{A-1}^{*2}+p_0^2}$, $-\vec{p_0}$.}
\label{qe_pwia}
\end{figure*}

Quasi-elastic scattering describes the knock-out of a nucleon from the
nucleus by an electron.  The final state is made up of the
scattered electron, the ejected nucleon, and the remainder of the
nucleus, (A-1), which can possibly be in an excited state (see Fig.~\ref{qe_pwia}).


%
%
The most general expression for electron scattering in the one-photon
exchange approximation is ~\cite{DeForest:1967ab}
\begin{equation}
\frac{d^4 \sigma}{(d \Omega dE')_{N} \: (d \Omega dE')_{e}}=\frac{2\alpha ^2}{Q^4} \frac{E'_e}{E_e}\:|\vec{p_N}| L^{\mu \nu}W^A_{\mu \nu}
\label{generic_sigma}
\end{equation}
where the subscripts $e$ and $N$ denote the electron and the ejected
  particle, respectively, $Q$ is the 4-momentum transfer and $L^{\mu \nu}$ and $W^A_{\mu
  \nu}$ are the leptonic and nuclear response tensors. 
While the leptonic tensor is completely determined by kinematics, in
order to calculate the
nuclear response tensor for the case of multi-GeV incident electrons, one must make several simplifying assumptions.

One prescription for this is the Impulse Approximation (IA), where the dominant process is taken to be scattering from
individual nucleons and only the motion of the struck nucleon must be
  treated relativistically.  In the IA, the cross section is written down in terms of the
nuclear spectral function $S'(E_0,p_0)$, which represents the probability
of finding a nucleon of energy $E_0$ and momentum $\vec{p_0}$ in the nucleus.  The Plane Wave Impulse
Approximation (PWIA) is the simplest implementation of this scheme and
includes the additional assumption that one neglects the final
state interactions of the knocked out nucleon with the residual nucleus.
These interactions are small and should decrease with increased values of Q$^2$,
corresponding to shorter interaction times.

  This allows the nuclear response tensor to be written as ~\cite{Benhar:2006hh}
\begin{equation}
W^A_{\mu \nu} (q) =\int{d^3 p \:dE \: S'(E_0,p_0) \: \tilde{W}_{\mu \nu} (p_0,q)}
\end{equation}
where $\tilde{W}_{\mu \nu} (p_0,q)$ is the electromagnetic tensor for a
bound nucleon.  Within the IA, binding effects can be accounted for, and
one can use the free nucleon tensor ($
\tilde{W}_{\mu \nu}(p_0,q) \rightarrow W_{\mu \nu} (p_0,\tilde{q})$)
written in terms of $\tilde{q} \equiv (\tilde{\nu},\vec{q})$  where the
energy loss is redefined in the following way, accounting for partial
transfer of energy to the spectator system ~\cite{Benhar:1993ja}:
\begin{equation}
\tilde{\nu}=\nu+M_A+E_{A-1}+\sqrt{|\vec{p_0}|^2+m^2}
\end{equation}
The terms describing the electromagnetic interaction can be
factorized from the spectral function and the total cross section can
be written as a sum of the off-shell electron-nucleon cross sections
weighted by the spectral function:
\begin{equation}
\frac{d^5\sigma}{dE' d\Omega d^3\vec{p}'}=\sum_{nucleons}\sigma_{eN}\cdot S'_N(E_o,p_0)
\end{equation}

With an unpolarized target, $S'_N$ can be taken to be
spherically symmetric, with the following normalization condition:
\begin{equation}
4 \pi \int ^{\infty}_{0} dE \int ^{\infty}_{0} p^2 S'_N(E_0,p_0) dp=1.
\end{equation}
To get the inclusive cross section, we separate the proton and neutron
contributions and integrate over the final state of the unobserved
nucleon: 

\begin{equation}
\frac{d^2\sigma}{dE' d\Omega}=\int (Z\sigma_{ep}S'_p(E_0,p_0)+N\sigma_{en}S'_n(E_0,p_0))d^3\vec{p}
\end{equation}

By neglecting any differences in the spectral functions for protons
and neutrons, we can replace the nucleon-specific spectral functions
with the more general $S'(E_0,p_0)$, and then factor it out.  The 3-momentum transfer, $\vec{q}$, is fixed
by measuring the incoming and scattering electrons, giving
$\vec{p}=\vec{p_0}+\vec{q}$ and  $d^3\vec{p}=d^3\vec{p_0}$.  Using
spherical coordinates, where $\theta$ is the angle between $\vec{p_0}$
and $\vec{q}$ and $\phi$ is the angle between the electron scattering
plane and the $(\vec{p_0},\vec{q})$ scattering plane and noting that
$S'$ has no $\phi$ dependence, we can rewrite the inclusive cross
section as:

\begin{equation}
\frac{d^2\sigma}{dE' d\Omega}=2\pi \int\tilde{\sigma_0}\cdot S'(E_0,
p_0)\cdot p_0^2\:dp_0\:d(cos\theta)
\label{shortcs}
\end{equation}

where $\tilde{\sigma_0}$ is defined as
\begin{equation}
\tilde{\sigma_0}=\frac{1}{2\pi}\int^{2\pi}_0 (Z\sigma_{ep}+N\sigma_{en})d\phi
\end{equation}

The initial and final state particles must be on mass-shell, which
when combined with energy and momentum conservation, gives the
constraints:
\begin{equation}
(M_A-E_0)^2=M_{A-1}^{*2}+p_0^2
\label{cons1}
\end{equation}
as well as
\begin{equation}
M_A+\nu=\sqrt{M^2+(\vec{p_0}+\vec{q})^2}+\sqrt{M_{A-1}^{*2}+\vec{p_0}^2}
\label{cons2}
\end{equation}
where we define the following:
\begin{eqnarray}
M_A \: &=& \: \rm{mass\: of\: the\: target\: nucleus}\nonumber \\
M_{A-1}^* \: &=& \: \rm{mass\: of\: the\: recoiling \:(A-1)\: nucleus}\nonumber\\
M\: &=& \:\rm{mass \: of \: the \: ejected \: nucleon}\nonumber\\
E_s\equiv M_{A-1}^*+M- M_A \ &=& \: \rm{separation \: energy \:(see\:Table.~\ref{sep_es}).}
\end{eqnarray}
Using Eq.~\ref{cons1}, solving for $M_A$ and substituting that
expression into Eq.~\ref{cons2}, we get a simplified expression:
\begin{equation}
E_0+\nu-\sqrt{M_{A-1}^{*2}+p_0^2+q^2+2pq\cos{\theta}}=0.
\label{deltaarg}
\end{equation}

\begin{table}[h!]
\begin{center}
\caption{Separation Energies used in F(y,$|$\textbf{q}$|$)
  extraction~\cite{sep_energy_table}.  Each of these is the minimum
  energy to remove a nucleon from the given nucleus.}
\vspace*{0.25in}
\begin{tabular}{|c|c|}
\hline
Target &
$E_s^{min}$ (MeV) \\
\hline
$^2$H & 2.2 \\
$^3$He & 5.5 \\
$^4$He & 20 \\
$^9$Be & 16.9 \\
$^{12}$C & 16 \\
$^{63}$Cu & 6.1 \\
$^{197}$Au & 5.8 \\
\hline
\end{tabular}
\label{sep_es}
\end{center}
\end{table}
For any given set of $\vec{p_0}$, $\vec{q}$, and $\nu$ values, we can
determine $E_0$.  This allows us to rewrite the expression in
~\ref{shortcs} as follows:

\begin{equation}
\frac{d^2\sigma}{dE' d\Omega}=2\pi \int\tilde{\sigma_0}\cdot S'(E_0,
p_0)\cdot \delta(Arg)\cdot p_0^2\:dp_0\:d(\cos{\theta})\:dE
\label{deltacs}
\end{equation}
where the argument of the delta function is the left-hand side of
Eq.~\ref{deltaarg}.  Next, we can use the $\delta$-function to perform
the integral over $\cos \theta$, the result of which is:
\begin{equation}
\frac{d^2\sigma}{dE' d\Omega}=2\pi \int\tilde{\sigma_0}\cdot
\frac{E_N}{|{p_0}||{q}|} \cdot S'(E_0,
p_0)\cdot p_0^2\:dp_0\:dE_0
\end{equation}
where the energy of the final state nucleon is
$E_N=\sqrt{M^2+(\vec{p}+\vec{q})^2}$.

\begin{figure*}
\center
\includegraphics[scale=0.45, angle=270]{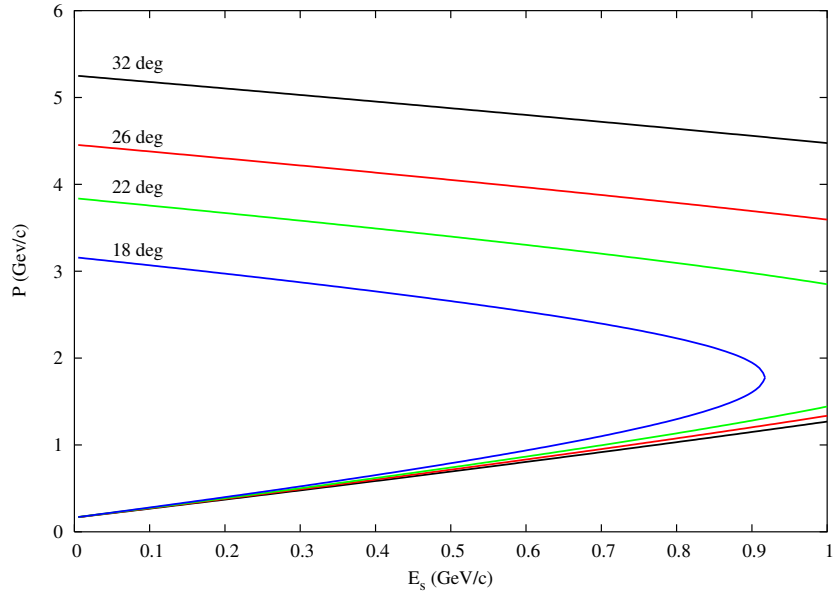}
\caption[F($|$\textbf{q}$|$,y) region of integration for
  $^{12}$C.]{The region of integration (to the left of each curve) for Eq.~\ref{qecs_limits} shown for
selected angle settings, $^{12}$C target.  The $E'$ value for each angle was calculated
with $y_1$=-0.2 GeV/c, E$_s^{min}$ of 0.016 GeV, and initial energy of 5.766 GeV.}
\label{integration_region}
\end{figure*}

It is more convenient to rewrite the spectral function in terms of
$E_s$, the separation energy, rather than $E_0$, the nucleon's
initial energy.  The spectral function becomes:
\begin{equation}
S(E_s, p_0)dE_s=-S'(E_0, p_0)dE_0
\end{equation}
where the Jacobian from the $E_0\rightarrow E_s$ transformation has
been absorbed into the new definition of $S$.  Next, we define
$\tilde{\sigma}=\tilde{\sigma_0}\cdot K$, where $K=E_N/|q|$, and the cross section
becomes:
\begin{equation}
\label{qecs_limits}
\frac{d^2\sigma}{dE' d\Omega}=2\pi \int^{E^{max}_s}_{E^{min}_s} \int^{p_0^{max}(E_s)}_{p_0^{min}(E_s)}\tilde{\sigma}\cdot S(E_s,p_0)\cdot p_0\:dp_0\:dE_s
\end{equation}
The limits of the $p_0$ integral are given by  $|y_1|$ and $|y_2|$,
the solutions of $p_0$ to
Eq.~\ref{cons2}, for the case where $\vec{p_0}$ and $\vec{q}$ are
parallel.  An example of the integration region of Eq.~\ref{qecs_limits} is shown in Fig.~\ref{integration_region}.  The $E_s=E_s^{max}=\sqrt{(M_A+\nu)^2-q^2}-M_A$ limit (for
which $p_0^{max}=p_0^{min}$) occurs when the struck
nucleon is at rest in its final state and $E_s=E_s^{min}$ occurs when
the recoiling nucleus is in its ground state.  In the latter case, the
minimal longitudinal momentum also defines the $y$ scaling variable.
\begin{equation}
y=\frac{\vec{q}( \vec{q}\: ^2-W')+\sqrt{\vec{q}\: ^2
    (W'-\vec{q}\:^2)^2-(W^2-\vec{q}\:
    ^2)\cdot(4W^2M^2_{A-1}-\vec{q}\:^4+2W'\vec{q}
    \:^2-W'^2)}}{2(W^2-\vec{q}\: ^2)}
\end{equation}
where $W$ and $W'$ are defined as:
\begin{eqnarray}
W &=& M_A+E-E'-E_s \nonumber\\
W'&=& W^2+M^2_{A-1}-M^2\nonumber
\end{eqnarray}
\begin{figure*}
\center
\includegraphics[width=.7\textwidth]{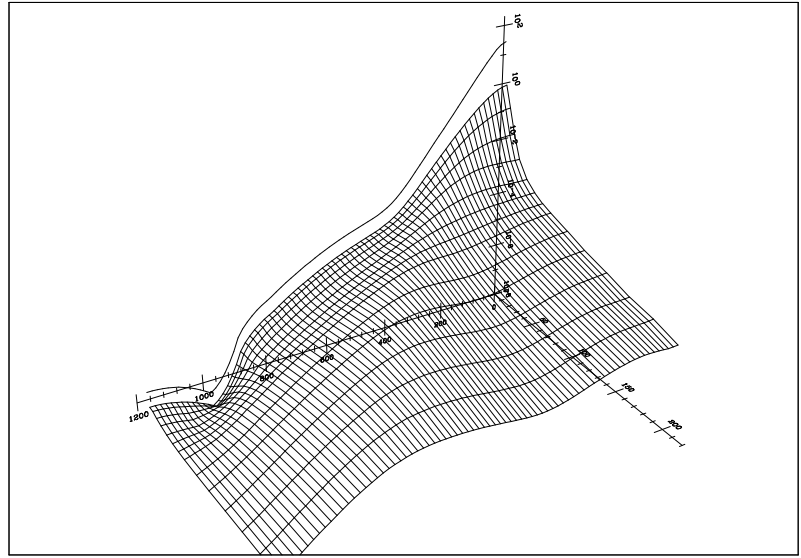}
\caption[Theoretical calculation for the $^3$He spectral function.]{Theoretical calculation by the Hanover group of the spectral
  function for $^3$He, with separation energy (GeV) on the right-hand
  axis and momentum (GeV/c) on the left-hand axis. The calculation was performed using the Fadeev
  method and the Paris potential. Contributions from correlated nucleon pairs
  with isospin=0 in the final state are included~\cite{MeierHajduk:1983gd}.}
\label{he3_spect_func}
\end{figure*}
The expression in Eq.~\ref{qecs_limits} can be further simplified.  Since
$\tilde{\sigma}$ varies slowly over the breadth of the spectral
function,  we can remove it from the integral and evaluate it at the
kinematics corresponding to the lower integration limits, when the spectral function is at maximum within the limits.  It is useful to be able to visualize a
nuclear spectral function, an example of which is seen in
Fig.~\ref{he3_spect_func} for $^3$He, and the curve corresponding to
single nucleon knockout is clearly visible along with a continuum
corresponding to higher removal energies.  The rapid
decrease of the spectral function with energy and momentum also lets us extend the upper limits
of the integrals to infinity. The error associated with this decreases
quickly with increasing values of Q$^2$. Finally, since the spectral
function is peaked around $E_s=E_s^{min}$ for all momenta, we can also approximate the
lower limit of the momentum integral with a constant value, 
$|y_1(E_s^{min})| = |y|$.   These simplifications yield:
\begin{equation}
\frac{d^2\sigma}{dE' d\Omega}=2\pi\:\tilde{\sigma_0}\:
K \int^{\infty}_{E^{min}_s}
\int^{\infty}_{|y|}S(E_s,p_0)\cdot p_0\:dp_0\:dE_s
\label{pwia_sigma_simplified}
\end{equation}
\subsection{\label{yscaling}y-scaling}
\begin{figure*}
\center
\includegraphics[angle=270,clip=true, trim=2.5in 0in 0in 0in, width=\textwidth]{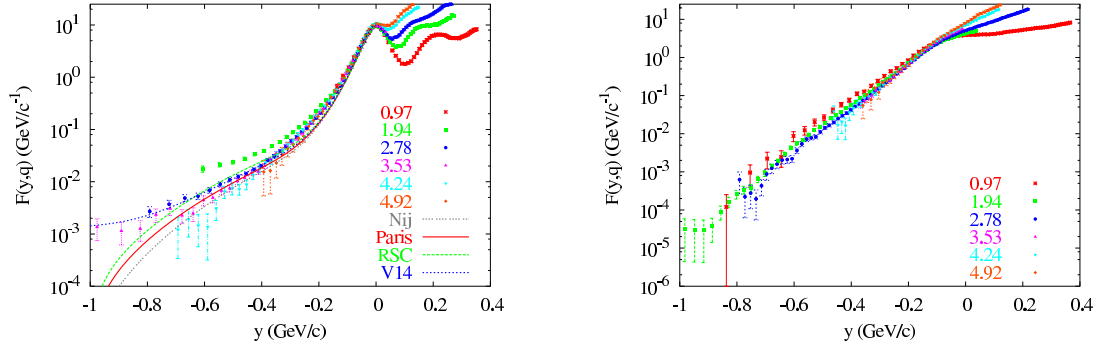}
\caption[F(y,$|$\textbf{q}$|$) for $^{2}$H and $^{12}$C from Jlab experiment
E89-008.]{F(y,$|$\textbf{q}$|$) extracted from deuterium cross
  sections for E89-008.~\cite{Arrington:1998ps}.  The numbers indicate
  the Q$^2$ values of the different data sets at the quasielastic
  peak.  The inelastic
  contribution has not been subtracted, but should be small for $y<$0.}
\label{yscaling_4gev}
\end{figure*}

Having reduced the nuclear quasielastic cross section to the form
given by Eq.~\ref{pwia_sigma_simplified}, it's now possible to examine
its scaling behavior and the approach to it as a function of Q$^2$.
Scaling is the name given to the reduced cross
section when its behavior becomes independent of Q$^2$~\cite{Day:1990mf}.
The double integral over the spectral function in
Eq.~\ref{pwia_sigma_simplified} defines the nuclear scaling function
$F(y,|\textbf{q}|)$:
\begin{equation}
F(y,|\textbf{q}|)=2\pi\int^{\infty}_{E^{min}_s}
\int^{\infty}_{|y|}S(E_s,p_0)\cdot p_0\:dp_0\:dE_s
\label{nk_fy}
\end{equation}
Note that the energy integral over the spectral function alone yields
a momentum distribution of the nucleons inside the nucleus.  This
simplified expression is a common definition of $F(y,|\textbf{q}|)$ and is written as:
\begin{equation}
F(y,|\textbf{q}|)=2\pi\ \int^{\infty}_{|y|}n(p_0)\cdot p_0\:dp_0
\label{fy_nk_simple}
\end{equation}
This scaling function can be extracted from experimental data provided
that the assumptions that went into its derivation apply:  the
remainder nucleus is in its ground state, there's no contribution from
inelastic processes or Final State Interactions, and there's no effect
from the nuclear medium modification.
Scaling functions for $^2$H and $^{12}$C extracted from previous data~\cite{Arrington:1998ps}
can be seen in Fig.~\ref{yscaling_4gev}.

%
%
It should be noted that the spectral function can be broken into 2 parts, with one,
$S_{gr}(E_s,p_0)$ corresponding to the probability distribution of having
the recoil nucleus in the ground state, and the other, $S^*(E_s,p_0)$
corresponding to an excited state ~\cite{CiofidegliAtti:1988ar}.  The
scaling function is then rewritten to reflect this:
\begin{equation}
F(y,|\textbf{q}|)=2\pi \int^{\infty}_{|y|} n_{gr}(p_0)\: p_0\: dp_0-B(y)
\end{equation}
where $B(y)$ is the contribution given by the integral over $S^*(E_s, p_0)$ and is
referred to as the \textit{binding correction} to the
scaling function.  This contribution has a Q$^2$-dependence for small
momentum transfers and breaks the direct relationship between $F(y)$
and $n(p_0)$. When the binding correction is absent, as in the
case of deuterium, then the simplified relationship of
Eq.~\ref{fy_nk_simple} between the
scaling function and the momentum distribution holds.  

\begin{figure*}[h!]
\center
\includegraphics[width=\textwidth]{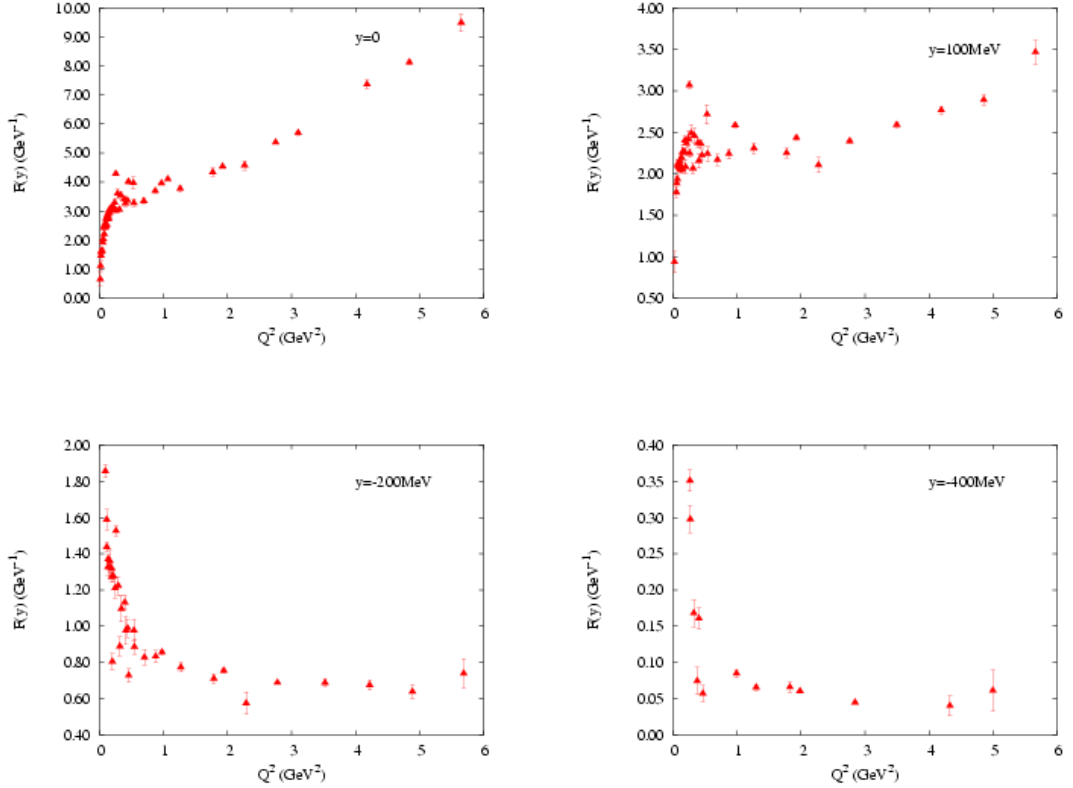}
\caption[Approach to scaling for $^{12}C$.]{Approach to scaling of F(y,$|$\textbf{q}$|$) for $^{12}$C.  F(y,$|$\textbf{q}$|$)
 was extracted from available inclusive data ~\cite{Benhar:2006er} for a variety of
 $y$ values. As $y$ becomes more negative, the inelastic contribution,
 which is largest at high Q$^2$, decreases rapidly and the data appear
 to approach an asymptotic limit.  The steep fall-off with Q$^2$
 reflects the fact that FSIs also fall with Q$^2$. }
\label{scaling_limit_q2}
\end{figure*}

A more serious roadblock to extracting momentum distributions is the
contribution from Final State Interactions between the struck nucleon
and the remainder nucleus, within a distance of 1/$|\textbf{q}|$.  In the presence FSIs, one can still
extract nucleon momentum distributions from the data if the asymptotic
value of the scaling function can be determined from data taken at
finite momentum transfer.  This is done by examining the Q$^2$ limit
of the scaling function for constant $y$-values as shown in
Fig.~\ref{scaling_limit_q2}.

However, while the fall off at low values of Q$^2$ is steep, it
continues for high values of Q$^2$ as well, but at a slower rate.  The fast fall off at
low Q$^2$ is due to the effects of FSIs, whereas the further
falloff has been theorized to be the result of the near cancellation of 
contributions from FSIs and the high removal energy tail of the
spectral function ~\cite{CiofidegliAtti:1990rw}.  

\begin{figure*}
\center
\includegraphics[angle=270,width=\textwidth]{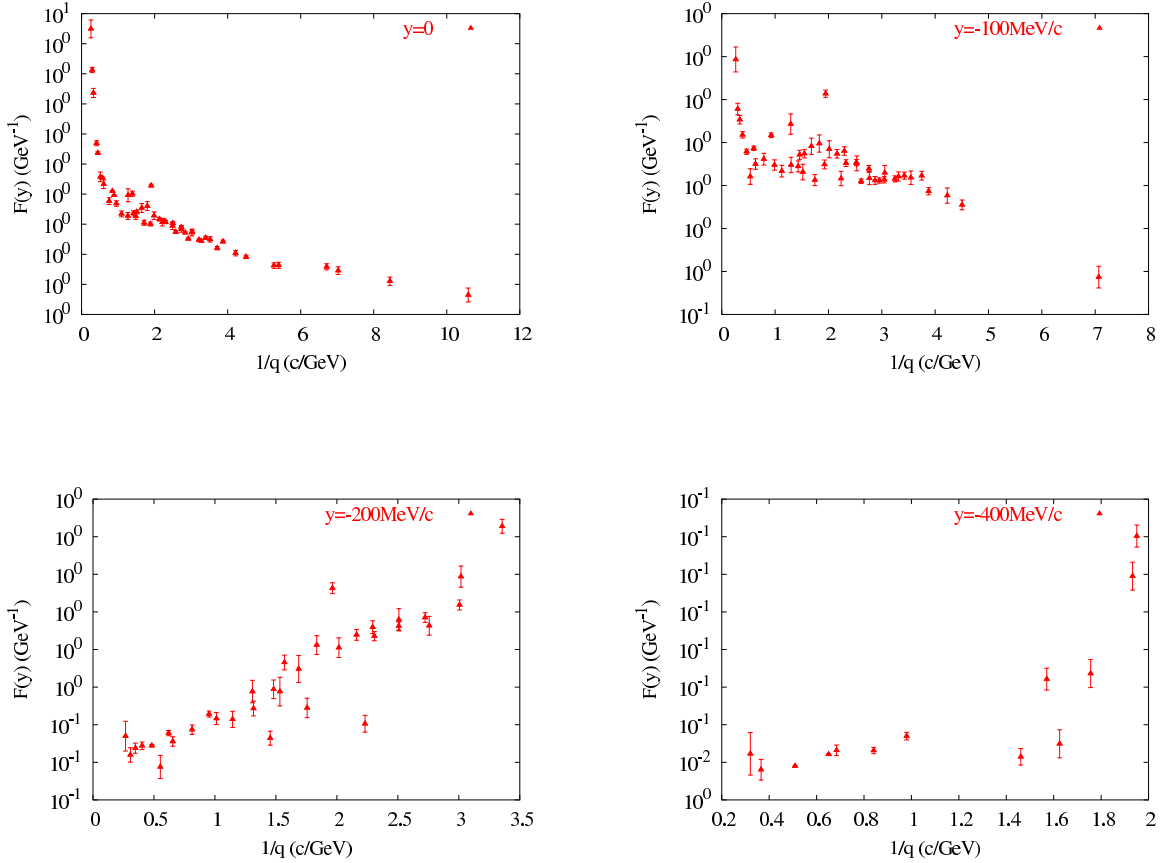}
\caption[F($|$\textbf{q}$|$,y) as a function of 1/q for $^{12}$C.]{Experimental scaling function, F(y,$|$\textbf{q}$|$) for $^{12}$C
  as a function of the inverse of momentum transfer for a variety of
 $y$ values. For the low $y$ values, the inelastic contribution is not
  negligible.}
\label{scaling_limit_1q_linear}
\end{figure*}

In order to get the asymptotic limit of the scaling function from the experimental data
in the presence of FSIs, one can follow the approach described in
~\cite{CiofidegliAtti:1989qs} and rewrite F($|$\textbf{q}$|$, y) as a power
series of $1/q$.  
 This expansion in $1/q$ yields:
\begin{equation}
\label{fy_expanded}
F(y,|\textbf{q}|)=F(y)+\frac{F_{(-1)}(y)}{q}+\frac{F_{(-2)}(y)}{q^2}+\frac{F_{(-3)}(y)}{q^3}+...
\end{equation}
For large values of $q$, the scaling function can be approximated by
the first two terms on the right hand side, with the first term being
the asymptotic limit of the scaling function and the second term
representing the effects of FSIs (as do the other $q$-dependent terms,
which can be neglected in this limit).  If we now plot the
experimental scaling function versus $1/q$ for fixed values of $y$ (Fig.~\ref{scaling_limit_1q_linear}),
the extrapolated y-intercept should represent the asymptotic scaling function for
that $y$ value.
\begin{figure*}
\center
\includegraphics[width=\textwidth]{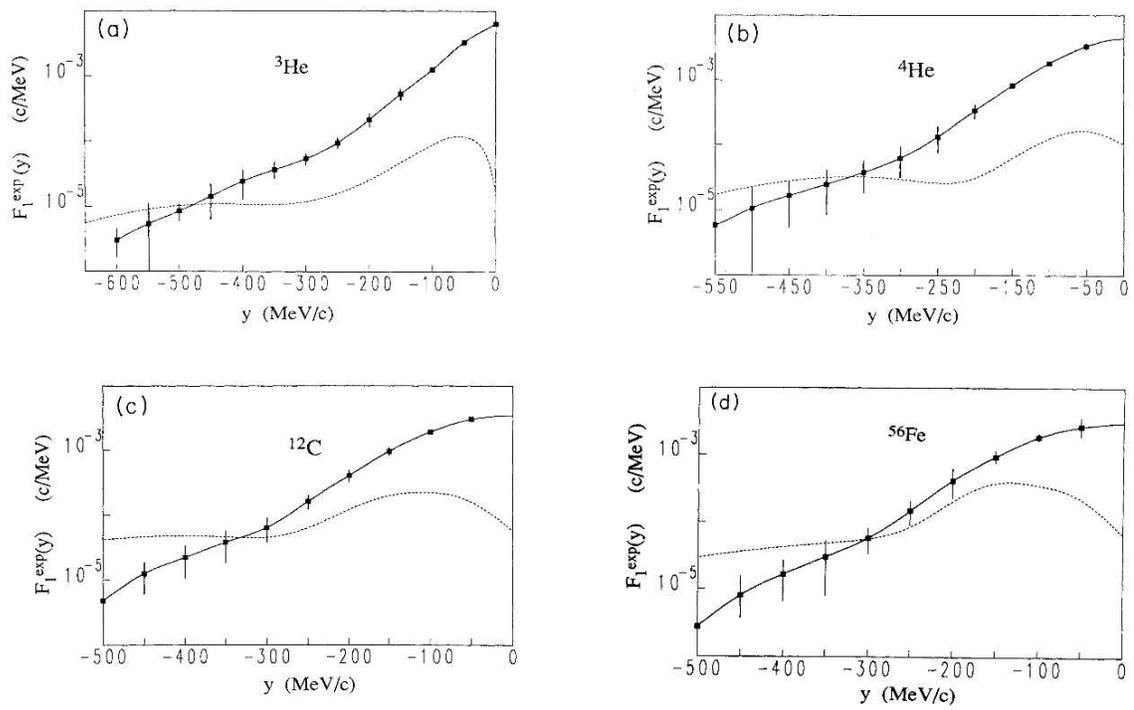}
\caption[Momentum Distribution for $^3$He, extracted from the
  asymptotic limit of F($|$\textbf{q}$|$,y)]{Momentum Distribution for $^3$He, extracted from the
  asymptotic limit of F($|$\textbf{q}$|$,y) ~\cite{CiofidegliAtti:1989qs}.
  The data were analyzed ignoring binding effects. See reference for more details.}
\label{scaling_limit_q2_linear}
\end{figure*}

Finally, with the asymptotic value of the scaling function in hand,
one can go on to extract the momentum distributions provided that the
binding correction, can be calculated or at least, estimated.
A calculation of the spectral function is needed, the accuracy of
which decreases with the increasing complexity and size of the
nuclei~\cite{CiofidegliAtti:1989qs}.
\paragraph{Extracting Scaling Functions from Data \label{extract_fy_section}}
To extract the scaling function $F(y,|\textbf{q}|)$ from the measured data cross
sections, we need a model of off-shell electron-nucleon cross
sections.  For this analysis, the we used the $\sigma^{cc}_1$ formalism
described by DeForest in ~\cite{DeForest:1983vc}.  In this
prescription, the contribution to the cross section from a single
bound and moving
nucleon is given by:
\begin{eqnarray}
\label{sigcc}
\lefteqn{\sigma_{(ep|en)}=\frac{\sigma_{mott}}{\bar{E}E_N}\{(F_1+F_2) \cdot
\left[\frac{\bar{Q}^2}{2} \tan ^2
  \frac{\theta}{2}+\frac{Q^2}{4q^2}(\bar{Q}^2-Q^2)\right]+} \nonumber \\
& & (F_1+\frac{\bar{Q}^2}{4M^2}
F_2)\cdot\left[\frac{Q^4}{4q^4}(\bar{E}+E_N)^2+(\frac{Q^2}{q^2}+\tan^2\frac{\theta}{2})p'^2\sin^2\theta\right]\}
\end{eqnarray}
where $\bar{E}=\sqrt{(\vec{p}-\vec{q})^2+M^2}$,  $Q^2=q^2-\nu ^2$,
$\bar{Q}^2=Q^2-(E'-\bar{E})^2$, and the Mott cross section is given
by:
\begin{equation}
\sigma_{mott}=\frac{\alpha^2(\hbar c)^2\cos^2\frac{\theta}{2}}{4E^2\sin^4\frac{\theta}{2}}
\end{equation}
and $F_1$ and $F_2$, the Pauli and Dirac form factors, are related to
the Sachs nucleon form factors through
the following relationships:
\begin{eqnarray}
F^{(p|n)}_1 &=& \frac{G_e^{(p|n)}+\tau G_m^{(p|n)}}{1+\tau} \nonumber \\
F^{(p|n)}_2 &=& \frac{G_m^{(p|n)}- G_e^{(p|n) }}{1+\tau}  \\
\tau &=& \frac{Q^2}{4M^2}
\end{eqnarray}

To extract scaling functions from data, we use Eq. \ref{pwia_sigma_simplified} and
solve for $F(y,|\textbf{q}|)$:
\begin{equation}
F(y,|\textbf{q}|)=\frac{d^2 \sigma}{dE'd \Omega} \cdot \frac{1}{Z\sigma_{ep}+N\sigma_{en}}\frac{|\textbf{q}|}{\sqrt{M^2+(\vec{p}+|\textbf{q}|)^2}}
\end{equation}
Table \ref{sep_es} lists the separation energies used in this data
analysis for the various targets.  This derivation of $y$-scaling
using the PWIA demands the use of the minimum, rather than
the average separation energies~\cite{Moniz:1971mt}, which have been used in past analyses.

\subsection{Other Scaling functions - Superscaling}

The $y$-scaling described above has been called scaling of the first
kind in literature ~\cite{Donnelly:1999sw}.  This refers to the
\textit{q}-independence of the reduced response (in this case, the
scaling function $F(y,|\textbf{q}|)$, obtained from the measured cross section).  To observe scaling of the second kind, it
is necessary to incorporate the momentum scale of a given nucleus into
the definition of the scaling variable and into the reduced response.
Subsequently, the scaling that is observed is independent of the
target nucleus - i.e. all nuclei lie on the same scaling curve.
Observation of both kinds of scaling is referred to as \textit{superscaling}. Two
superscaling variables are defined in Ref.~\cite{Maieron:2001it}, $\psi$ and
$\psi '$.  The first variable, $\psi$, is equivalent to $y(E_s=0)/k_F$ when
the separation energy (Table~\ref{sep_es}) is not included in the
calculation.  On the other hand, $\psi'$ does take the shift in energy (adjusted $E_s$
value), which is chosen ``empirically'' (Table ~\ref{eshift}) by the
authors of~\cite{Maieron:2001it}, into
account and is then defined as:
\begin{eqnarray}
\psi ' &=&\frac{1}{\sqrt{\xi _F}} \frac{\lambda ' -\tau
  '}{\sqrt{(1+\lambda)\tau '+\kappa \sqrt{\tau ' (\tau
  '+1)}}} \\
&\rm{with} \nonumber \\
\xi _F&=&\sqrt{1+\eta_F^2}-1, \:\:  \eta_F\equiv \frac{k_F}{m_N} \nonumber\\
\lambda ' & \equiv &\frac{\nu-E_{shift}}{2m_N} \nonumber\\
\vec{\kappa} &\equiv & \frac{\vec{q}}{2m_N} \nonumber\\
\tau ' & \equiv & \kappa ^2 -{\lambda '} ^2 \nonumber
\end{eqnarray}
where the primed quantities have had the energy shift applied to
them  and $\xi_F$ is a dimensionless scale of the fermi momentum
relative to the nucleon mass, $m_N$. Similar to the $y$-scaling analysis, the superscaling function
$F(\kappa, \psi)$ is extracted by taking the measured differential
cross section and dividing out the single-nucleon elastic
cross section (see Eq.13 in~\cite{Maieron:2001it}).  

\begin{figure*}
\center
\includegraphics[width=\textwidth,clip]{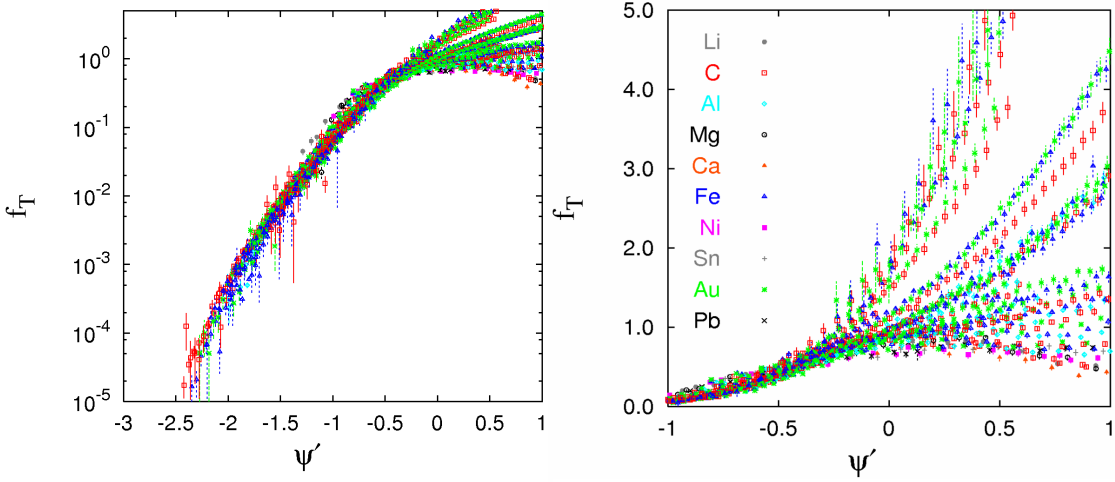}
\caption[Transverse component of the superscaling function, $f_T$.]{Transverse Scaling function, $f_T$ for a variety of nuclei
  and kinematics extracted here ~\cite{Maieron:2001it}.  The longitudinal contribution has been subtracted
  using a superscaling model discussed in the same publication.}
\label{superscale_transverse}
\end{figure*}

It is then also possible to extract additional superscaling
functions, $f_T(\psi ')$ and $f_L(\psi ')$, when the separated
transverse and longitudinal contributions to the cross section are available.
  Fig.~\ref{superscale_transverse} shows the transverse superscaling
  function extracted for data from a variety of experiments ~\cite{Maieron:2001it}.
\begin{table}[h!]
\begin{center}
\caption{Adjusted Separation Energies and Fermi momenta used in the
  extraction of superscaling functions ~\cite{Maieron:2001it}.  The
  answers are especially sensitive to $k_f^{adj}$.}
\vspace*{0.25in}
\begin{tabular}{|c|c|c|}
\hline
Target &
$E_s^{adjusted}$ (MeV) & $k_f^{adjusted}$ (MeV/c) \\
\hline
$^{4}$He & 15 & 200\\
$^{12}$C & 20 & 228\\
$^{63}$Cu & 23& 241 \\
$^{197}$Au & 25 & 245\\
\hline
\end{tabular}
\label{eshift}
\end{center}
\end{table}
For the kinematic coverage of E02-019, the longitudinal contribution
is negligible, so the entire response is taken to be transverse.  It is important to note that the extracted superscaling
functions are extremely sensitive to the input values of $k_F$ and the
use of conventional values, which differ by as little as 10 MeV from
the values given in Ref.~\cite{Maieron:2001it} will prevent
superscaling from being observed.
\subsection{Other Scaling variables - $y_{cw}$ \label{ycw_section}}
As was discussed in Sec.~\ref{yscaling}, the (A-1) nucleus can be
left in an excited state and therefore a non-zero binding correction
exists. This means that $y$, which is the longitudinal momentum, is
quite different for weakly bound nucleons than for strongly bound
ones.  As a result, the scaling function is not necessarily related to
the longitudinal momentum for large negative values of $y$, where the
response is dominated by the electron scattering from strongly bound, correlated nucleons.
An alternate scaling variable, $y_{cw}$, was introduced by Ciofi degli Atti
and Faralli
~\cite{Faralli:1999dk} with the hope of being able to represent the
longitudinal momenta of weakly and strongly bound nucleons equally
well.  This definition includes the excitation energy of the daughter
nucleus (A-1), which is a
function of the relative momentum of the correlated pair and its
momentum in the CM frame and is given by~\cite{Faralli:1999dk}:
\begin{equation}
\langle E^*_{A-1}(k) \rangle \simeq \left(\frac{A-2}{A-1}\right)\frac{\rm{\textbf{k}}
^2}{2M}+b_A-c_A \mid \rm{\textbf{k}} \mid
\label{y_cw_excit}
\end{equation}
where the parameters $b_A$ and $c_A$ describe the CM motion of the
correlated pair.
Additionally, this energy is shifted by the average shell-model
removal energy, $\langle E_{gr}\rangle$, which is obtained using the Koltun~\cite{Koltun:1972kh} sum
rule.  The new scaling variable, $y_{cw}$, is then
\begin{equation}
y_{cw}=\frac{\tilde{q}}{2}+\sqrt{\frac{\tilde{q} ^2}{4}-\frac{4\nu
    _A^2 M^2-W_A^4}{4W_A^2}}
\label{ycw}
\end{equation}
where $\nu_A=\nu+2M-E_{th}^{(2)}-b_A+\langle E_{gr} \rangle$,  $E_{th}^{(2)}$ is the
two-body break-up energy, $W_A^2 \equiv \nu _A^2 -q^2$, and
$\tilde{q}=q+c_a \nu _A$.  At low values, y$_{cw}$ approximately reduces to the
usual scaling variable, $y$.

\begin{figure*}[htpt]
\center
\includegraphics[width=\textwidth]{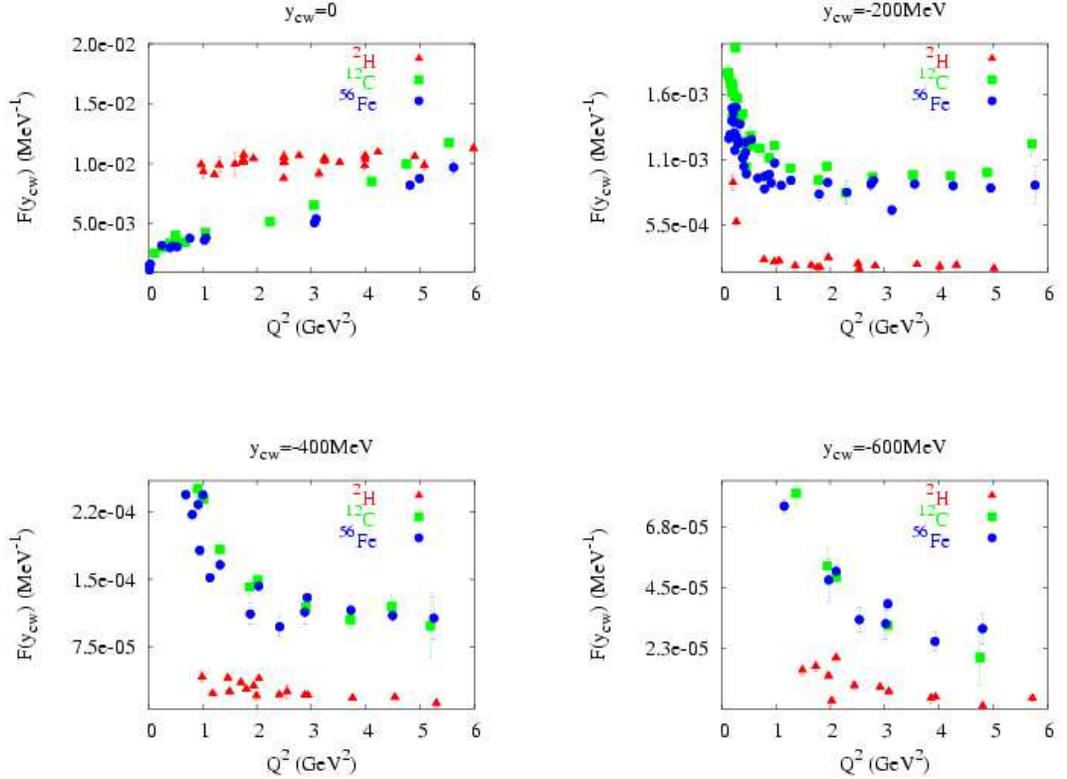}
\caption[F($|$\textbf{q}$|$,y) as a function of $y_{cw}$ for several targets.]{The scaling functions of $^2$H, $^{12}$C, and $^{56}$Fe for
  existing quasi-elastic world data plotted for fixed values of
  y$_{cw}$ as a function of Q$^2$.  Scaling is approached from above
  (decreasing F($|$\textbf{q}$|$,y) with increasing Q$^2$),
  which is also the case with $y$, suggesting that the main scale-breaking
  mechanism is FSIs. }
\label{ycw_example} 
\end{figure*}

Using $y_{cw}$ allows us to establish a more direct link between the
scaling function and the nucleon momentum distributions as well as to
disentangle the contributions from binding effects and FSIs.  The binding effects are taken
into account in the definition of the new scaling variable, $y_{cw}$, rather than as
an additional calculated correction to the scaling function.

 Another useful feature of this
scaling variable is that it allows one to obtain scaling functions for
heavier nuclei by
scaling up the deuteron's F($|$\textbf{q}$|$,y$_{cw}$) by a
target-specific constant, $C_A$~\cite{Faralli:1999dk}.
 
 The data, when analyzed using y$_{cw}$, Fig.~\ref{ycw_example},
 clearly shows the approach to scaling from above, indicating the fall-off of FSI effects with
increasing values of Q$^2$, as is the case with $y$.

\section{Inelastic Cross section\label{inel_theory_section}}

In inclusive inelastic electron nucleus scattering, the only
information available about the final state is its
invariant mass,
\begin{equation}
W^2=M^2+2M\nu-Q^2.
\end{equation}
The most general expression for an inclusive cross section is:
%
\begin{equation}
\frac{d^2\sigma}{dE' d\Omega}=\frac{\alpha ^2}{Q^4}\frac{E'}{E}L_{\mu
  \nu}W^{\mu \nu},
\label{sigma_general_inclusive}
\end{equation}
where $L_{\mu \nu}$ and $W^{\mu \nu}$ are the leptonic and
hadronic tensors, respectively.
Using the target tensor as given in~\cite{Itzykson:1980rh}, it can be
shown that the
unpolarized cross section for electron scattering from a nuclear
target is then:
\begin{equation}
\frac{d^2\sigma}{dE'd\Omega}=\frac{d\sigma}
{d\Omega}_{Mott}\left[W_2^{A}(\nu,Q^2)+2W_1^{A}(\nu,Q^2)\tan
  ^2\frac{\theta}{2}\right]
\label{inelastic_sigma_simple}
\end{equation}
Thus, the inelastic cross section is characterized by two structure
functions, $W^A_1$ and $W^A_2$, which are functions of momentum and energy
transfers:
\begin{eqnarray}
\nu &=& E-E'=\frac{\vec{q}\cdot\vec{p}}{M} \nonumber \\
Q^2 &=& -q^2=4EE' \sin ^2 \frac{\theta}{2},
\end{eqnarray} 
both of which are evaluated in the lab frame. 

 The nuclear structure functions can be
related to the spectral function and the nucleon structure functions in
the following way:~\cite{Benhar:1997vy}
\begin{eqnarray}
\lefteqn{W_1^A(Q^2,\nu)=\int
d^4p_0\:S'(E_0,p_0)\left(\frac{M}{\tilde{p_0}}\right)\left[W_1^N+\frac{W^N_2}{2M^2}\frac{|\textbf{p$_0$}\times\textbf{q}|^2}{|\textbf{p$_0$}|^2}\right]}
\\
\lefteqn{
W_2^A(Q^2,\nu)=\int
d^4p_0\:S'(E_0,p_0)\nonumber} \\
&&\times\left\{W_1^N\frac{q^2}{|\textbf{q}|^2}\left(\frac{q^2}{\tilde{q}^2}-1\right)+\frac{W_2^N}{M^2}\left[\frac{q^4}{|\textbf{q}|^4}
\left(\tilde{E_0}-\tilde{\nu}\frac{\tilde{q} \cdot
  \tilde{p_0}}{\tilde{q^2}}\right) -\frac{q^2}{|\textbf{q}|^2}\frac{|\textbf{p$_0$}\times\textbf{q}|^2}{|\textbf{p$_0$}|^2}\
\right]\right\}\nonumber
\end{eqnarray}
The quantities denoted with tilde refer to the free nucleon, rather
than the bound one.  For a detailed explanation, refer to ~\cite{Benhar:1997vy}.
In the limit of high energy and momentum transfers, the nucleon structure functions 
 are reduced to being functions
of $x=\frac{Q^2}{2M\nu}$, which is the fraction of the total
momentum of the nucleon that is carried by the struck quark.  In this
limit, the structure function is related to the momentum distribution
of the quarks, which is different than that of quarks in a free
 nucleon due to presence of other nucleons.

Our analysis of the inelastic electron-nucleon cross section is done through the study
of the F$_2$ structure function, which can be extracted from
the data provided that the ratio, $R$, of the longitudinal cross
section to the transverse cross section is known.  Using a simple
parametrization of $R=0.32\rm{GeV}^2/Q^2$~\cite{Bosted:1992fy}, we can extract $F_2^A$:
\begin{equation}
F_2^A=\frac{d^2 \sigma}{d\Omega dE '} \cdot
\frac{\nu}{\sigma_{mott} [1+2 \tan ^2(\frac{\theta}{2})\frac{1+\nu ^2/Q^2}{1+R}]}
\end{equation}
Scaling of $F_2^A$ has been observed as a function of $x$ in the
deep inelastic region, as discussed previously (see Sec.~\ref{inelastic_intro}).

%
%

%
\subsection{$\xi$-scaling\label{xsi_scaling_section}}
Inclusive electron nucleus data can also be analyzed in terms of 
the Nachtman variable, $\xi=\frac{2x}{1+\sqrt{1+4M^2x^2/Q^2}}$, which
is analogous to $x$ in that it is the fraction of the nucleon momentum
carried by the struck quark, but target mass
effects are not neglected ~\cite{Nachtmann:1973mr}.  In the $Q^2 \rightarrow \infty$ limit,
$\xi$ reduces to $x$.  This means that the scaling of the structure
functions that is observed in $x$ in the deep inelastic region should
also be observed in $\xi$ in the same kinematic region.  When the data
were examined in this way~\cite{Filippone:1992iz}, approximate $\xi$-scaling was observed for all values of $\xi$, including the
quasielastic region where it was not expected.  One explanation ~\cite{Benhar:1995rh} of this scaling in $\xi$ was
through the observation that $\xi$ can
be expanded as a function of $y$ at high values of Q$^2$, so data can
be expected to show the same kind of scaling in $\xi$ as it does in
$y$ in a purely quasi-elastic kinematic region.  The authors suggest
that the relationship
between the two scaling variables masks the presence of Final State
Interactions at high $x$ and the observed $\xi$ scaling is an
accident.  This explanation is not very satisfying, as the reduced
responses (F($|\textbf{q}|$,y) and F$_2^A$) are obtained in different
ways and do not represent the same quantity, so the fact that the
variables in which they scale can be related to each other is not very illuminating.
\begin{figure*}[h!]
\center
\includegraphics[width=0.7\textwidth]{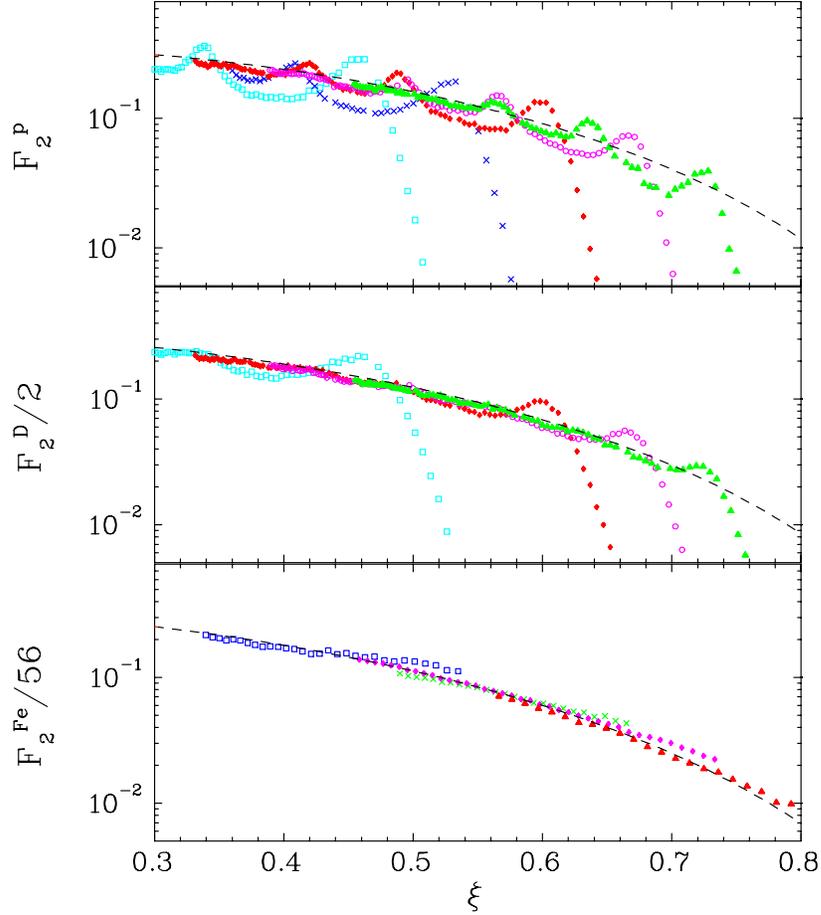}
\caption[F$_2$ vs $\xi$ for H, $^2$H, and $^{56}$Fe along with the
  MRST fit.]{F$_2$/A vs $\xi$ for H (top), $^2$H (middle), and $^{56}$Fe
  (bottom) along with the
  MRST~\cite{Martin:2002aw} parametrization of the structure functions.  For
  $^{56}$Fe, the quasielastic peak was removed with a $W^2>1.2$
  GeV$^2$ cut.  Figure from ~\cite{Arrington:plot}}
\label{xi_scaling_john} 
\end{figure*}
Local duality~\cite{Bloom:1971ye,Arrington:2003nt} has been suggested as another
explanation for the observation of $\xi$ scaling. When the structure function $\nu W_2$ is
examined in terms of the dimensionless variable $w
'=1/x+M_N^2/q^2$ or $\xi$, it
is found to have the same Q$^2$ behavior as the resonance form
factors.  In fact, averaging over the resonance peaks produces the
scaling curve for the structure functions.  The resonance peaks do not
disappear with increasing values of Q$^2$, but decrease in strength
and move to lower values of $w'$ (higher $\xi$), following the magnitude of the
scaling limit curve.  For $\xi$-scaling, this
means that the scaling curve observed at large values of $\xi$ is due to
the local averaging of the resonances by the momentum distributions of
the nucleons in the nucleus.  Recall that Bloom and Gilman proposed
that the resonances are not a separate entity, but are an intrinsic
part of the scaling behavior and there's a common origin for both
phenomena.   

\begin{figure*}[h!]
\center
\includegraphics[angle=270,width=0.7\textwidth]{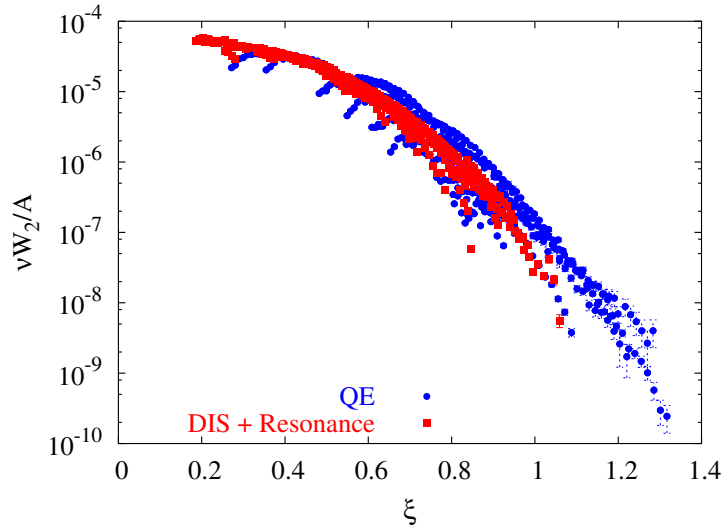}
\caption[Relative contributions to F$_2$ for $^{197}$Au as a function
  of $\xi$.]{$\nu W^A_2$ per nucleon for $^{197}$Au as a function of $\xi$ with the
  quasielastic contribution subtracted (red squares) and with the
  DIS/resonance contribution subtracted (blue circles).
  0.5$<Q^2<$5.75 (GeV$^2$) at the quasielastic peak.  Note that
  neither contribution scales over the whole range in $\xi$.  Data
  shown is from SLAC experiment NE3 and from JLab experiment E89-008~\cite{Benhar:2006er}.}
\label{no_xsi_scaling_for_you} 
\end{figure*}

And, the final explanation for the observed $\xi$-scaling
~\cite{Day:2003eb} proposes that it's purely accidental.  At high
$\xi$ and low Q$^2$, the
DIS and resonance contributions fall at approximately the same rate as the quasielastic
contribution grows and as a result, scaling is observed over the whole
range of $\xi$. 
At high values of $\xi$, the growing contribution from DIS and the
resonances compensates for the fall-off of the quasielastic contribution, which has a
strong Q$^2$-dependence coming from the form factors.  These are
unrelated reaction mechanisms and, when their contributions are examined separately, they
do not scale over the entire range of
$\xi$, as can be seen in Fig~\ref{no_xsi_scaling_for_you}.

\subsection{Final State Interactions}
In the limit of large energy and momentum transfers, the reduced cross
section is expected to scale in both the quasi-elastic and inelastic
regimes, although in different variables.  This expectation is partly based on the assumption that the
reaction is well described by the exchange of a single virtual
photon with a single nucleon that does not react with the A-1 nucleus.
However, final state interaction contributions to the cross section
can arise from the interaction between the struck object and the
remainder nucleus.  It is thought that the inclusive cross section is
only sensitive to these FSIs taken place within a distance of $\approx
1/|\textbf{q}|$, implying that the inclusive cross sections at large momentum
transfer should have little effect from FSIs.  The data tell a
different story.

Existing data show $y$-scaling violations for heavy targets, which
suggests that the PWIA regime has not been reached.  Previous analyses ~\cite{Benhar:1995te}
have suggested that most of the strength in the $x>1.4$ region 
is due to the electron scattering from a low-momentum nucleon and that nucleon's
FSIs with the A-1 nucleus.   O.Benhar's
calculations~\cite{Benhar:2007dd}
show that the contributions from FSIs decrease and approach a limit with increasing Q$^2$
values, but never vanish.

Another school of thought~\cite{Frankfurt:1981mk} states that since
scattering in the $x>1.4$ region takes place from correlated nucleons,
the FSIs are confined to the correlation: i.e. the struck nucleon
interacts only with the other nucleon(s) in the correlation.  This
means that while FSIs exist, they will cancel out in the cross section
ratios of heavy nuclei to light ones.

Consensus has not been reached on the Q$^2$- and $A$-dependence of the
FSIs.  While E02-019 will not resolve this issue, its vast data set
will contribute greatly to the discussion.
\subsection{Short Range Correlations \label{src_section}}
\begin{figure*}[htpt]
\center
\includegraphics[width=0.6\textwidth]{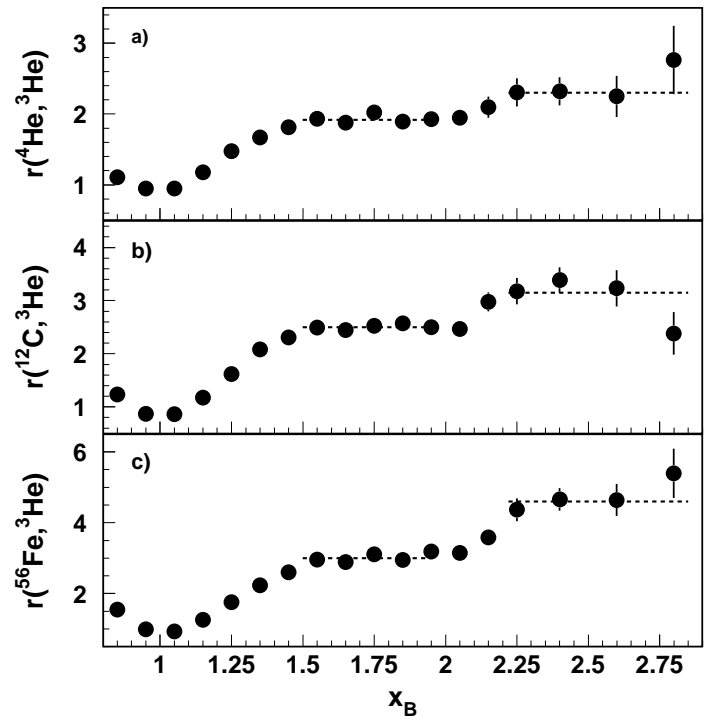}
\caption[Ratios of $^4$He, $^{12}$C, and $^{56}$Fe to $^3$He.]{Ratios of $^4$He, $^{12}$C, and $^{56}$Fe to $^3$He,
  respectively~\cite{Egiyan:2005hs}. The plateaus at 1.5$<x<$2 and 2.25$<x<$3 correspond to
  scattering from 2N- and 3N- correlations, respectively.}
\label{hallb_srcs}
\end{figure*}
A large contribution to the cross section at $x>1$ comes from the
interaction of the virtual photon with a nucleon in a correlated pair.
When the interaction distance between the nucleons becomes smaller than the
average inter-nucleon spacing, the repulsive $NN$ force imparts high
momenta on these nucleons.
The ideal regime for studying SRCs is one in which scattering from low
momentum nucleons is suppressed and the energy transfer is higher than
the kinetic energies of correlated nucleons, i.e. at $x>1$ and
$Q^2>1$.  At the high momentum kinematics, where the mean field
contribution is negligible, the inclusive cross section can be
approximated with terms that describe scattering from nucleons in
multi-nucleon correlations:
\begin{equation}
\sigma_A(x,Q^2)=\sum_{j=2}^{A} \frac{a_{j(A)}}{j} \sigma_j (x,Q^2)
\end{equation}
where $\sigma_A(x,Q^2)$ is the electron-nucleus cross section and $\sigma_j(x,Q^2)$
is the electron-$j$-nucleon-correlation cross section ($\sigma_j(x,Q^2)=0$
for $x>j$), and $a_j(A)$ is proportional to the probability of finding
a nucleon in a $j$-nucleon correlation~\cite{Frankfurt:1981mk}.  For normalization purposes,
$\sigma_2$ and $\sigma_3$ are taken to be the cross sections for
electron scattering from deuterium and $^3$He, respectively and are expected
 to be closely related to the number of 2- and 3- nucleon pairs in
nuclei.  Cross
section ratios of heavy nuclei to light nuclei are expected to scale
(show no dependence on Q$^2$ or $x$ in a given $x$-range)
if the process is dominated by scattering from $j$-nucleon
correlations and the $j$-nucleon correlations in A$>j$ nuclei are
similar to those in the A$=j$ nuclei.

Previous measurements have shown the presence of both 2- and 3-nucleon
correlations at low values of Q$^2$ as is shown in Fig.~\ref{hallb_srcs}.

\chapter{Experimental Details}
\section{Overview}

Experiment E02-019, ``Inclusive Electron Scattering from Nuclei at $x > 1$ and
High $Q^2$ with a 6 GeV beam'', was performed in experimental Hall C
of the Thomas Jefferson National Accelerator Facility (TJNAF) in the fall of 2004. 
All three experimental halls were operational during the running of
E02-019.  Inclusive  scattering of 5.767 GeV electrons
from Deuterium, $^3$Helium, $^4$Helium, Beryllium, Carbon, Copper, and
Gold was measured in the High Momentum Spectrometer (HMS) at 6 scattering angles and
several HMS momentum settings.  Hydrogen data was taken for the purposes of
calibration and normalization.

\section{Accelerator}

\begin{figure*}[h!]
\center
\includegraphics[width=6in]{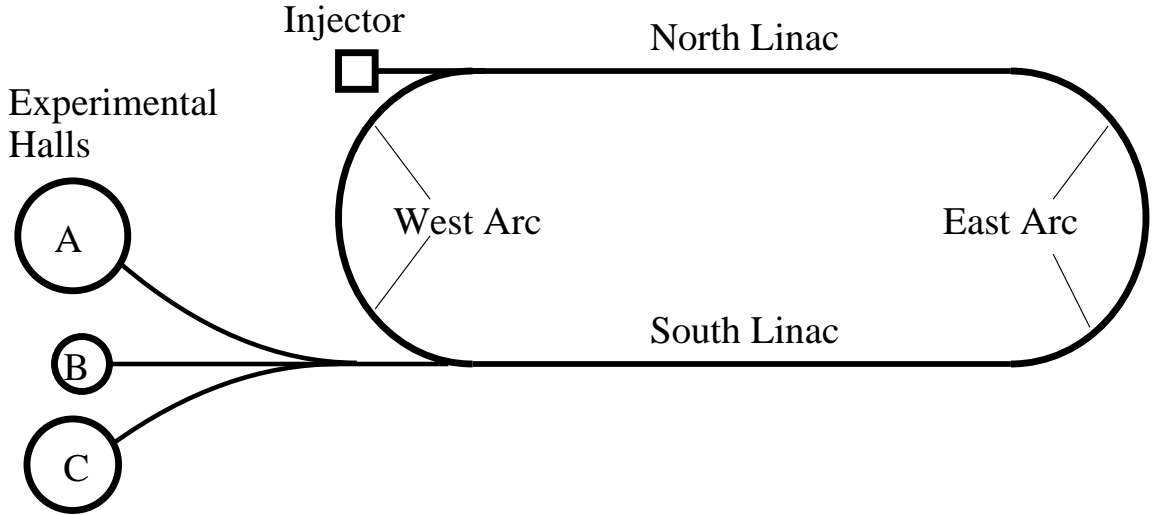}
\caption{Overhead Schematic of the Accelerator and the experimental halls}
\label{machine}
\end{figure*}

For E02-019, CEBAF supplied Hall C with a continuous wave (CW) electron
beam of 5.767 GeV (highest energy available) at currents as high as 80$\mu$A.  The
accelerator complex (Fig. \ref{machine}) consists of an injector, two superconducting linacs, 9
recirculation arcs, a Beam Switch Yard (BSY) and the three
experimental halls (A, B, and C).
The electrons are injected into the North Linac at 67 MeV, where
they're accelerated 570 MeV by superconducting radio frequency
cavities, bent through the east arc, and accelerated an additional 570
MeV in the South Linac.  At this point, the electron beam can be extracted to any one of the three
experimental halls through the BSY or be sent back through the west arc for as many
as 5 total acceleration passes and a maximum beam energy of 5.767
GeV.  The electrons can be extracted at the end of any pass, providing
a beam energy equal to the injector energy plus a multiple of the
linac energy.

As mentioned above, the electrons are accelerated in the linacs using
superconducting radio frequency cavities tuned to 1497 MHz.  The cavities, which are phased for maximum acceleration,  are kept at 2 Kelvin using liquid helium from
the Central Helium Liquefier (CHL).  The beam
has a structure that consists of 1.67 ps pulses coming at 1497
MHz, meaning that each hall receives a 499 MHz beam, or a third of all
the pulses.  Three beams of
different energies and different currents can be delivered into the
halls simultaneously.

\section{Hall C Beamline}

\begin{figure*}
\center
\includegraphics[width=4in, angle=270]{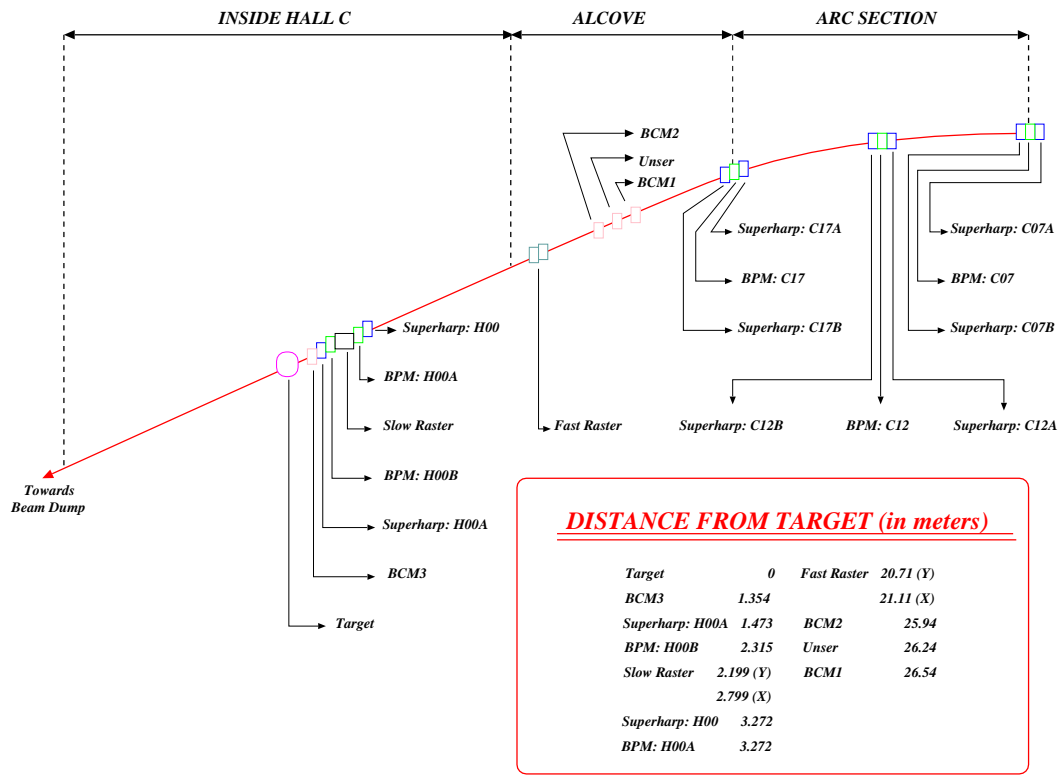}
\caption{Hall C Beamline and Hardware.  See text for details.}
\label{arc}
\end{figure*}

Once the beam arrives at the BSY, it is sent through the Hall C arc
and into the hall.  The Hall C beamline consists of 8 dipoles, 12
quadrupoles, 8 sextupoles, which are used to focus and steer the beam.
The arc is equipped with several devices to measure the profile,
energy, position, and the current of the beam.  Figure
 \ref{arc}
shows the described hardware in the Hall C Arc and beamline.

\subsection{Beam Energy Measurement}

The beam energy measurement was performed by using the Hall C
Arc magnets as a spectrometer ~\cite{Yan:1993fd}.   The position and
direction of the beam are determined at the entrance and
at the exit of the 34.3$^{\circ}$ bend of the Hall C Arc
line by 2 pairs of high resolution harps (wire scanners).  A harp is made up of 3 wires, two of which
are oriented vertically and one horizontally.  All magnetic elements except
for the dipoles are turned off during this measurement.  The beam
is steered so that it follows the central ray of the magnets in
the arc.  Then, the beam momentum (and therefore, energy) can be
determined as a function of the current in the arc dipoles:
\begin{equation}
E\simeq p=\frac{e}{\theta}\int\:\vec{B}\cdot\vec{dl}, 
\label{beamenergy}
\end{equation}
where $\theta$ is the arc bend angle , $e$ is the electric charge, and
$\int\:\vec{B}\cdot\vec{dl}$ is the magnetic field integral over the path of the beam.  For
this measurement, a precise field map of at least one arc magnet is
required.  One of the arc dipoles has been precisely mapped and the
others are assumed to have the same field map, normalized to the
central field value. With this information in hand, the incident beam
energy can be measured with an uncertainty of $\delta p/p \approx$ 2
$\times$ 10$^{-4}$.

The beam energy measurement was performed twice during the running of
E02-019 with agreement of $<0.0007\%$ between the two results.  The
measurements gave beam energies of 5.76713 GeV$\pm$1.65 MeV and
5.76717$\pm$1.65 MeV, respectively.
\subsection{Beam Position Monitors}

During the running of E02-019, the position and trajectory of the beam
were monitoring using 3 Beam Position
Monitors (BPMs): H$\emptyset\emptyset$A, H$\emptyset\emptyset$B, and H$\emptyset\emptyset$C. However, only 2 of them performed reliably (H$\emptyset\emptyset$C
did not).
The BPMs are cavities with four antennae (rotated $\pm 45^\circ$
from the horizontal and vertical directions to minimize synchrotron
damage), sensitive to the relevant harmonic frequency.  The beam
pulse frequencies in the accelerator and in Hall C are
harmonics of the fundamental frequencies of the BPMs.  The amplitude
of the signal from each antenna can be related to the distance between
the beam and that antenna.  Relative beam positions in combinations with superharp
measurements can be used to calculate the absolute beam position.
More information is available in Ref. ~\cite{Gueye:1992wn}.
\begin{table}[h!]
\begin{center}
\caption{Nominal Beam positions for E02-019}
\vspace*{0.25in}

\begin{tabular}{|c|c|c|c|}
\hline
 BPM &
Nominal x-pos (mm) &
Nominal y-pos (mm) &
Distance from target (m) \\
\hline
H00A & 0.80 & -1.83 & 3.272 \\
H00B & 0.48 & -1.50 & 2.315 \\
H00C & 0.48 & -1.08 & 1.360 \\
\hline
\end{tabular}
\label{bpms}
\end{center}
\end{table}

\begin{figure*}[h!]
\center
\includegraphics[height=4in]{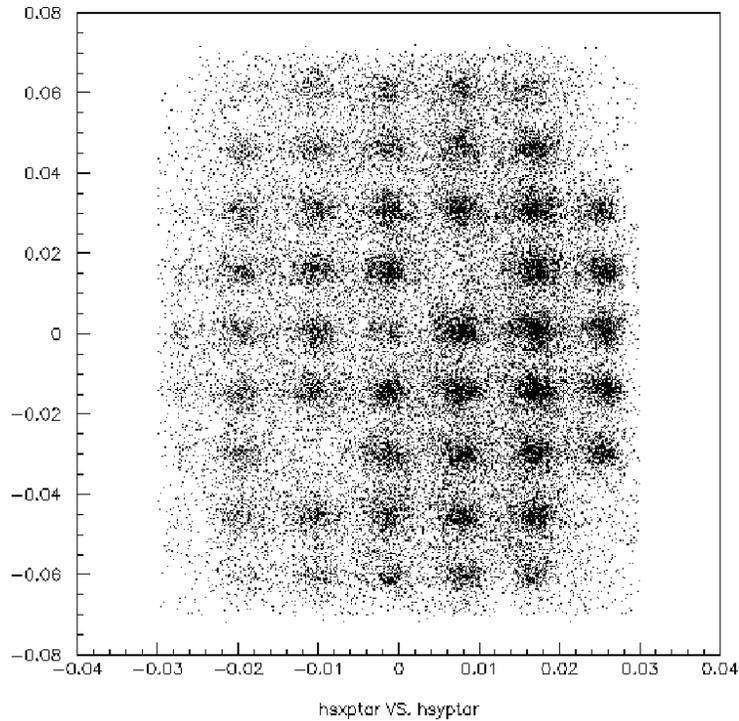}
\caption[Sieve Slit Pattern.]{Sieve Slit Pattern.  The nominal beam position was achieved
  so that the pattern is centered about the small aperture at
  $xptar,yptar=0$, where $xptar$ and $yptar$ are the out-of-plane and
  in-plane angles with respect to the central ray.  Detailed
  descriptions of the target quantities can be
  found in Sec.~\ref{coordinates}. }
\label{sieve}
\end{figure*}

The BPM readouts were closely monitored by the shift workers during
the running of the experiment.  The positions were kept to within
0.1mm of the nominal positions for each BPM (Fig.\ref{bpms}) to make sure that the beam
went through the center of the target and the target thickness
(traversed by the beam) remained constant from run to run.  The
nominal position of the beam was calculated using a sieve slit run, where a HEAVYMET collimator (Tungsten with 10$\%$ CuNi) with grid of apertures is placed in front of the HMS to let the particles through. This
was done by steering the beam until the HMS axis and the beam axis
intersected which was achieved when the sieve slit pattern is centered
about the central aperture, which is smaller than the others, see
Fig.\ref{sieve}.  Using the position of the central aperture from the
calibration run in
combination with the locations of the BPMs, the nominal beam position
coordinates for the BPMs can be calculated.

\subsection{\label{bcmcal_sec}Beam Current Monitors}

The current of the electron beam in the hall was measured using 2
microwave cavity Beam Current Monitors (BCM1 and BCM2) and a
parametric current transformer (Unser monitor).   The BCMs are resonant cavities
of a design similar to the previously described BPMs.  The antennae
pick up the signal that is proportional to the beam current for all
resonant modes of the cavity, but for certain modes (TM$_{010}$ mode),
the signal is relatively insensitive to beam position.  One can make
the cavity sensitive to this mode by adjusting its size and
selecting the resonant frequency  to be identical to the accelerator
RF frequency.  Then, the BCM cavity is excited to this mode when the
electron beam passes through it and the beam current can be measured
by analyzing the signal picked up by the cavity antennae.

\begin{figure*}[h!]
\center
\includegraphics[height=3in]{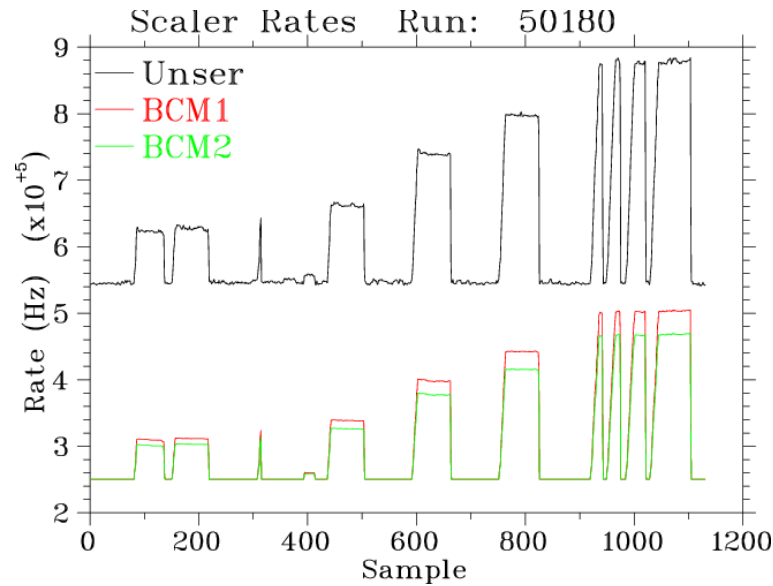}
\caption[BCM Calibration Run.]{BCM Calibration run consisting of alternating periods of beam on and off
  with increasing current during each beam on period.  Short beam-on
  periods centered around sample 1000 are not used in the calibration
  as they're too short and a result of beam trips.}
\label{bcmcal}
\end{figure*}
The Unser monitor consists of an active current transformer and a
magnetic modulator, linked together in a feedback loop.  
 The toroid which is the main current sensor, carries
two windings which close the feedback loop of the operational
amplifier.  This amplifier maintains the balance between the primary
and feedback currents.  The error in the balance between these currents
is detected and corrected for by a magnetic modulator-demodulator circuit consisting of
a pair of toroidal cores. The output signal voltage,
proportional to the beam current is measured with a precision resistor
from the feedback current~\cite{Unser:1981fh}.

\begin{figure*}[h!]
\center
\includegraphics[height=3in]{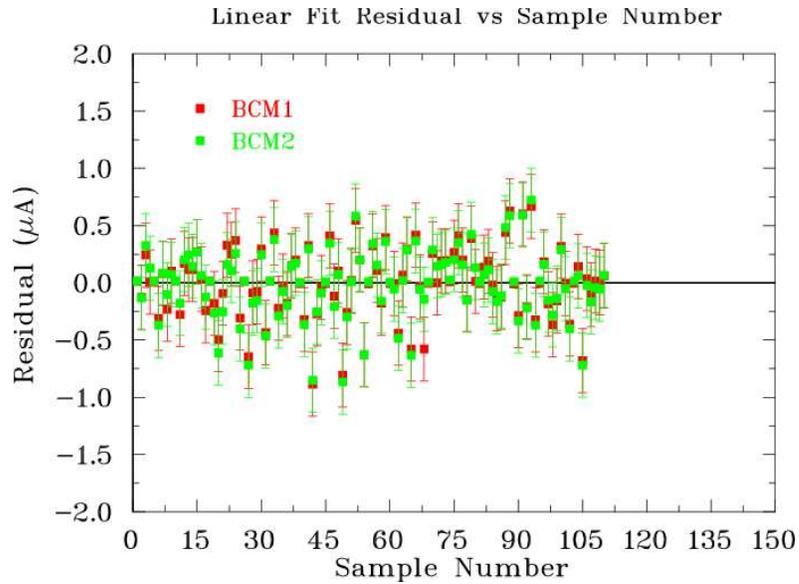}
\caption[BCM Calibration Residuals]{BCM Calibration: Residuals for each run compared to the
  global fit as a function of sample number (equivalent to time).}
\label{bcmres}
\end{figure*}

The Unser has a stable and well measured
gain, but large drifts in its offset and therefore, is not used to
measure the beam current.  There were 17 BCM calibration runs taken
over the course of the experiment.  The calibration runs were about an
hour long and consisted of alternating beam on and beam off periods of
2 minutes each, with the beam current increasing for every beam on period until a
current of $\approx$100
$\mu$A was reached (Fig. \ref{bcmcal}), and then the current was decreased during the
remaining beam on periods.  During the beam off periods, the offsets
of the Unser and the BCMs can be determined, and the beam on periods
can be used to calibrated the gains of the BCMs using the well-known
gain of the Unser and its measured offset.  The data from all
the calibration runs were combined into one calibration file, and a
Hall C calibration program was used to
select stable intervals within each beam on and beam off period
.  Normally, all the beam off periods are averaged over,
but since there was concern about the Unser zero drifting over the
hour-long run, the beam off periods were used on either side of a
beam-on period. The residuals from each
calibration run when compared to the global fit can be seen in
Fig.\ref{bcmres} and display a nice scatter around zero without any outliers or
trends.  Residuals vs current also showed no systematic trends.

A more detailed and complete description of both the BCM and Unser
monitors can be found in C.S. Armstrong's thesis ~\cite{Armstrong:1998hx}.

There are several sources of systematic uncertainties on the
measurement of the charge.  The stability of the gain on BCM 2 (which was used in data analysis) is taken to be good to 0.2$\%$.
The uncertainty in the Unser zero is taken to be 250nA (see
Sec.~\ref{target_boiling_section}), for an additional 0.3$\%$
uncertainty.  Finally, the measurement is sensitive to temperature,
which yield another 0.2$\%$ uncertainty~\cite{john_thesis}.  The total
systematic uncertainty on the charge measurement is 0.5$\%$, with a
0.4$\%$ scale uncertainty and 0.3$\%$ relative uncertainty.
  
\subsection{Beam Raster System}

The E02-019 experiment was performed with a beam current of
$\sim$80$\mu$A, with the exception of Aluminum data, which
were taken at $\sim$35$\mu$A.  This is large enough that the
cryogenic liquid targets can be overheated by the deposited energy of
the beam and bubbles can be formed, resulting in non-uniform target
density~\cite{Yan:1992wn}.  To minimize this effect, as well as to
avoid damaging the beam dump window, the beam was rastered,
which increases the effective spot size and reduces the energy density.

\begin{figure*}[h!]
\center
\includegraphics[height=4in]{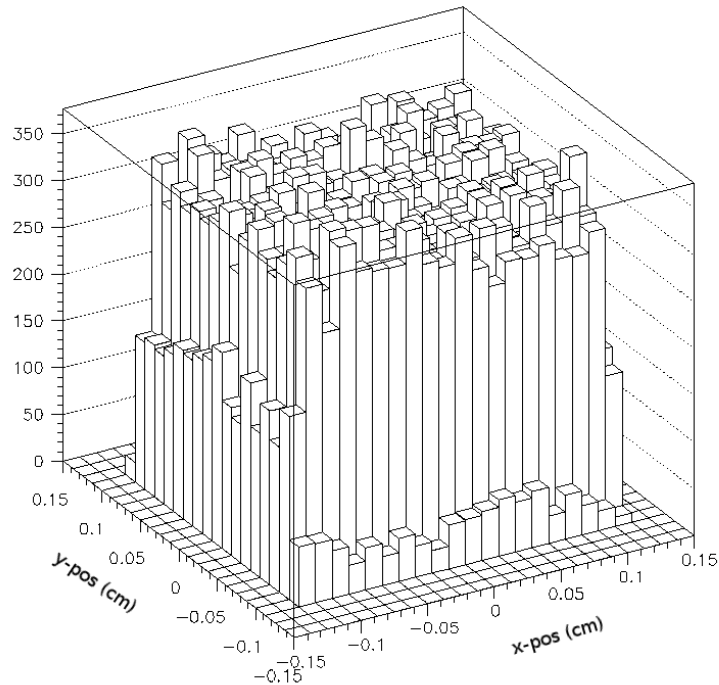}
\caption[Beam Profile with the fast raster turned on.]{Beam Profile with the fast raster turned on.  The beam was
  rastered in both horizontal and vertical directions to avoid melting
  or boiling of the target.}
\label{raster}
\end{figure*}
The fast raster system consists of 2 sets of steering magnets and is
located 20m upstream of the target.  The first set rasters the beam in
the vertical direction and the second in the horizontal.  
Until recently, a Lissajous raster pattern was in place.
The waveform of the magnet current was sinusoidal, and it slowed down as
it approached its peak before reversing direction at the edge of the
scan region.  This caused more energy to be deposited at the edges,
reducing the effectiveness of the raster.  In August of 2002, a linear raster
system~\cite{Yan:2005vi} went into operation.  The waveform of the
new raster system is triangular, resulting in a uniform rectangular shape on
the target.  The design of the new system had two main
goals:  to maximize linear velocity and to minimize the turning time.
In comparison to the Lissajous raster, the linear raster has a highly
homogeneous density distribution over the entire raster region with
98$\%$ linearity, 95$\%$ uniformity, and 1000m/s linear sweep
velocity.  The turning time at the vertex of the raster pattern is
$\approx$ 200ns and the beam traversal time from one edge to another
is 20 $\mu$s.  The beam 2mm$\times$2mm  raster pattern can be
seen in Fig.~\ref{raster} and is highly uniform.
\section{Targets\label{reg_target_section}}
Experiment E02-019 was run with 4 cryogenic ( H, $^2$H, $^3$He,
$^4$He) and
4 solid ($^{12}$C, Be, Cu, and Au) targets.  The targets were arranged
on a ladder that was remotely controlled by the target operator during
the run.  The cryogenic targets were contained in Aluminum cells, so
the detected electrons could have scattered from either the cryogenic
target or the cell.  In order to allow a measurement
 of the contribution from the container cell, a dummy target
 was also placed onto the target ladder.  Since the
target ladder only had space for 3 cryo target cells and one was kept as a spare, there were 2
running periods, with helium data taken during one and hydrogen
and deuterium data taken during the other. 
\begin{figure*}[h!]
\center
\includegraphics[height=3in]{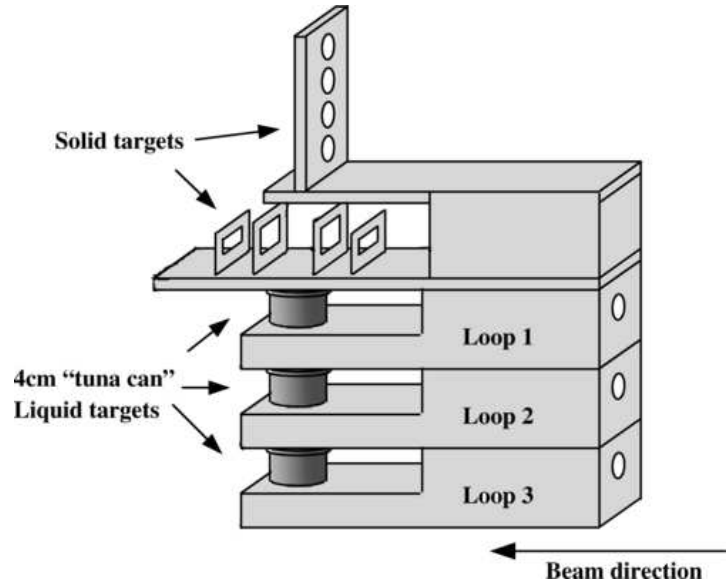}
\caption{Hall C Target ladder.}
\label{ladder}
\end{figure*}
The systematic uncertainties in the thicknesses of the solid targets
are given in Table~\ref{solid_sys_err}.  These are normalization, or
scale uncertainties and have no kinematic or time dependence, unlike
those for the cryogenic targets, discussed in the next section.
\begin{table}[h!]
\begin{center}
\caption{Systematic uncertainties in the target thicknesses for solid targets.}
\vspace*{0.25in}
\begin{tabular}{|c|c|c|c|}
\hline
Target &
Uncertainty \\
\hline
$^{9}$Be & 0.5$\%$\\
$^{12}$C & 0.5$\%$\\
$^{63}$Cu & 1.0$\%$\\
$^{197}$Au & 2.0$\%$\\
\hline
\end{tabular}
\label{solid_sys_err}
\end{center}
\end{table}

\subsection{Cryogenic Targets \label{cryo_target_section}}

The cryogenic targets were each held inside of a cylindrical ``tuna-can'' Al
cell with thin walls.  The can was
oriented vertically (Fig.~\ref{ladder}), so that horizontal changes in the beam position correspond
to changes in the effective target thickness.  Accurate knowledge of
the effective target thickness is vital to the extraction of the cross sections. To measure the
target thickness, we need precise measurements of the dimensions of
the target cell as well as corrections for the thermal contraction as
the dimensions were measured at room temperature.
The target ladder as well as the elements on it can move relative to each other
during the cooling process and the beam positions may be offset from
the center-beam line.  Finally, the densities of the cryogenic targets
can change with time if there are any leaks in the loops.

\begin{figure*}[h!]
\center
\includegraphics[height=3in]{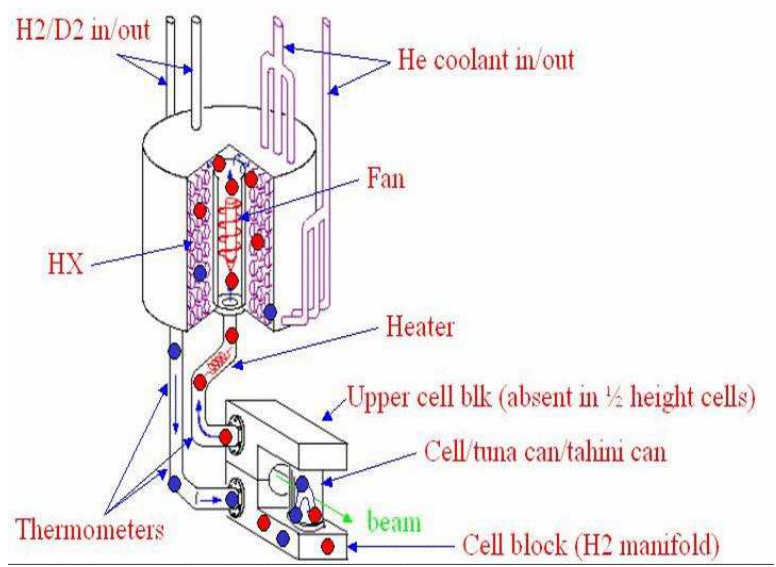}
\caption[Cryogenic Target]{Detailed view of a cryotarget loop and its components}
\label{sweet_cryo}
\end{figure*}
The cryotarget has three separate loops in it, each attached to short
(4 cm diamter) cell.  Even
though all three loops can be used at any given time, only 2 were
filled with cryogenic materials for this experiment, with the third
being left empty as a possible spare.  A loop consists of a target
cell, a re-circulation fan, heat exchangers and 700W heater (Fig.~\ref{sweet_cryo}).  The
coolant flow to the loops is controlled by the target operator through
Joule Thompson (JT) valves.  The coolant, 15K helium for H/$^2$H and
4K helium for $^3$He/$^4$He running, comes through
the heat exchanger, where the target material is cooled.  The cooled
material moves continuously between the heat exchanger and the target
cell. The target
heater is used to maintain a constant heat load on the target, so that
the cooling power doesn't change as the beam current changes.

For E02-019, the H and $^2$H targets were operated at 19K, 23 Psia and
22K, 23 Psia, respectively.  The helium targets had small leaks, which
were corrected for, but
their nominal operating parameters were 5.8K at 117 Psia and 6.2K at
182.5 Psia for $^3$He and $^4$He, respectively.

The dimensions of the three cryogenic cells are listed in Table
\ref{targetgeo} and were measured and compiled by David Meekins
~\cite{tgt_report} from
the JLab target group.  The thermal contraction of the cells is
determined from the final temperature of the loops and the contraction
of Aluminum, whose contraction factor becomes relatively independent
of temperature below 25K.  The cryotargets were operated below this
threshold temperature and the contraction factor was found to be 0.996
for the three loops.

\begin{table}[h!]
\begin{center}
\caption{Geometry of the Cryogenic Target Loops.  The outer diameter (OD) includes the
cell walls. }
\vspace*{0.25in}
\begin{tabular}{|c|c|c|c|c|c|}
\hline
 Loop &
OD(mm) &
Wall 1 (mm) &
Wall 2 (mm) &
ID - warm (mm) &
ID - cold (mm)\\
\hline
1 & 40.13$\pm$0.08 & 0.1384$\pm$0.0013 & 0.1270$\pm$0.0013 &
 39.86$\pm$0.08 & 39.69$\pm$0.08 \\
2 & 40.18$\pm$0.08 & 0.1219$\pm$0.0013 & 0.1219$\pm$0.0013 &
 39.94$\pm$0.08 & 39.77$\pm$0.08 \\
3 & 40.16$\pm$0.08 & 0.1232$\pm$0.0013 & 0.1194$\pm$0.0013 &
 39.92$\pm$0.08 & 39.75$\pm$0.08 \\
\hline
\end{tabular}
\label{targetgeo}
\end{center}
\end{table}

Once cooled down, the target ladder was offset by 2.5 mm beam right,
with an additional 1 mm offset that occurred after the target
chamber was evacuated.  The nominal beam position was calculated to be 1.1 mm beam
left.  These offsets (4.6 mm)
were taken into account when the target thickness was being calculated
in addition to each cryo-cell's offset from the axis of the target ladder.
(Table \ref{targetoffsets}).

\begin{table}[h!]
\begin{center}
\caption{Target Thicknesses for E02-019. Note that the numbers
  listed  for the helium target
  thicknesses assume no leaks and the actual thicknesses were calculated on a
  run-by-run basis.  Radiation lengths include only the target
  materials, not their containers. }
\vspace*{0.25in}
\begin{tabular}{|c|c|c|}
\hline
 Target &
Thickness (g/cm$^2$) &
Radiation Length ($\%$) \\
\hline
H & 0.2828 & 0.463 \\
$^2$H & 0.6525 & 0.535\\
$^3$He & 0.2769 & 0.429\\
$^4$He & 0.5285 & 0.429\\
$^9$Be & 1.8703 & 2.870\\
$^{12}$C & 0.6667 & 1.561\\
$^{63}$Cu & 0.7986 & 6.210\\
$^{197}$Au & 0.3795 & 5.875\\
\hline
\end{tabular}
\label{targets}
\end{center}
\end{table}

\begin{table}[h!]
\begin{center}
\caption{Cryo-loop offsets and adjusted quantities.  The offsets in
  target position are
  relative to the center of the target ladder.   The minus sign denotes
  beam-right directions.  The final quantities are calculated using
  raster-averaged target length and the nominal target density and
  take into account the cryo-loop offsets, target ladder and vacuum
  motion offsets, as well as the nominal beam position offset.}
\vspace*{0.25in}
\begin{tabular}{|c|c|c|c|c|c|c|}
\hline
Target &
 Loop &
cell offset(mm) &
y$_{ave}$ (cm) &
$\tau$ (g/cm$^2$) &
$\rho$ (gm/cm$^3$)\\
\hline
H & 2 & -0.03$\pm$0.02 &  3.894$\pm$0.026 &
0.2794$\pm$0.0023 & 0.0723\\
$^2$H & 3 & -0.10$\pm$0.02 &  3.861$\pm$0.026 &
0.6446$\pm$0.0052 & 0.167\\
$^3$He & 2 & -0.03$\pm$0.02 &  3.894$\pm$0.026 &
0.2736$\pm$0.0022 & 0.0708\\
$^4$He & 1 & 0.31$\pm$0.02 &  3.890$\pm$0.026 &
0.5229$\pm$0.0042 & 0.135\\

\hline
\end{tabular}
\label{targetoffsets}
\end{center}
\end{table}

If the beam path does not go directly through the center of the target
of radius $r$, but rather traverses a path displaced by $x$ from the
center of the target, then the actual length traversed is
\begin{equation}
y=2 \: \sqrt{r^2-x^2}.
\label{thickness}
\end{equation}

The relative position of the beam is given relative to the sum of all
the offsets.  Also, the fact that the beam is rastered over a small
area on the target must be accounted for when the target thickness is determined.  The
final target thickness is calculated by multiplying the target density
by the raster-averaged target length:
\begin{equation}
\overline{y}=2 \: \frac{\int^{x_0+w}_{x_0-w} \sqrt{r^2-x^2} dx}{\Delta x}, 
\end{equation}
where the limits are defined by the edges of the raster pattern.  For
E02-019, the size of the raster was $\pm$1.2mm about the center of the
beam, $x_0$. The nominal and raster-averaged thicknesses are listed in
Table \ref{targetoffsets}. 

\begin{figure*}[h!]
\center
\includegraphics[height=2.5in]{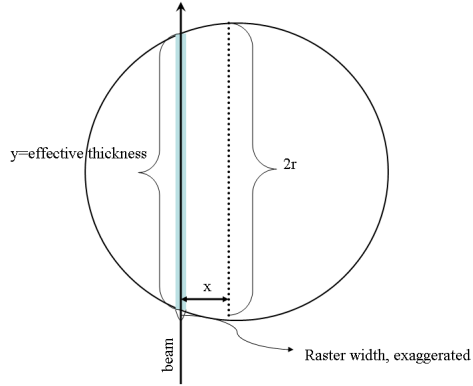}
\caption[Beam Offset.]{Beam path is offset from the center of cryotarget with shaded
region indicating the raster pattern. $y$ is the actual thickness when
the beam is offset by $x$ from the center of the cryotarget, of
thickness 2r.}
\label{tagetoffcenter}
\end{figure*}

\begin{figure*}[h!]
\center
\includegraphics[width=\textwidth]{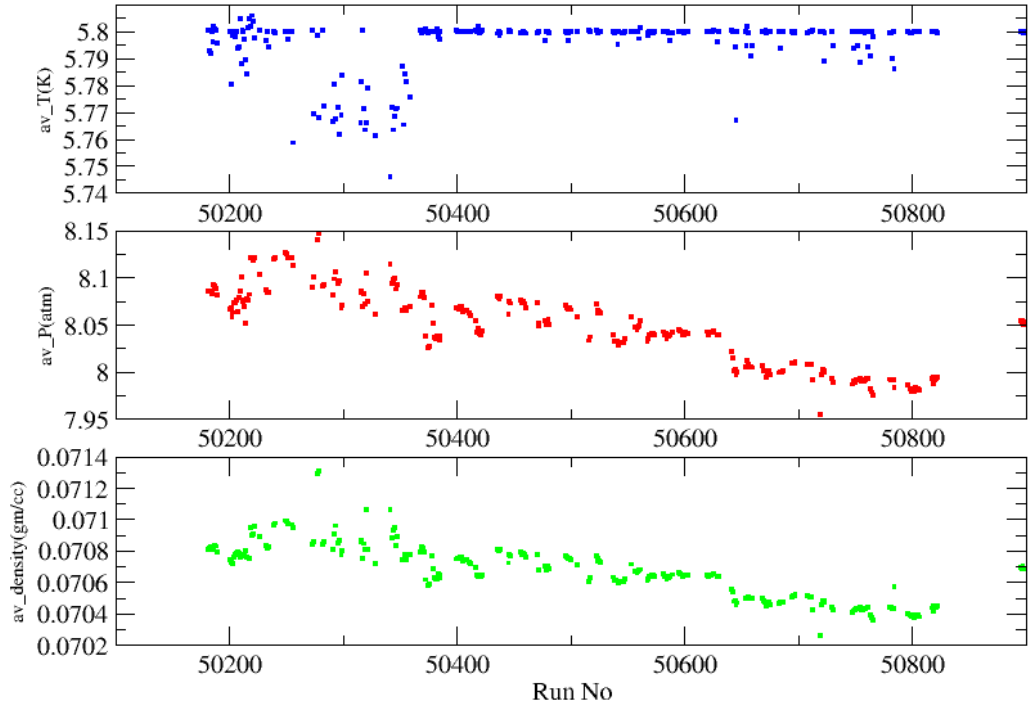}
\caption[Pressure, Temperature, and density for each $^3$He run.]{The average values for pressure, temperature and calculated
  density for each $^3$He run. The decreasing density indicates a leak
  in the cryocell.}
\label{he3_dens}
\end{figure*}
Another fact that leads to a correction to the target thickness is that
the beam position is not constant, but deviates by as much as
0.1 mm from the target center.  EPICS data (Sec.~\ref{daq}) from the BPMs was taken in 30-second intervals for
every run.  An weighted average for the position of the beam was
calculated for each run and converted into a correction to the target
thickness using Eq. \ref{thickness}.  A systematic uncertainty of
1.3$\%$ was assigned to this correction based on the maximum deviation
from the nominal beam position.

There was an additional factor to account for during the Helium
running period. Both Helium loops had small leaks, so that the density
did not remain constant over the data-taking period.  To account for
this, the EPICS (sec. \ref{daq}) temperature and pressure readouts were used to
calculate the density for each run and the correction applied on a
run-by-run basis in the data analysis procedure.  The EPICS readouts
are done in 2 second increments, which means that a given run has many
pressure, temperature readouts.  A density was calculated for each EPICS
event and then averaged for the whole run.  For $^3$He, tables~\cite{Gibbons:1967ab} listing
density as a function of temperature for a variety of pressures were used and for $^4$He, a NIST report was used
~\cite{nist_table}. Fig. \ref{he3_dens} shows
the EPICS quantities for the $^3$He runs as well as the extracted
density as a function of run number.  From the decreasing density, it
is evident that there was, in fact, a leak.  No such leak was detected
for the hydrogen/deuterium running period, and therefore a constant
density value was used for all the runs.

There are systematic uncertainties associated with the knowledge of
the equation of state for each cryogenic target, and the measurements
of temperature and pressure.  The temperature is known to within 0.1K
and the pressure is known to within 1.5Psi.  This translates to different
uncertainties in the thickness for different targets, which are listed
in Table~\ref{target_sys_err}.  Additionally, there's an uncertainty
of 1.3$\%$ due to the variation in the average beam position on target
and a 0.2$\%$ uncertainty in the knowledge of the cell diameter.
These are the same for all cryogenic targets.
\begin{table}[h!]
\begin{center}
\caption{Target-dependent systematic error sources for the cryogenic targets.}
\vspace*{0.25in}
\begin{tabular}{|c|c|c|c|}
\hline
Source &
$^2$H &
$^3$He &
$^4$He \\
\hline
Equation of State & 0.5$\%$ & 1.0$\%$ & 1.0$\%$ \\
Pressure & 0.01$\%$ & 0.6$\%$ & 0.2$\%$ \\
Temperature & $<$0.1$\%$ & 1.7$\%$ & 0.9$\%$ \\
Cell Diameter & 0.2$\%$& 0.2$\%$& 0.2$\%$\\
Beam position on target & 1.3$\%$& 1.3$\%$& 1.3$\%$ \\
\hline
Total & 1.4$\%$ & 2.4$\%$ & 1.9$\%$ \\
\hline
\end{tabular}
\label{target_sys_err}
\end{center}
\end{table}

\section{High Momentum Spectrometer}

\begin{figure*}[h!]
\center
\includegraphics[width=\textwidth]{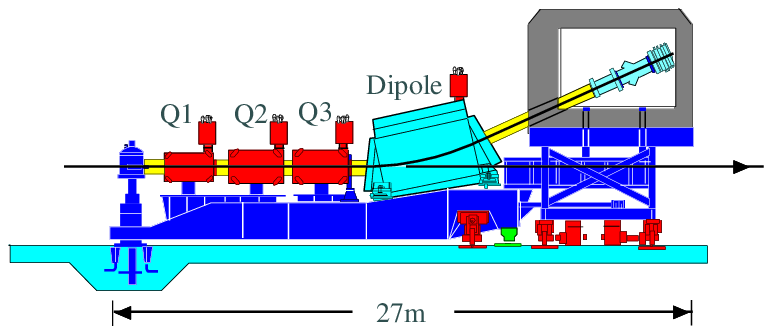}
\caption{HMS: Side view.}
\label{hms}
\end{figure*}

The scattered electrons were deflected and detected using Hall C's
High Momentum Spectrometer (Fig. \ref{hms}). The HMS is a 
QQQD system with a bend angle of 25$^{\circ}$ and a 10.5$^{\circ}$ minimum closing angle with respect to the
beam line ~\cite{design}.  The HMS magnets are supported on a common carriage which
rotates around a central bearing, rigidly mounted to the floor.  
 All four magnets are superconducting and are cooled with
4K Liquid Helium from the End Station
Refrigerator (ESR).   The quadrupole magnets determine the
acceptance and focusing properties of the HMS, while the dipole magnet
is used to set its central momentum.  In point-to-point tune
(optimized for vertex reconstruction), the Q1
and Q3 magnets focus in the dispersive directions and Q2 focuses in
the transverse direction.  The dipole is also a dispersive magnet.
The HMS is referred to as a 'software' spectrometer, meaning that the
momentum of a detected particle is determined by software reconstruction.
The HMS can be set to a maximum central momentum of 7.4 GeV/c with a
solid angle of $\sim$6.7mSr and a momentum acceptance of $\sim\pm$12$\%$ around
the central setting.  However, to limit the analysis to the area
where the acceptance was well known, only events within  $\sim\pm$9$\%$
central momentum were used in the analysis.  The parameters of the
quadrupole and dipole
magnets can be found in Tables \ref{Quadrupoles} and \ref{Dipole}, respectively.

\begin{table}[h]
\begin{center}
\caption{HMS Quadrupole Parameters}
\vspace*{0.25in}

\begin{tabular}{|c|c|c|c|c|}
\hline
& Warm Radius (cm) &
Pole Radius (cm) & Length (m) &
Pole Field (T) \\
\hline\hline\hline
Q1 & 22 & 25 & 1.89 & 1.5 \\
Q2 & 30 & 35 & 2.10 & 1.56 \\
Q3 & 30 & 35 & 2.10 & 0.86 \\
\hline
\end{tabular}
\label{Quadrupoles}
\end{center}
\end{table}

\begin{table}[h!]
\begin{center}
\caption{HMS Dipole Parameters}
\vspace*{0.25in}

\begin{tabular}{|c|c|}
\hline\hline\hline\hline
Gap & 42 cm \\
Good Field Width & $\pm$30 cm  \\
Length & 5.26 m  \\
Pole face Rotations & -6$^\circ$,  -6$^\circ$ \\
\hline
\end{tabular}
\label{Dipole}
\end{center}
\end{table}

The field in the quadrupole magnets was set by current and the field
in the dipole was set using an NMR probe inserted in the magnet.  To
minimize hysteresis effects, the magnets were always ramped 100A above the
highest field first, and then ramped down to each data setting.  The
fields in the magnets were monitored using Hall probes.

\begin{figure*}[h!]
\center
\includegraphics[height=1.5in]{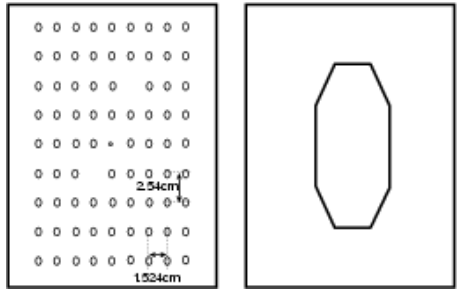}
\caption{HMS Collimators: Sieve and Pion}
\label{slits}
\end{figure*}

Before entering the first magnet, the scattered particle passes
through a collimator, mounted on the front of the Q1 magnet.  There
was a choice of 3 collimators: pion, large, and sieve.   The sieve
slit (Fig. \ref{slits}) is used to test the optics of the HMS and
consists of an array of small apertures separated by 2.54 cm in the
vertical direction and 1.524 cm in the horizontal direction with 2
holes missing in order to verify
the orientation of the slit.  The pion collimator (Fig. \ref{slits}) was used to take
production data.  It is an extended octagon in shape and sets the HMS
angular acceptance.

\section{Detector Package}

\begin{figure*}[htbp]
\center
\includegraphics[width=3in,angle=270,clip]{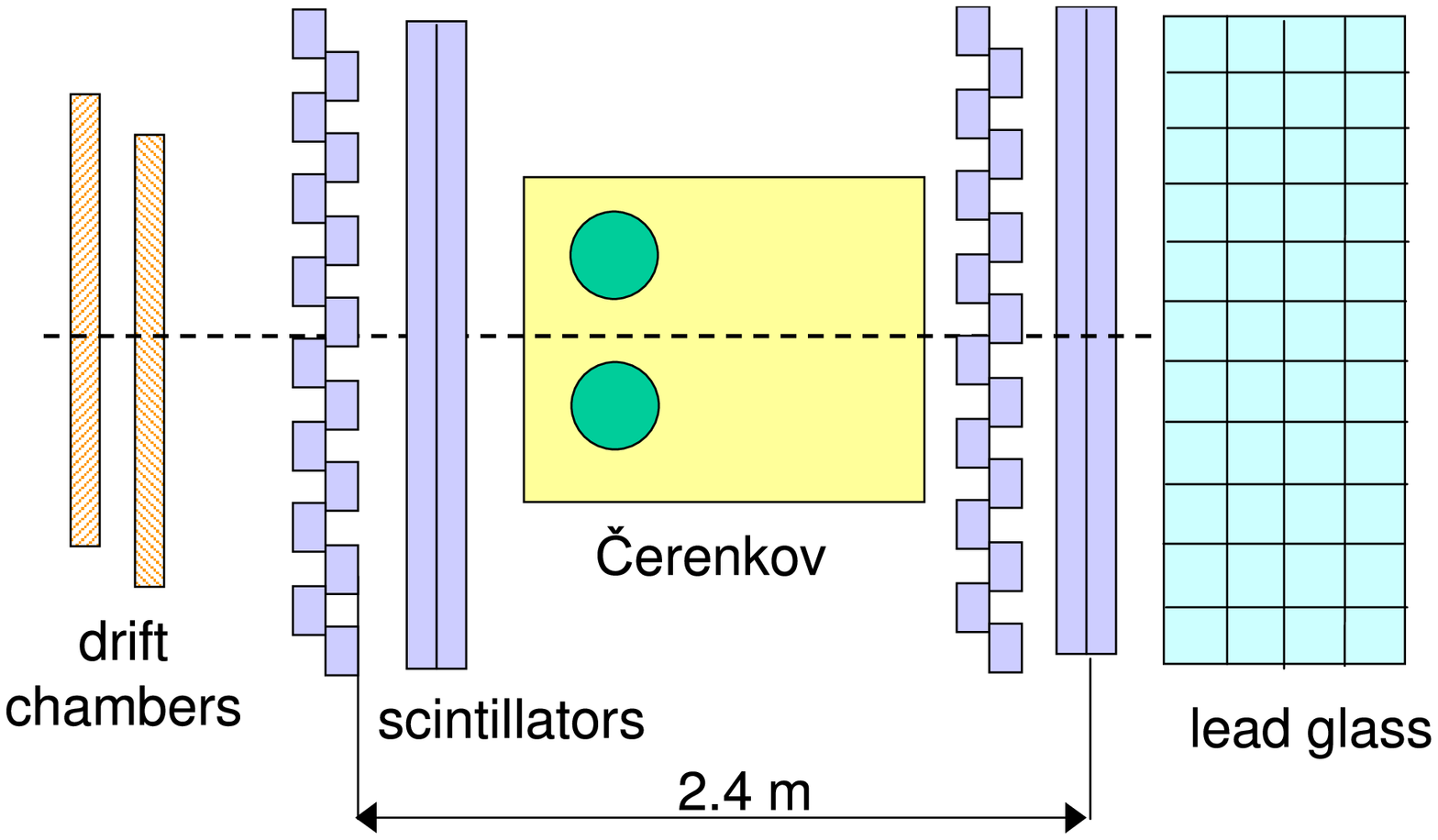}
\caption[HMS Detector Stack.]{Detector Stack in the HMS}
\label{detectors}
\end{figure*}

The detector package used in the HMS is
pictured in Fig. \ref{detectors}.  The Aerogel \v{C}erenkov detector
is not shown as it was taken out for this
experiment.  The detector stack contains two drift chambers, two sets of x-y
hodoscopes, a gas \v{C}erenkov and a lead glass calorimeter.  The
drift chambers are used for tracking particles, while the
hodoscopes are used to form the primary trigger. The calorimeter and
the \v{C}erenkov are used to form several legs of the electron
trigger and are also used for particle identification (and pion
rejection) in the offline analysis.

\subsection{Drift Chambers}
The drift chambers provide tracking information.  When a charged
particle passes through a gas-filled chamber and ionizes the gas, the
ejected electrons ``drift'' to the wires in the chamber. By measuring
the drift time, we can calculate the drift distance and the position
of the track.

\begin{figure*}[h!]
\center
\includegraphics[width=\textwidth]{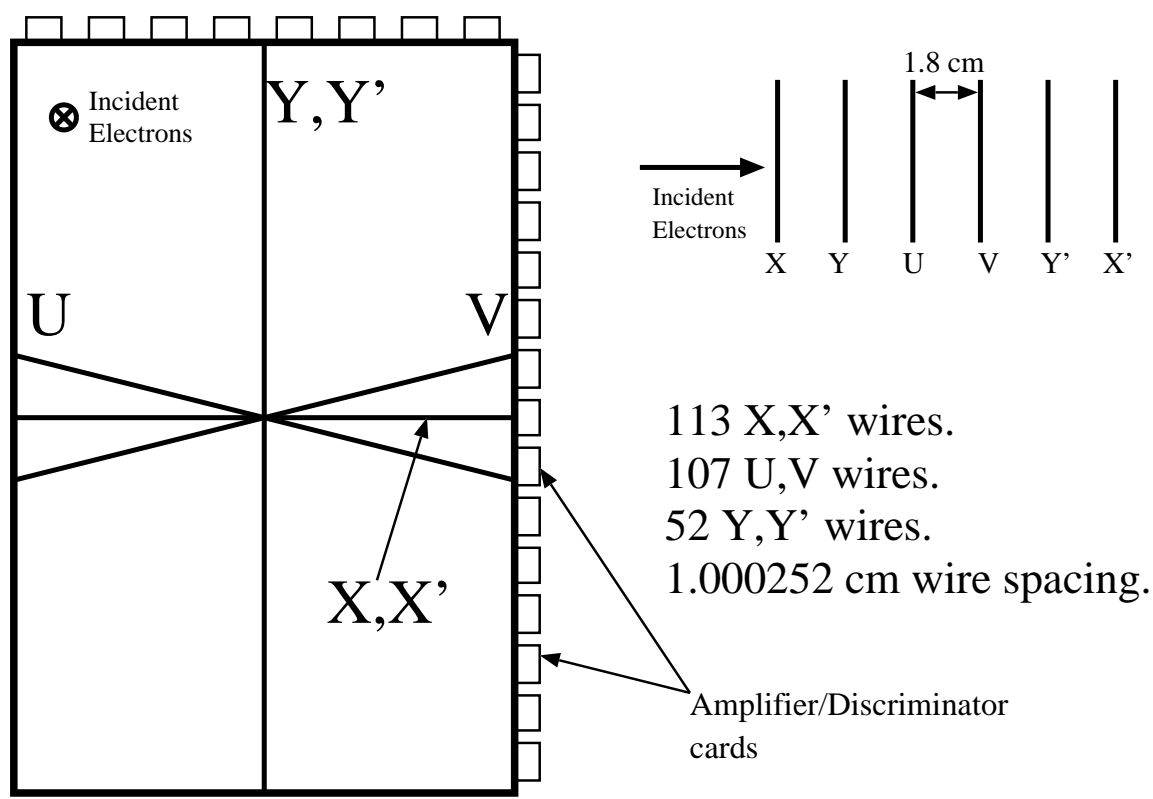}
\caption[HMS Drift Chamber.]{HMS Drift Chamber, a front view}
\label{hdc}
\end{figure*}

There are 2 drift chambers in the HMS, one located in the front and
the other in the back of the focal plane.  There are 6 planes in the
drift chambers: $X, Y, U, V, Y', X'$ with spacings of 1.4 cm between
them.  The $X$ and $Y$ planes are orthogonal and the $U, V$  planes
are inclined 15$^{\circ}$  with respect to the $X$ and $X'$ planes.  The
$X'$ and $Y'$ planes are offset from the $X$ and $Y$ planes by 5cm.
The sense wires (anodes), which reside at ground potential, detect the ionized electrons and the field
wires (cathodes) produce the electric field that guides the ionized
electrons to the sense wires.  A schematic of the drift chambers can
be seen in Fig. \ref{hdc}.

The signals from each anode wire are amplified and read out in
Nanometrics and LeCroy 2735DC discriminator cards.  The discriminated
signals are carried to the LeCroy 1879 Time-to-Digital Converters
(TDCs) where a stop signal is formed. The TDCs were programmed to read
out events within a $\sim$4$\mu$s window, since the trigger signal
from a particle in the spectrometer arrives at the TDC in  $\sim$2$\mu$s.

The drift chambers are filled with a mixture of argon/ethane gas.  When a charged particle passes through the chamber, it ionizes the
gas.  The ejected electrons are accelerated by the electric field and
drift from the track to the closest sense wire.   The electrons
are detected at the sense wires and the signal is read out as a
current over some period of time.  The sense wires are
positioned 1 cm apart from each other, meaning the particle's position can
be measured with a 0.5 cm precision, half the spacing by this method
alone.  However, precision of 0.5 cm is not nearly
good enough for tracking and instead the time that it took for the ionized
electrons to reach the sense wires is used.  This time is calculated by
using the hodoscope TDCs to determine the time that the particle passed
through the focal plane and the time that the sense wire detected the
ejected electron.  This time is then
converted to a drift distance which is added to the position of the
wire, giving the position of the track.
  Making use of the different orientations of the sense
wires in the 6 planes, the trajectory of the particle can then be reconstructed.

\subsection{Drift Chamber Calibration}

\begin{figure*}[htpt]
\center
\includegraphics[height=4in]{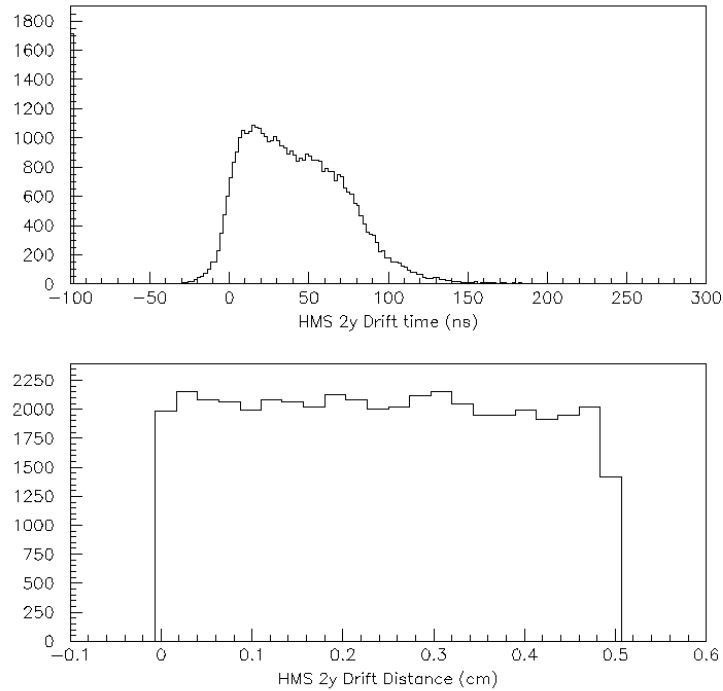}
\caption[Drift time and drift distance Spectra.]{Drift time and drift distance spectra for one plane of one of
the HMS drift chambers.  A negative drift time is recorded because
there's an
overall offset between times measured in the drift chambers and by the
hodoscope.}
\label{drift}
\end{figure*}

Each hit in the drift chambers has a TDC value associated with it
which is read out.  These values can be converted to a drift time
using timing information from the hodoscopes.  The final product, a
time-to-distance map, a driftmap,  gives the calculated distance from
the wire where the event occurred.  To calculate a driftmap, we assume
a direct relationship between the drift distance and drift time,
$x=f(T)$, and that the number of events in a given interval of
distance must be equal to that in some corresponding interval in time.  We use the
drift time distribution in each plane using a large number events and
the TDC values for all the wires.   Integrating
over the time spectrum, we get the drift distance:
\begin{equation}
\label{drifttime}
d=d_{max}\frac{\int^T_{t_{min}} F(t)dt}{\int^{t_{max}}_{t_{min}}
  F(t)dt}, 
\end{equation}
where $t_{min}$ and $t_{max}$ define the times to be used in the fit,
$d_{max}$ is them maximum distance from the wire, $T$ is the time
recorded by the TDC, $F(t)$ is the drift time distribution.  When properly
calibrated, the drift distance histogram should be flat.  A
driftmap is generated for each of the planes in the drift chambers.  A
sample drift time histogram and the distance histogram calculated from
it can be see in Fig. \ref{drift}.  A single driftmap was used for all
the runs in the analysis.

\subsection{\label{hodohardware}Hodoscopes}
The HMS has two pairs of ``x-y'' scintillator planes - one placed
before and one after the drift chambers.  Each x-plane had 10 paddles
oriented in the dispersive direction and each y-plane had 16 paddles
oriented in the non-dispersive direction.  The paddles are 1 cm thick
and 8 cm wide and are positioned so that there's 0.5cm overlap between
adjacent paddles preventing any particles from passing between the paddles.
The paddles are made of strips of BC404 scintillator with UVT lucite
light guides and Philips XP2282B photo-multiplier tubes on each end.

Charged particles ionize atoms in the scintillator paddles when they
pass through them. The
``liberated'' electrons excite molecular energy levels, which, upon
their decay, emit
light along the length of the paddle.  The light that
hits the surface of the scintillator at $\theta > \theta_{c}$ is
completely reflected ($\theta_c=\sin ^{-1} (1.0/1.58)=39.3^\circ$).  The reflected light is detected with
the PMTs mounted on the ends of the scintillator paddles.  The signal
is then converted into an electrical pulse, split, and sent to ADCs
and TDCs, as well as scalers and trigger logic modules in the electronics room of the counting house.  The signals
from all the tubes on a one side of a given plane (for example: S1X=S1X+ and S1X-) are ORed together.
Then, the outputs of the ORs for each plane were ANDed together and
then ORed with the remaining hodoscope plane in the pair (for example: S1=S1X and S1Y,
S2=S2X and S2Y), giving the primary hodoscope signals used in the trigger.



Timing information from the hodoscope had to be corrected for timing
and pulse height offsets in each element.  Different particles deposit
different amounts of energy, but since the electronics use constant
threshold discriminators, which turn on once the received pulse
reaches a set point on the discriminator, the timing information for
particles that deposit more energy would be biased to earlier time
values.  

The timing correction was a consequence of the fact that light that
arrives at the PMT could have undergone multiple internal
reflections.  This effect was corrected by using the distance,
$d_{prop}$, to determine the time of propagation
from the point where particle passed through the scintillator paddle
to the PMT.  The effective velocity of light in the paddle is a
function of both the index of refraction and the dimensions of the
paddle.  The propagation time for event $e$ at PMT $i$,
$t^{i,e}_{prop}=d^e_{prop}/v^i_{scin}$, was then subtracted from the
hit time recorded at the PMT.  The timing correction was on the order
of a few ns.

The need for a pulse height correction arises from the correlation
between the ADC value of a PMT signal,  $A^{i,e}_{pmt}$, and its
arrival time at the TDC.  The pulse shape of the analog signal from a PMT
is a weak function of amplitude, so the greater the amplitude of the
ADC signal, $A^{i,e}_{pmt}$, the sooner it crosses the threshold
voltage of the discriminator. This is corrected for by assuming a
mathematical form for the leading edge of the analog PMT signal.  The
form used was a Lorentzian, which fit parameters determined for each
PMT.  The pulse height correction was on the order of a few ns and was
subtracted from the hit time at each PMT.  A more detailed explanation
of both corrections can be found in Ref. ~\cite{Armstrong:1998hx}.

Prior to the beginning of the experiment, the hodoscope PMTs were gain-matched using a $^{60}$Co gamma ray
source placed in the center of each paddle.   The voltages on all the tubes were
set so that the Compton edge of the gamma rays gave a pulse height
of 175 mV at the discriminator inputs in the electronics room.

\subsection{Gas \v{C}erenkov}

The gas \v{C}erenkov detector is named after the man who first
observed and characterized \v{C}erenkov radiation.
\v{C}erenkov radiation is emitted when the speed of a particle is
greater than the phase velocity of the electromagnetic fields at a given
frequency $\omega$, i.e. $v>\frac{c}{\epsilon(\omega)}$.  In other words, the speed of the particle in a medium must
exceed the speed of light in the same medium, $v>\frac{c}{n}$.
The angle $\theta_{c}$ of emission of \v{C}erenkov radiation relative
to the velocity of the particle is given by 
\begin{equation}
\cos{\theta_{c}} = \frac{1}{\beta n}
\end{equation}
Mirrors inside the detector focus the \v{C}erenkov light onto
photo-multiplier tubes, which output a current that is subsequently converted into an
electronic signal.
The pressure inside the \v{C}erenkov was chosen so that electrons with a
given energy would emit \v{C}erenkov radiation, but heavier particles, such as
pions, would not.  Some pions will still produce a \v{C}erenkov signal
if the pion produces a knock-on electron of sufficient energy to emit
\v{C}erenkov light.  To minimize the rate of the knock-on electrons,
the entrance window to the detector is made as thin as possible.
\begin{figure*}[htpt]
\center
\includegraphics[height=3in]{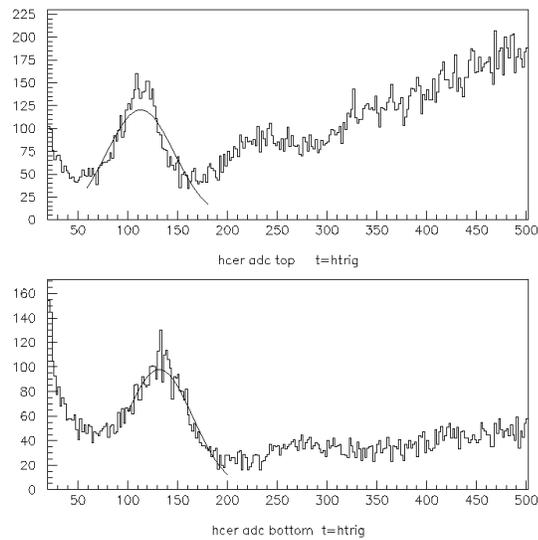}
\caption[ADC spectra for the HMS \v{C}erenkov PMTs]{ADC spectra for the HMS \v{C}erenkov PMTs, bottom and top.  The single
  photoelectron peak was fit with a Gaussian for calibration. Run
  50103 was used.}
\label{cerfit}
\end{figure*}

The HMS \v{C}erenkov is a cylindrical tank with an inner diameter of
150cm and a length of 165cm. The \v{C}erenkov detector was filled with
a C$_4$F$_{10}$ gas ($n$=1.00143 at STP),
pressurized to 0.35Atm (5.15Psi) giving a refractive index of
1.0005. The resulting pion threshold was $\approx$ 4.4GeV/c and the electron
threshold was $\approx$ 16 MeV.  There are 2 spherical mirrors  at the back of
the tank which reflect and focus the \v{C}erenkov light into 2
5'' Burle 8854 photo-multiplier tubes.    
\begin{figure*}[htpt]
\center
\includegraphics[height=3in, bb= 20 140 550 655]{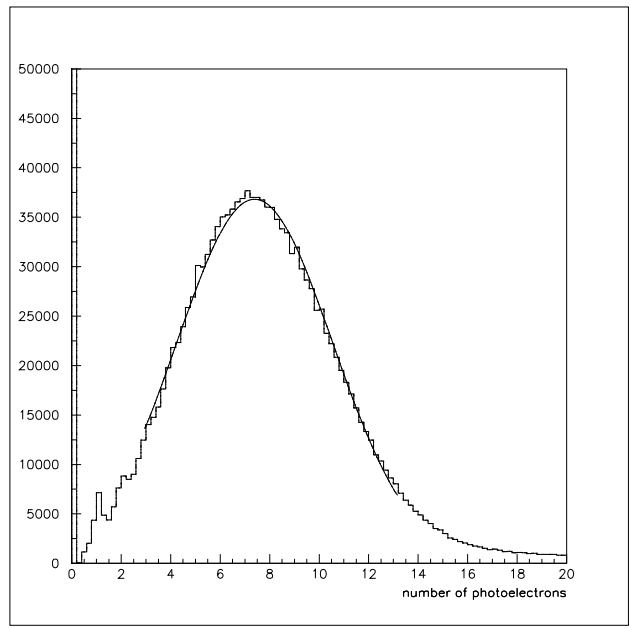}
\caption[Typical Spectrum of the HMS \v{C}erenkov.]{A typical spectrum for the HMS \v{C}erenkov. Run
  50553 was used, which was taken at p$_{HMS}$=3.74 GeV/c, below the
  pion threshold.}
\label{ave_pe}
\end{figure*}
The signals were relayed into a LeCroy
1881M ADC as well as a discriminator that was subsequently an input to
the trigger.
 The minimum signal results when one
photoelectron is ejected from the photocathode.  The calibration
involved locating the single photoelectron in the ADC spectrum of each
tube and fitting a Gaussian to it (see Fig.\ref{cerfit}), the location of the peak was used
as the calibration constant to calculate the number of photoelectrons
produced by particles in production running (summed over the
tubes). The average yield was 6-8 photoelectrons.  A typical spectrum
of photoelectrons can be seen in Fig. \ref{ave_pe}.

\subsection{Lead Glass Calorimeter}

The HMS calorimeter was used for particle identification in
conjunction with the \v{C}erenkov. The
calorimeter is the last detector in the HMS hut and consists of four
layers of 13 10cm x 10cm x 70cm TF1 lead glass blocks with Philips XP3462G Photomultiplier tubes attached to
them. The 2 front layers have the PMTs on both ends of each block,
whereas the last 2 layers have PMTs on only one side.

The calorimeter measures the energy deposited by a charged particle
when it passes through it.  An accelerating high energy electron emits photons through Bremsstrahlung.  These photons form
electron-positron pairs which radiate more photons and the
process is repeated.  Therefore, the number of particles increases exponentially 
with depth in the calorimeter. This is referred to as a ``cascade
shower'' of particles earning the calorimeter a nickname  ``shower
counter''.

Electrons and positrons deposit all of their energy in the detector,
while pions and protons deposit a relatively constant amount of
energy, due to ionization and direct \v{C}erenkov light ($\sim$300MeV
for pions).  So, as long as the energies of the particles are well above this constant, the calorimeter can be used to separate electrons
from other particles by measuring the deposited energy.  However,
pions can undergo nuclear interactions in the lead glass, which can
lead to a shower similar to those created by electrons.  The pion
signal then has a long tail that extends under the electron peak.  For
this reason, the calorimeter is not used alone for particle
identification, but rather in combination with the \v{C}erenkov.

The signal from each PMT is split and the outputs are sent into a
LeCroy 1881M ADC and a Philips 740 linear fan-in module where they're
summed.  The sum in the first layer (PRSUM) and the sum of all the
layers (SHSUM) are discriminated to give three logic signals for the
trigger.  PRLO and PRHI are low and high threshold on the energy
deposited in the first layer, and SHLO is a threshold on the total energy
deposited in the calorimeter and is above the pion peak.  

Two corrections are applied to the raw ADC values. First, the signal is
corrected for attenuation through the block so that it is independent of
the distance from the PMT.  Once the attenuation correction is made, a
gain correction factor is applied to each channel.  This gain correction
factor is determined through a calorimeter
calibration procedure, such that the sum of the corrected ADCs gives
the electron energy.

The calorimeter calibration was performed more than once, using
different methods before settling on one approach.  The goals were to minimize the width of and the
position dependence of the electron peak.  Each calorimeter
block has a gain associated with it, which is used to calculate the
deposited energy during the event reconstruction process.  When calibrated properly,
the electron peak will be placed at electron energy, reflecting the
energy deposited in the calorimeter.

%
%

\begin{figure*}[h!]
\center
\includegraphics[width=\textwidth]{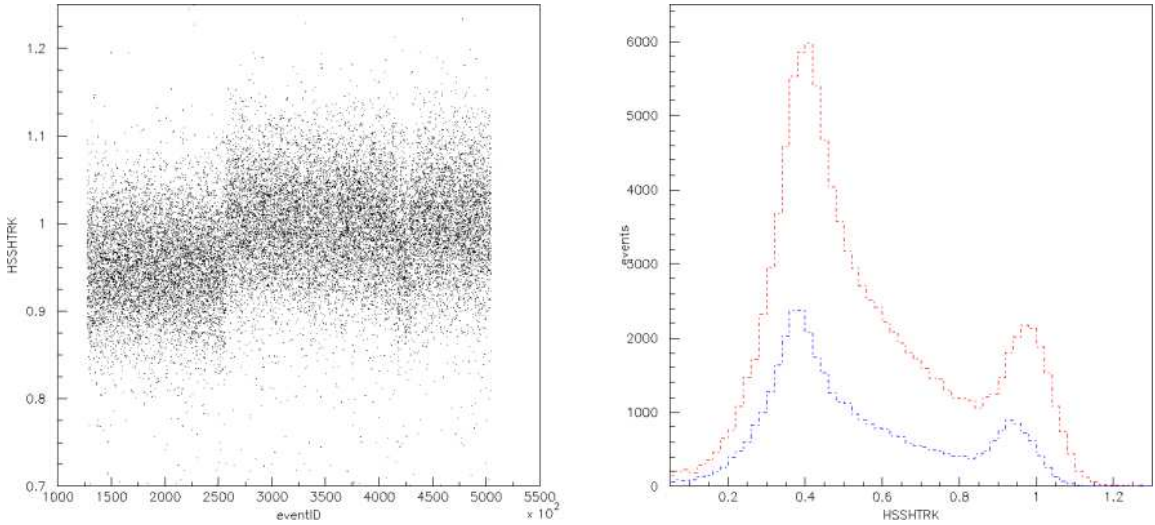}
\caption[Shifts in HMS shower counter spectrum.]{The left plot shows
  the electron spectrum in the calorimeter for run 51572 as a function
  of event number with sudden shifts.  The plot on the right shows the
  complete spectrum for event numbers corresponding to the nominal
  peak (red) and the shifted peak (blue).}
\label{shifting_peak}
\end{figure*}

After performing an initial calibration, the position of the peak was
examined as a function of time to check the stability of the calibration.  What 
was discovered instead was that some electron peaks had anomalous
widths (asymmetric, wider than expected).  The problem became obvious
when the position of the peak was examined as a function of time for a
given run.  Fig.~\ref{shifting_peak} shows fairly sudden changes in
the calorimeter spectrum (HSSHTRK=(energy deposited for a
track)$/p_{HMS}$) as a function of event number.  The right plot shows
the two electron peaks that correspond to shifted and unshifted spectra.  Since the pion peak (located at $\approx$0.4 in the HSSHTRK
spectrum) does not appear to be shifted by as much as the electron
peak, the shift cannot be a result of an offset in the pedestals.

This shift was determined to be the result of an intermittent
degradation of the leading edge of the discriminator signal used to gate the
ADC.  No correction was made in the end as the small loss in
resolution was insufficient to limit $\pi$ rejection.  For a detailed discussion, refer to~\cite{jason_thesis}.

\begin{figure*}[h!]
\center
\includegraphics[width=\textwidth]{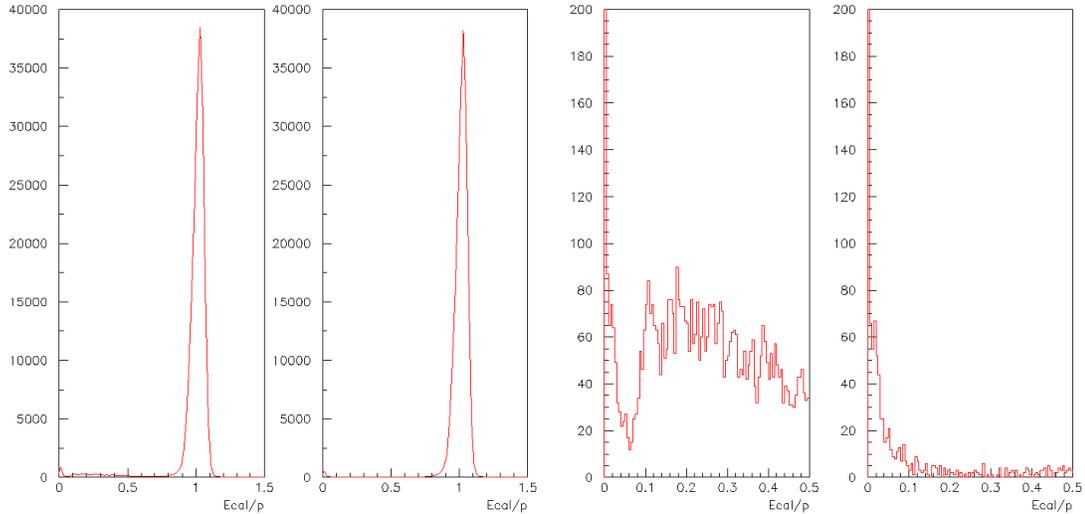}
\caption[HMS shower counter spectrum.]{HMS shower counter spectrum (Energy deposited in the
  calorimeter divided by the particle's momentum). The figures are for
  a deuterium run (51390) at 18$^{\circ}$ and p=3.74 GeV/c.  The two figures on the
  left show the electron peak before and after a \v{C}erenkov
  cut of 1.5 photoelectrons applied. The two figures on the right show
  the region 0-0.5 in more detail.} 
\label{calospectrum}
\end{figure*}

The final calibration was performed using a
stand-alone compiled program ~\cite{Arrington:hssh}. This code uses some initial set of gains and performs a random walk,
changing the gain on each block individually and checking to make sure
that the $\chi^2$ of the energy spectrum improves.  A sample HMS
shower spectrum is shown in Fig. \ref{calospectrum}.

One run was chosen to perform a calibration that would be used for all
the data.  This run was at our smallest angle (18$^{\circ}$) and a fairly high
momentum (3.74 GeV/c) so that the pion background was minimal, but the
momentum was still low enough so that the focal plane was well-populated.

\section{Trigger \label{trigger}}

There were several triggers in use during the running of E02-019 and
they were used for different purposes.  Choosing the correct trigger
is important since it's the first line of particle identification and
time and disk space can be saved if background events can be rejected at the
trigger level.  The time is of interest because it can contribute to
inefficiency by increasing the deadtime - periods when the data
acquisition system is unavailable and cannot accept new events because
it's processing/recording the previous event.  For E02-019, the
deadtime was kept below 20$\%$ by the occasional use of prescale
factors. Most of the time, the rates were low enough that a prescale
factor of 1 was used.

\begin{figure*}[htbp]
\center
\includegraphics[width=\textwidth,clip]{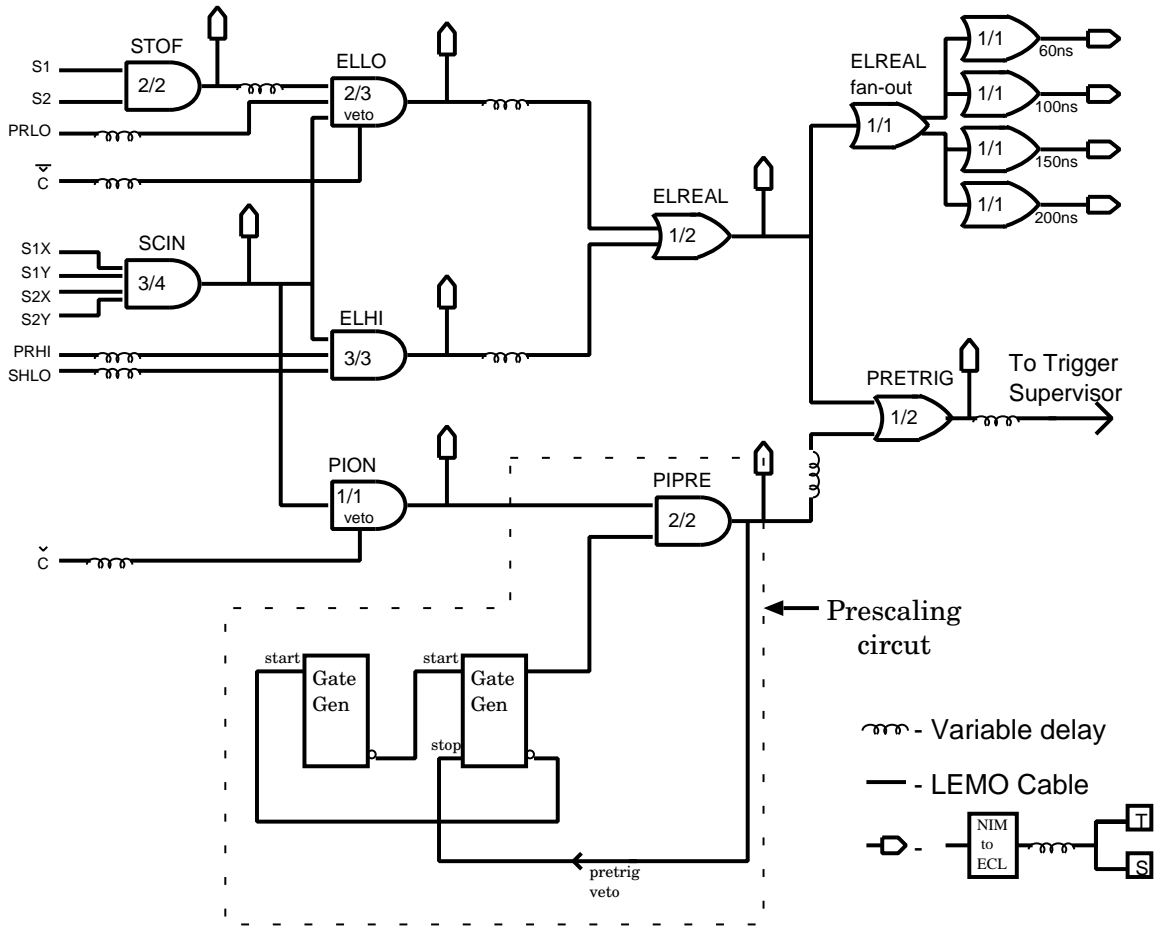}
\caption{HMS single arm electronics}
\label{elreal}
\end{figure*}

The trigger that was used for the main data taking (e${^-}$ detection)
 was ELREAL, which can be traced in Fig.~\ref{elreal}.  ELREAL is an electron trigger and requires
scintillator hits and user-defined particle identification signals.

The first part of the trigger was formed using the signals from the
hodoscope.  As described in Sec.~\ref{hodohardware}, a hit in a given plane
of the hodoscope is defined as a coincidence between 2 signals from
the opposite sides of a plane, not necessarily from the same
paddle.  Two scintillator triggers are then formed, STOF and SCIN
(a.k.a '3/4').  STOF requires a hit in one of the S1 (front) planes and
another in one of the S2 (rear) planes, the minimum number of planes for time of flight calculation.  SCIN requires hits in 3 of the 4
planes.  By that definition, if SCIN is satisfied then STOF is also
satisfied.

The other legs of the ELREAL trigger provide preliminary particle
identification. The analog signals from the first layer of the
calorimeter are sent to discriminators and if they exceed a low and/or a high
threshold, PRLO and/or PRHI logic signals are sent to the trigger.
SHLO is another logic signal that is sent to the trigger from the
calorimeter, but the input to the discriminator in this case is the
energy sum in all the layers of the calorimeter.
%
Finally, the CER signal requires the
\v{C}erenkov sum  to fire the
discriminator (threshold set to $\sim$0.5 photoelectrons) . 

ELLO requires the CER, and 2 of the following: STOF (loose hodoscope
cut), SCIN (tight hodoscope cut), or PRLO (loose shower counter cut).
The other leg of the trigger, ELHI, requires SCIN, PRHI, and SHLO to
be satisfied.  So, if an event satisfies either the ELLO or ELHI leg
of the trigger, it is a good candidate for an electron.   ELREAL, the
primary electron trigger, is
defined as the OR of ELLO and ELHI, and a more strict electron
trigger, ELCLEAN, is defined as their AND signal.  Note that the
trigger is robust against inefficiency in any given leg.  For example,
a trigger signal is formed in the absence of a \v{C}erenkov signal or
a calorimeter signal, as
long as the other detectors are performing well.

In addition to the ELREAL events, a sample of pions was also taken
in order to study the pion background and also to have a sample of events
that was not sensitive to inefficiencies associated with particle
identification requirements. This trigger, PION
(Fig. \ref{elreal}), was defined as a good hodoscope signal (SCIN) and
a \v{C}erenkov veto (note that PION events can still satisfy the ELREAL
trigger via the ELLO route).  These events were then prescaled by a dynamic
prescaling circuit so that the number of events recorded was a small
fraction of the number of ELREAL triggers.

Each trigger and most of the pre-trigger signals were output to a TDC
and read out by the DAQ system.  This made it possibly to cut on the 
various branches of the trigger during the data analysis and
efficiency calculations.

\subsection{\label{daq} Data Acquisition System }

\begin{figure*}[h!]
\center
\includegraphics[height=4in,bb=20 300 550 755,clip]{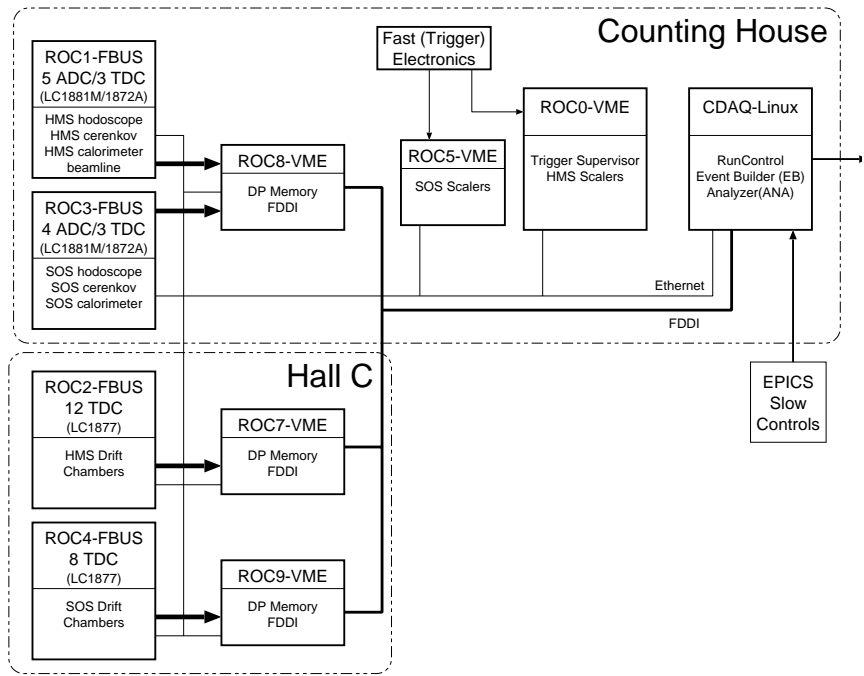}
\caption[Schematic of Hall C data acquisition system]{Schematic of
  Hall C data acquisition system, showing the elements described in Sec.~\ref{daq}}
\label{coda}
\end{figure*}

The data acquisition for E02-019 was handled by the CEBAF Online Data
Acquisition System (CODA)~\cite{Abbott:1995ab}.  A schematic of the
CODA  software chart can be seen in Fig. \ref{coda}. 
Information from ADCs and TDCs is read out from the
Read-Out Controllers (ROCs) which are CPUs in the
FASTBUS and VME crates.  The ROCs communicate with the Trigger
Supervisor, which generates the triggers that start the ROC read
outs.  Then, the ROCs send the data event fragments to the event
Event Builder, where all the header information and
data is compiled into an event.  Once the event is built, it is placed
into a buffer, then tested (and possibly rejected) and finally
written to disk.  CODA also contains a graphical user interface
(RunControl) that allows the shift worker to start and stop runs and
change run parameters.

Three types of information are read out and written to disk during each run: event
information from the ADCs and TDCs (read out for event, including
raster/BPM information), scaler information (read out every
2 seconds), and slow control variables from the Experimental
and Physics Industrial Control System (EPICS) database (read out
every 30 seconds).   EPICS quantities are not directly associated with
data acquisition and include information on spectrometer magnet
settings, target status variables, beamline controls, etc.

\chapter{Data Analysis}
The first stage of analysis of the raw data was done using the Hall C
reconstruction software, the ``Replay Engine.'' The code reads in the
raw detector signals, generates tracks as well as particle
identification information for each event.  In addition, the engine
also keeps track of software and hardware scaler quantities that are
written to
report files at the end of the replay.  The data is output in two
formats: PAW HBOOK files and PAW ntuples. The HBOOK files contain a
large set of histograms that are used to check detector performance as
well as to perform some calibrations.
The ntuples are organized event-by-event, with a set of quantities
calculated for each event.  The ntuples are then used as the input to
the second stage of the analysis, where physics quantities like cross
sections are extracted.  

In this chapter, the methodology for extracting cross sections will be discussed, including how tracks are reconstructed, how good events are selected and the corrections that are applied.

\section{Spectrometer Optics/Coordinates \label{coordinates}}
Before discussing the detector quantities, it is important to define
the coordinate system used in measuring these quantities as well as
the different reference frames used.
The spectrometer coordinates (see Fig.~\ref{tarcoords}) are defined with $\hat{z}$ along the optical axis (see below) inside
the spectrometer, $\hat{x}$ points down in the dispersive direction, and $\hat{y}$
points out in the non-dispersive direction (beam-left).

The central ray is defined as the trajectory of a particle that enters
the spectrometer through the center of the entrance aperture or the
optical axis of the first magnet.  The plane approximately half-way between the two
drift chambers is the detection or focal plane.  The central rays are
those that pass through the center of the detection plane and follow the optical
axis.  The momentum of the particles whose trajectories lie along the
optical axis defines the central momentum of the spectrometer.  The
detection/focal plane is perpendicular to this central trajectory and tilted
85$^{\circ}$ with respect to the 'true' focal plane of the
spectrometer.  The 'true' focal plane is an approximation to the
surface defined by the initial rays at various angles, used to
determine the position at which they're focused by the magnets.
\begin{figure*}[htbp]
\centerline{\includegraphics[width=0.7\textwidth]{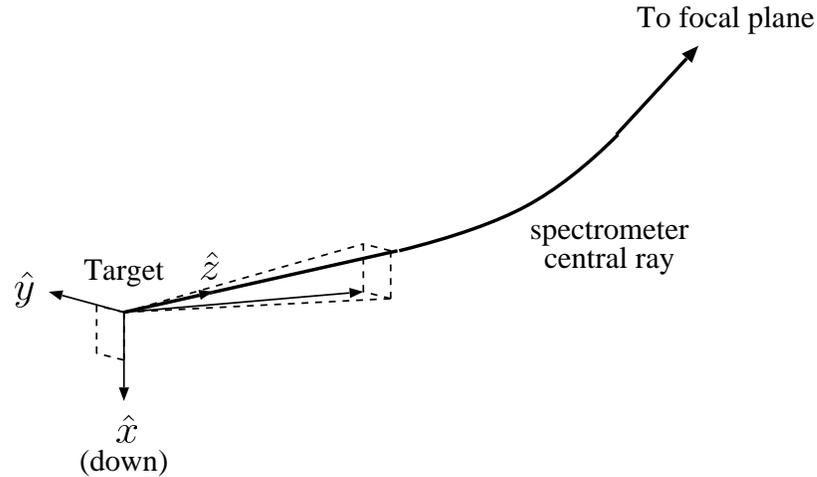}}
\caption[Target Coordinate system.]{Target coordinate system:
  $\hat{z}$ points along the optical axis inside
the spectrometer, $\hat{x}$ points down in the dispersive direction, and $\hat{y}$
points beam-left.}
\label{tarcoords}
\end{figure*}

Within the spectrometer coordinate system, two frames are used: the
focal plane frame and the target frame.  The focal plane coordinate
system has its origin in the center of the detection plane (but the
two planes do not coincide), and its
variables are given the subscript with ``fp''.  The target coordinate system is
centered at the target and the variables associated with it are
have ``tar'' subscripts (see Fig.~\ref{tarcoords}).

\begin{figure*}[htbp]
\center
\includegraphics[height=4in,clip]{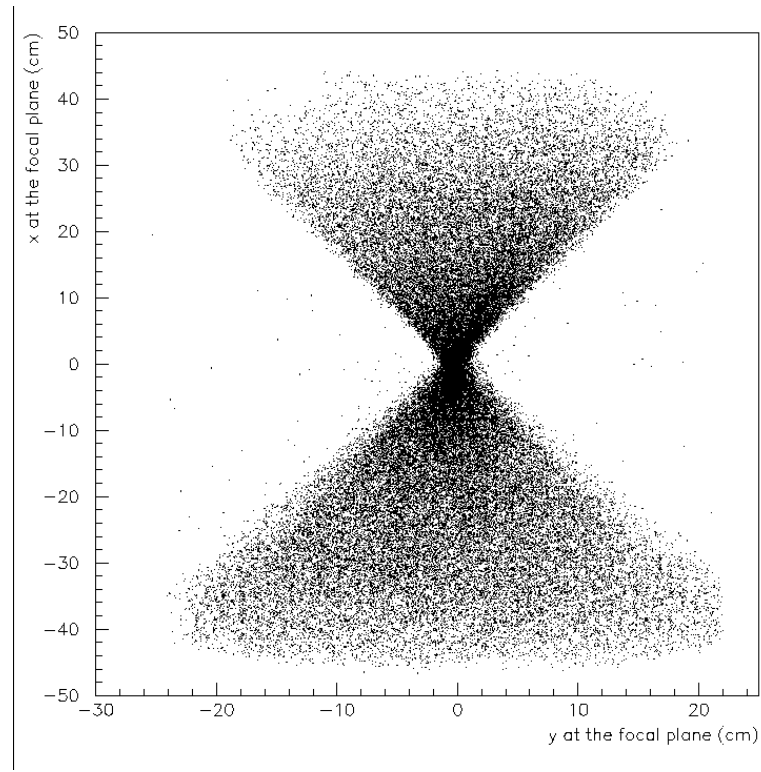}
\caption[Typical ``hourglass'' distribution of electrons in the
  HMS.]{Typical ``hourglass'' distribution of electrons in the
  HMS.  Data from a carbon run at $\theta _{HMS}=50^{\circ}$,
  $p_{HMS}$=-1.00 GeV/c is shown.}
\label{hourglass}
\end{figure*}

The typical distribution of electron events in the focal plane is
shown in Fig. ~\ref{hourglass}.  This shape, commonly referred to as
an 'hourglass', is the result of focusing in the dispersive
direction by the Q1 and Q3 quadrupole magnets and in the transverse direction by
the Q2 quadrupole.

\section{Event Reconstruction}
%


The trajectory of the particle is reconstructed using information from the two drift
chambers.  The hits in the first drift chamber are used to identify
clusters of hits, called ``space points''.  The hits in each of these
clusters are then fit to a
mini-track, called a ``stub''.   Within each stub, it is necessary to
determine whether the particle passed the wire on the left or the
right.  The brute force approach is to fit all the left-right
combinations and to choose the best one based on $\chi ^2$, but as this method is time
consuming, it is the last resort.  Instead, for planes with parallel wires ($y$ and $y'$, for
example), a small angle approximation is used. If the same
position is reconstructed for the two planes, since the planes are
offset by 1/2 cell, the particle is assumed
to have passed between the two wires. 

  After all the fits have been
performed for the first drift chamber, this procedure is repeated for
the second one.  The tracking code then fits one track through
them.  If more than one track is found, the one with the smaller
$\chi^2$ is chosen.  The track is then projected to the focal plane,
defined as being approximately half-way between the two chambers, and the focal
plane variables ($x_{fp}, y_{fp}, x'_{fp}, y'_{fp}$)  are determined.

Once the position and angles of the particle are known in the
focal plane, we can reconstruct the same quantities at the target.
This is done using a transformation matrix of the form:
\begin{equation}
\label{matrix}
a^i_{0}=\sum_{j,k,l,m} M^i_{jklm} (x_{fp})^j (x'_{fp})^k (y_{fp})^l (y'_{fp})^m\;\;\rm{for}\;\;(1 \leq j+k+l+m \leq N)
\end{equation}
where  $a^i_0$
is a target quantity, $M^i_{jklm}$ is one column of the reconstruction matrix, and $N$ is the order of the transformation, which for the HMS is 5.
The quantities given by this transformation are $x'_{tar}=a^1_{0}$, which is the
slope of the track ($\frac{dx}{dz}$) in the $x$ (dispersive) direction, $y'_{tar}=a^3_{0}$, which is the
slope of the track ($\frac{dy}{dz}$) in the $y$ direction,  $y_{tar}=a^2_{0}$,
which is the position of the interaction point, and $\delta=a^4_{0}$, which is
the deviation from the central momentum in $\%$.  $x_{tar}$ is not given
by this transformation (only 4 quantities can be extracted since only 4 are measured) and is taken to be 0.  While the
primed quantities correspond to slopes, the acceptance of the HMS is
small enough ($<$100mr), that it's a reasonable approximation to treat
those slopes as angles.

The reconstruction matrix elements were obtained previously using an iterative
fitting procedure, starting with a model of the spectrometer optics,
from COSY INFINITY ~\cite{Makino:2006sx}.  The matrix elements are
then refit with HMS data taken with the sieve slit, whose pattern
makes it possible to determine which hole the particle went through.
The matrix elements are then modified so that each event in the sieve
slit data is
reconstructed to the correct position.  Different targets are used for
optimization of different variables, and many fits are performed
before the matrix elements are finalized.  The details of the fitting
procedure can be obtained from J. Volmer's
thesis~\cite{Volmer:2000ab}.

\section{Particle Identification and Electron Selection}

For the majority of the experiment, the HMS was run with the
magnets in negative polarity, accepting all negatively charged particles.
Since we're interested only in electrons, it is important to be able
to select them and reject the pions.  The gas \v{C}erenkov and the
lead glass Calorimeter were used to select electrons both at the
trigger level and with software cuts in the analysis.

For the analysis, a \v{C}erenkov cut of 1.5 photoelectrons and a
calorimeter cut of 0.7 times the value of the electron peak (E$_{cal}$/p$\approx$1) were used
to select electrons.  If the electron peak were perfectly centered at
1.0 in the normalized calorimeter energy spectrum (E$_{cal}$/p), then we could use a constant value cut, but since calibration is
not perfect, the reconstructed peak can be shifted a fraction of a percent
away from 1.0.  The electron peak was fit to a Gaussian, and the
calorimeter cut was defined as the peak position fit value times 0.7.
Those events that deposited less energy were rejected.  In reality, there is only a small
fraction of a per cent difference in the yield if a constant value cut for E$_{cal}$/p
is used.

There are several types of background that are present in the data
sample.  Some are unwanted electrons that are a result of pair
production from a photon emitted either through Bremsstrahlung or as a result of $\pi^{\emptyset}$ decay.  These are charge symmetric and are eliminated
through subtraction of a $e^+$ yield from the positive polarity
running.  Another background source is
electrons scattered from the container cell walls in the case of the
cryogenic targets and these are eliminated by subtracting the empty
aluminum can data yield.  Finally, some pions pass all the software cuts
designed to reject them and get
into the electron sample by producing knock-on
electrons of energy sufficient to produce a signal in the \v{C}erenkov detector.  Knock-on electrons are emitted from atoms when a charged particle passes through matter and transfers energy to the constituent atoms.  Even
though these pions deposit only $\sim$300MeV of energy in the
calorimeter, forming a secondary peak in the spectrum at $\sim$0.3/p,
the tail of this peak extends beyond the cut and into the electron sample.
\subsection{Charge Symmetric Background \label{csbg_subtraction_section}}
Some scattered electrons emit high energy photons in the target, which
in turn form electron-positron pairs.  These pairs can also be produced by the $\pi^\circ \rightarrow 2\gamma \rightarrow e^+e^-$ decay process of the neutral pions produced in the target.  Some of the electrons from
these pairs fire the detectors and can form a trigger.  Since these
events come from a different process than the one we're interested in,
it's important to remove their contribution from the data sample.  The background yield
is charge symmetric because the positrons and electrons are produced
at equal rates.  This means that an accurate measurement of the
positron yield can be used to determine the background
electron yield from pair production.  The positron background was
measured directly by running the HMS in positive polarity (reversing
the polarity of the magnets to bend positively charged particles into
the detector hut
instead of the negatively charged ones) for all targets.  Measurements
were taken only at those angle and momentum settings where the
background was known to be significant.

\begin{figure*}[h!]
\center
\includegraphics[height=4in,clip]{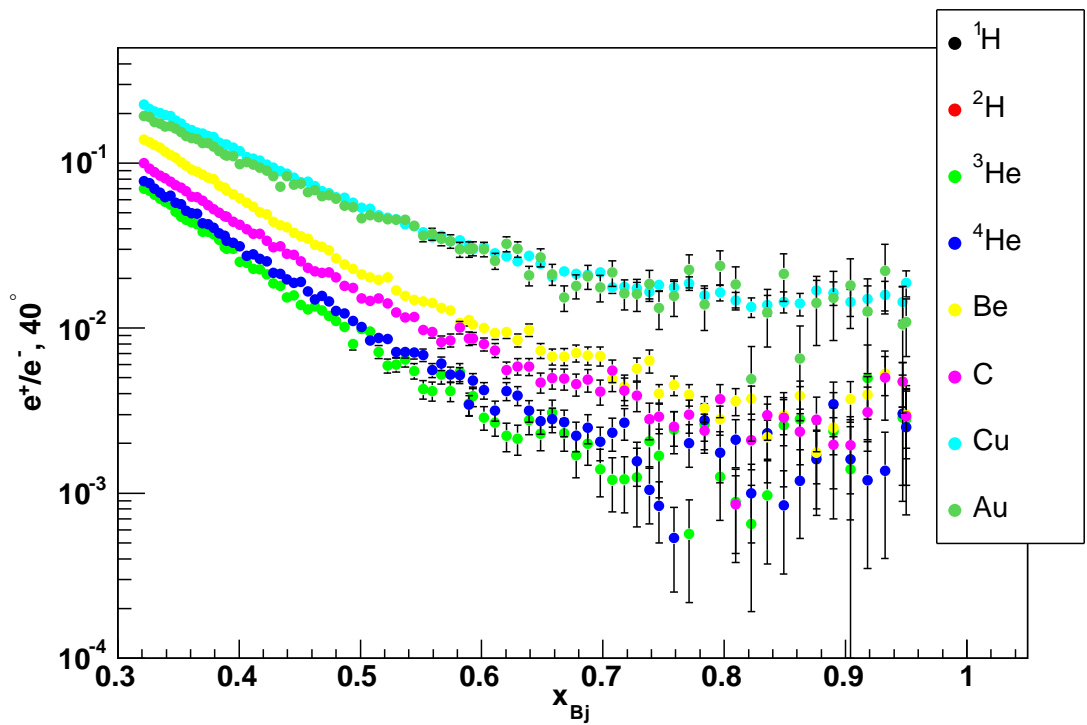}
\caption[Charge symmetric background for all targets at 40
  $^\circ$.]{The ratio of $e^+/e^-$ yields (a.k.a. CSBG)for all targets at 40
  $^\circ$ as a function of $x$}
\label{csbg}
\end{figure*}

At small angles and high momentum settings, positron data were not taken since
the background yield was negligible.  Fig.~\ref{csbg} shows the
relative positron yield when compared to the electron yield (including
the CSBG) at the
same setting.  The relative yield is quite small for high $x$ and as a result, no
positron data was taken for 18$^{\circ}$ or 22$^{\circ}$ for any
targets, and no positron data was taken for the highest momenta
settings at 26$^{\circ}$, except for the Copper target.

The same set of cuts are applied to the positron sample as the electron sample, but
since the rates are higher for positive polarity running due to the
presence of hadrons, a different trigger is used to keep the dead time
low.  ELREAL, described
in Sec.~\ref{trigger} is an OR of two signals, ELLO and ELHI.  For the
charge symmetric background measurement, the ELCLEAN trigger was used
instead, which is the AND of ELLO and ELHI, requiring both a calorimeter and a \v{C}erenkov signal.  For the computation of the charge symmetric
electron background, we also required the ELCLEAN signal for the
electron data sample (both signals were available).  The ratio is
then computed using 
\begin{equation}
R_{\frac{e^+}{e^-}}=\frac{e^+_{elcl}}{e^-_{elcl}}=\frac{e^+_{bg}}{e^-_{data}+e^-_{bg}}
\end{equation}
where $e^{+/-}_{bg}$ are the background yields from pair production
and $e^-_{data}$ is the electron data yield from the reaction we're
interested in. Then, using the fact that the
yield is charge symmetric, we subtract the background and the net
electron yield is then
\begin{equation}
Y^{-}=Y^{-}_{elrl}\cdot (1-\frac{e^+_{elcl}}{e^-_{elcl}}).
\end{equation}

Since the size of the charge symmetric background varies strongly with
kinematics, the systematic uncertainty associated with it is
angle-dependent.  The size was determined by comparing the $F_2^A$
structure function for copper at the two highest angles (40$^{\circ}$
and 50$^{\circ}$), where the correction is large.  As the structure
function is expected to scale and the size of the radiative correction
is known, we can ascribe any differences to the charge symmetric
background subtraction.  The uncertainty was determined to be 5$\%$ of
the size of the correction.  For example, the charge symmetric background for the
lowest momentum setting at 40$^{\circ}$ is about 25$\%$, which implies a
1.25$\%$ systematic uncertainty.   Since the contribution for each
angle falls with momentum, the $e^+/e^-$ ratio was parametrized and 
the systematic uncertainty was calculated for each final data point.

\begin{figure*}[h!]
\center
\includegraphics[angle=270,width=0.7\textwidth]{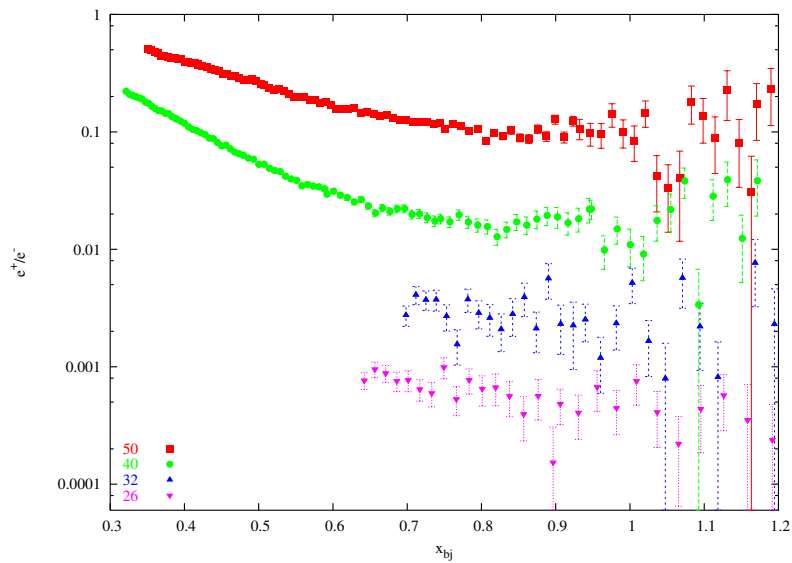}
\caption[Size of the charge symmetric background for copper.]{Size of
  the charge symmetric background for copper ($e^+/e^-$), for the angles where
  positron data was taken.  As this ratio falls quickly with smaller
  angles, no systematic uncertainty was assigned for $\theta <$ 26$^{\circ}$.}
\label{csbg_by_angle}
\end{figure*}

\begin{table}[h]
\begin{center}
\caption{Size of the systematic uncertainty due to the subtraction of
  the charge symmetric background for each of the HMS angle settings used.}
\vspace*{0.25in}

\begin{tabular}{|c|c|}
\hline
$\theta _{HMS}$ &
Uncertainty\\
 & due to CSBG($\%$)\\
\hline\hline
18$^{\circ}$ & 0\\
22$^{\circ}$ & 0\\
26$^{\circ}$ & 0.1\\
32$^{\circ}$ & 0.3\\
40$^{\circ}$ & 1.25\\
50$^{\circ}$ & 2\\
\hline
\end{tabular}
\label{size_csbg_table}
\end{center}
\end{table}
Table~\ref{size_csbg_table} shows
the maximum uncertainty associated with each angle setting of the HMS.  As the
contribution from the charge symmetric background falls quickly with
decreasing angle (see Fig.~\ref{csbg_by_angle}), the lowest angles are
assumed to have a negligible contribution.

\subsection{Pion Contamination \label{pion_contam_section}}

Another source of background events are the pions that make it past
all the software cuts.  This happens when the pions produce knock-on
electrons while passing through the entrance window of the gas \v{C}erenkov.
As mentioned earlier, these pions typically deposit about 300MeV of energy in the calorimeter forming
a second peak in the calorimeter spectrum.  This peak is well below
the calorimeter cut of $\sim$0.7 ($E_{cal}/p_{central}$), but the tail of this peak does extend
beyond the cut and into the electron peak.

\begin{figure*}[h!]
\center
\includegraphics[height=4in,clip]{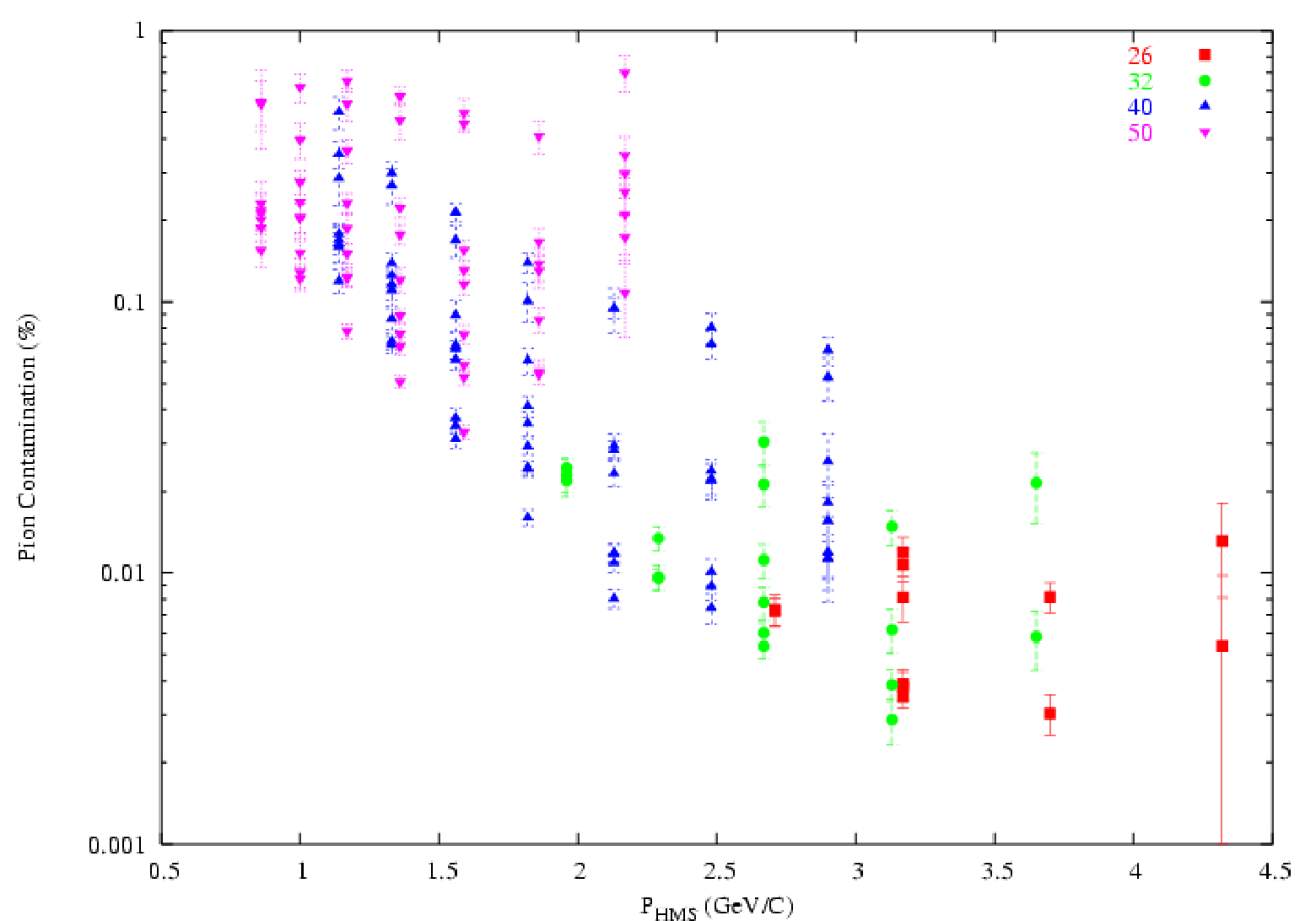}
\caption[Pion Contamination in the HMS.]{Pion Contamination in the HMS for all targets as a function of the central
  momentum for several $\theta$ settings.  There appears to be no dependence on angle.}
\label{pion_contam}
\end{figure*}

If the pion background is also charge-symmetric, the pions remaining
 after the software cuts
would be subtracted away
since they're also present in the positron sample.  However, a detailed
study was performed to see how many pions end up in the data sample.
Also, since no positron data was taken at the highest momentum settings at
the lowest angles where the $\pi/e$ ratio was deemed to be very low,
it is important to know what the maximum contamination is.

The $\pi/e$ ratio is largest at the lowest momentum setting of the HMS, where
it is 100:1 for heavy targets.
To calculate the pion contamination of the electron sample, it is
necessary to calculate the pion rejection of the \v{C}erenkov and of the
calorimeter.  This is done by using a set of acceptance cuts and a
strict particle ID cut on one of the detectors to get a clean pion
sample.  Then, the other detector's particle ID cut (used in
the analysis)  is applied and the remaining pions are counted.  

The problem is that even the ``pure'' pion sample is contaminated
with electrons no matter how tight the cuts.  The best solution is to
determine the efficiency of the cut at rejecting electrons and then to
calculate how many of the particles in the ``pure'' pion sample are
actually electrons and subtract them.  For example, to select a
``pure'' pion sample using the calorimeter, events with a signal of
$<$0.45 in the $E_{cal}/p$ spectrum are chosen.  In order to eliminate
possible electrons from this sample, a ``good'' electron sample is
selected (using a very tight \v{C}erenkov cut of 5 photoelectrons) and
the fraction of events in this sample that have a calorimeter signal
of $<$0.45 are then subtracted from the original pion sample.
Finally, the ratio of the pions obtained in this way to the number that
remains after the \v{C}erenkov cut used in the data analysis is applied
gives the pion rejection of the \v{C}erenkov.

For the setting with the worst $\pi/e$ ratio, the pion rejection rate in \v{C}erenkov and
calorimeter detectors are 500:1 and 100:1, respectively.  This results
in a 0.5$\%$ pion contamination.  At larger momenta, the $\pi/e$ ratio
improves as does the pion rejection of the calorimeter.   Since the
$\pi/e$ ratio falls off very quickly as a
function of the HMS momentum, positron data were not taken for the
highest momentum settings, since the pion contamination as well as the
contribution from the charge-symmetric background were thought be
to negligible.  The degree of pion contamination is shown in
Fig.~\ref{pion_contam} and it's small.

The results in Fig.~\ref{pion_contam} represent the upper limit on the
pion contamination and we expect that after the subtraction of the
charge-symmetric background, it is reduced even further. No correction was made in the analysis for the pion contamination.
Since the pion background is very small, we assign it a 50$\%$
uncertainty which results in a 0.2$\%$ systematic uncertainty in the
cross-section at the $x>1$ settings.
\subsection{Cryogenic Target Aluminum Cell \label{dummy_subtraction_section}}
\begin{figure*}[h!]
\center
\includegraphics[angle=270,width=\textwidth,clip]{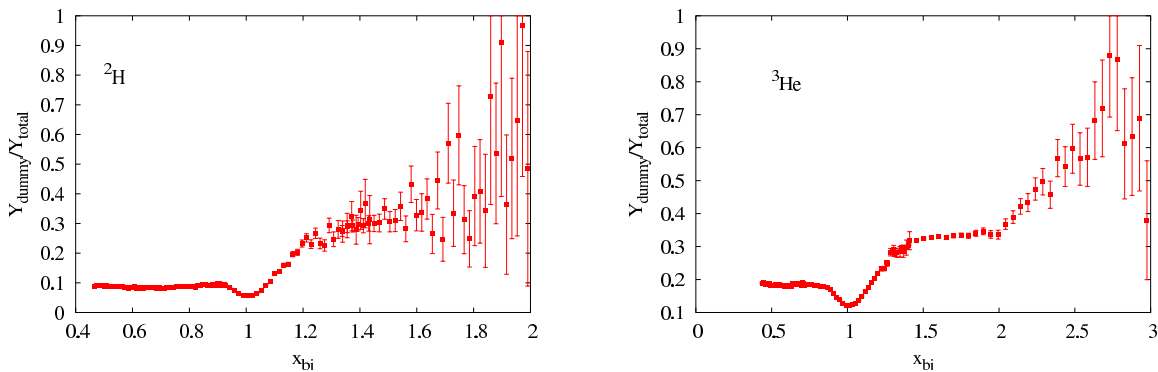}
\caption[Relative contribution from Al end-cap to the cryotarget cross
  section.]{Relative contribution from Al end-cap to the cryotarget cross
  sections for $^2$H and $^3$He. The fraction approaches 1 at
  $x\approx$A, as expected.}
\label{dummy_frac}
\end{figure*}

The cryogenic targets are liquid and contained inside of cylindrical
aluminum cells often called ``tuna cans'' because of their shape.  The cell walls are
$\approx$ 0.12 mm thick (see Table \ref{targetgeo}) and have a density
of 2.699 g/cm$^3$.
This constitutes a significant fraction of the total target thickness
and results in a large contribution to the measured yield:
$\approx$10$\%$ for $^3$He and $\approx$20$\%$ for $^4$He, $^2$H at
$x<1$ and as high as 80$\%$ for $x$ as it approaches $A$ (see Fig.~\ref{dummy_frac}).
This means that a significant fraction of the electron
counts detected come from electrons which scattered from the walls of
the cryogenic target container instead of the cryogenic target material.   This yield is measured
directly by taking data on an empty aluminum target, a.k.a. the
``dummy'' target.  There exists only one dummy target and the data
collected with it were for all the cryogenic targets. 
The dummy target  consists of two aluminum plates, thicker than the
walls of the
cryogenic target cylinders, which reduces the data collection time.  The
dummy target yields must be scaled by the dummy wall thickness before it is
subtracted from the cryogenic target yield.  The ratios of the wall
thicknesses of the dummy target relative to the cryogenic target cells are
listed in Table ~\ref{dummyratios}.


\begin{table}[h]
\begin{center}
\caption{Ratio of dummy to cryotarget wall thicknesses.}
\vspace*{0.25in}

\begin{tabular}{|c|c|c|}
\hline
Target &
Loop &
Ratio ($\frac{t_{dummy}}{t_{cryo}}$) \\
\hline\hline
$^1$H & 2 & 7.757$\pm$0.167 \\
$^2$H & 3 & 7.815$\pm$0.231 \\
$^3$He & 2 & 7.757$\pm$0.167 \\
$^4$He & 1 & 7.079$\pm$0.228 \\
\hline
\end{tabular}
\label{dummyratios}
\end{center}
\end{table}

These ratios are known to within 2-3$\%$. 
The dummy and cryogenic target data are subjected to identical cuts
and the charge symmetric background is subtracted from the dummy
data in the same way as it is from the cryogenic data.  Finally, the dummy yield is subtracted from the data yield.

\begin{figure*}[h!]
\center
\includegraphics[angle=270,width=0.65\textwidth,clip]{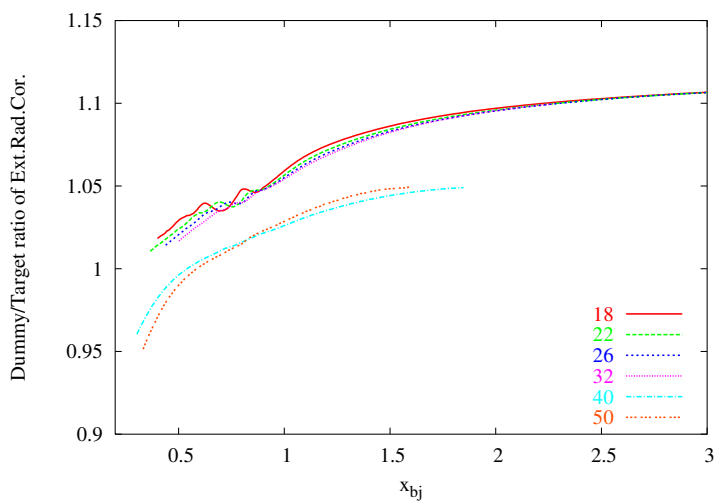}
\caption[External Radiative correction from end-caps of the cryogenic
  target.]{Ratio of the sizes of external radiative corrections for
  the aluminum dummy and the cryogenic target end-caps.  The
  noticeable decrease in the size of the correction at 40$^{\circ}$
  is due to the electron not passing through both dummy foils at large
  angles.}
\label{dummy_ext_radcor}
\end{figure*}

   An additional correction is applied to the dummy  to correct for energy loss due to
external radiative effects.  The dummy's thickness is several times
 the radiation length of the normal cryogenic target walls, which means
that the radiative effects need to be calculated separately for both cases.
The ratio of the two radiative corrections is applied as a weight to each event as it's
binned for runs taken with the dummy target.  A cross-section model is used to calculate this
correction and it is on the order of a few per cent.  We believe our
knowledge of this
correction to 10-20$\%$, implying a contribution 1$\%$
to the systematic uncertainty of the cross section.  The size of the
external radiative correction is shown in Fig.~\ref{dummy_ext_radcor}.

 Since the dummy contribution
to the cross section depends on $x$, a constant systematic uncertainty
could not be assigned. Instead, an error of 3$\%$ of the size of the
dummy contribution is used.  This ranges from 0.3$\%$ to 2.4$\%$.

\subsection{Boiling of the Cryogenic Targets \label{target_boiling_section}}

During production running with cryogenic targets, the 80$\mu$A beam is incident on a 2$\times$2
mm$^2$ area, depositing on the order of  100W of heat into the target
(130W for $^2$H, for example). If enough
energy is absorbed by the cryogenic target, its density can change
which has a direct effect on the normalization of the yield.  This effect
is referred to as 'target boiling.'  Data was collected specifically to measure the variation of the
cryogenic target densities as a function of beam current.  All
cryogenic targets were used as well as the carbon target,
 which should show no current-dependent changes in
yield, since it's solid and melts only at 3800K.

\begin{figure*}[h!]
\center
\includegraphics[height=4in,clip]{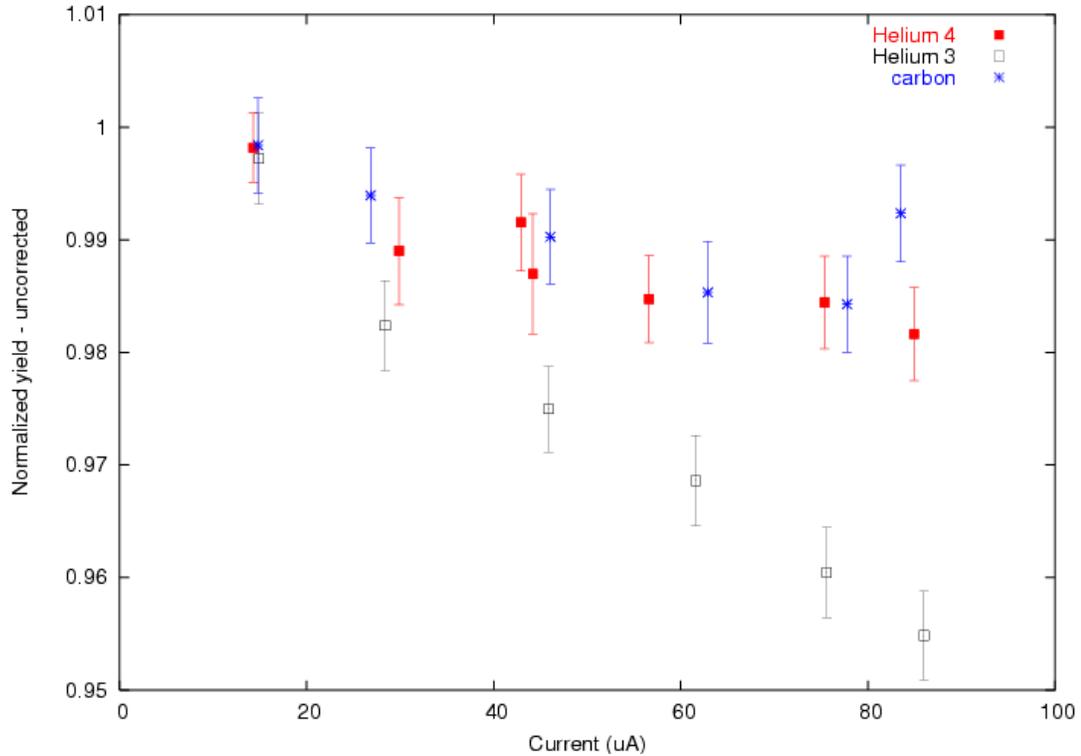}
\caption[Yields from the fourth luminosity scan.]{Yields from the fourth luminosity scan, in arbitrary units.  Target
  boiling observed for all three targets, including carbon.}
\label{precor}
\end{figure*}

Four luminosity scans were taken to obtain data for the
calculation of the target boiling correction.  There were three scans
for the running period with helium and one scan for the
deuterium/hydrogen period.
The yields of the targets for one of the scans before any corrections were made can be see
in Fig.~\ref{precor}.  There's an obvious problem: the normalized carbon yield decreases with increasing
current - carbon appears to boil when it shouldn't.

\begin{figure*}[h!]
\center
\includegraphics[height=3.5in,angle=270,clip]{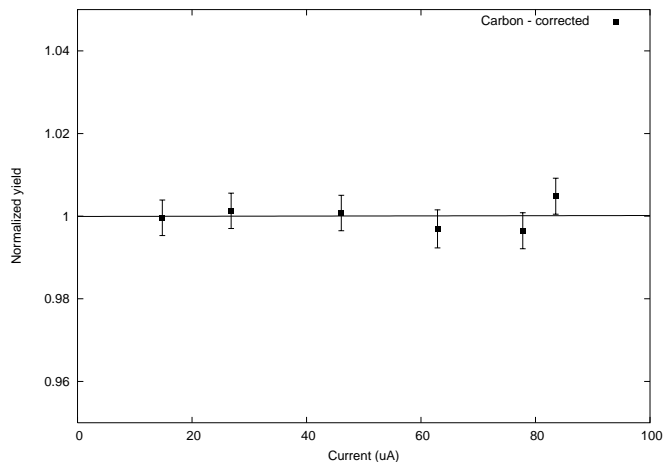}
\caption[Carbon yield for the fourth luminosity scan, corrected for
  BCM offset.]{Carbon yield for the fourth luminosity scan, corrected for BCM
  offset.  No residual slope remains, which means there's no local
  target boiling in this solid target.}
\label{carcor}
\end{figure*}

\begin{figure*}[h!]
\center
\includegraphics[height=3.5in,clip]{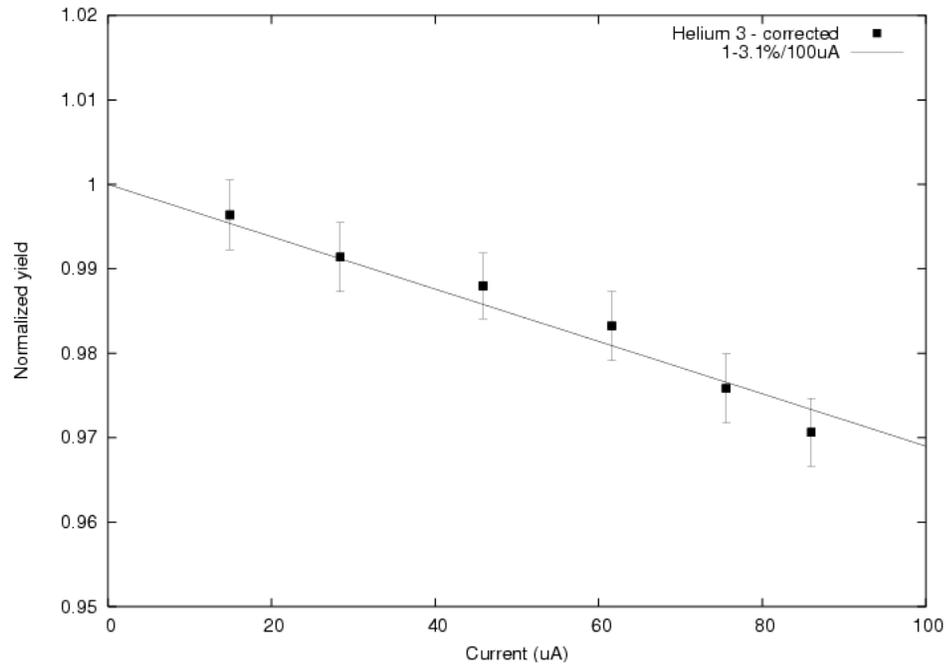}
\caption[$^3$He yield for the third luminosity scan, corrected for BCM
  offset.]{Yield for the fourth $^3$He scan, corrected for BCM offset.  The
  residual slope is the result of local target boiling}
\label{he3cor}
\end{figure*}

After studying this problem at length, this effect was explained by a shift in the BCM calibration.  As
discussed in Sec. ~\ref{bcmcal_sec}, one set of BCM calibration constants
was calculated using all the calibration runs, but some ``local''
fluctuations are allowed by the errors.  The
measured yield is related to the measured current, $I$, as follows:
\begin{equation}
\label{shiftcur}
Y_{measured}=\frac{N_{counts}}{Q}=\frac{N_{counts}}{I\Delta t}
\end{equation}
A shift in the BCM calibration results in an offset $\Delta I$ to the
current, thereby shifting the measured yield.  However, if we can
determine this offset, then we can extract the true yield:
\begin{equation}
\label{yieldcor}
Y_{actual}=\frac{Y_{measured}}{1+\frac{\Delta I}{I}}
\end{equation}
The carbon data were fit to this functional form and the necessary
offset, $\Delta I$ was determined and applied to the cryogenic target data.
Fig.~\ref{carcor} shows the yields for a carbon luminosity scan after
the offset was calculated and applied.  The offsets were on the order
of a few hundred nA, which is reasonable compared to the size of the
residuals for the BCM calibration seen in Fig.~\ref{bcmres}, and are
taken into account in the uncertainly of the charge measurement.
Once the cryogenic target luminosity runs were corrected for the BCM
offsets, the residual slope was taken to be the result of the target
boiling.  The data were fit to a straight line and the fit parameters (Fig.~\ref{he3cor}) were applied to the cryogenic targets in the data
analysis as a density correction.

\begin{table}[h!]
\begin{center}
\caption{Cryotarget Boiling Corrections}
\vspace*{0.25in}

\begin{tabular}{|c|c|}
\hline
Target &
Correction (\%/100uA) \\

\hline\hline
$^1$H & 0.0 \\
$^2$H & 0.0 \\
$^3$He & -3.10 \\
$^4$He & -1.27  \\
\hline
\end{tabular}
\label{boilcor}
\end{center}
\end{table}

The size of the boiling corrections can be seen in
Table.~\ref{boilcor}.  No target boiling was observed for $^1$H and $^2$H
after applying the BCM offset, which is in agreement with previous
experiments in Hall C that used very similar targets and raster. When analyzing the production data,
the cryogenic target yields were corrected by applying a factor
$C_{boil}=1-m_{boil} \cdot \bar{I}$, where
$\bar{I}$ is the average beam-on current for a given run and
$m_{boil}$ is the slope from the fit to the luminosity data.  This
correction accounted for the change in the thickness of the
cryogenic targets due to the energy deposited by the beam and adds 0.5$\%$
systematic uncertainty to the cryogenic target cross sections.  While
only the helium targets exhibit changes in density with higher
currents, the uncertainty is applied to data for all cryogenic targets.


\section{Extracting Cross Sections}

After running the Hall-C replay program, the data are now stored as a
series of CERNLIB HBOOK ntuples ~\cite{hbook}, which contain
event-by-event information on tracks, beam quantities, reconstructed target
quantities, timing and PID quantities, and
calculated kinematic variables.  The analysis code was written in Fortran which
is a natural complement to HBOOK ntuples and allows for easy access to and
manipulation of the data.

There was a series of central angle and corresponding central momentum
settings chosen for production data taking.  Data were collected on
every target for the same angle and momentum settings.
 From now on, ``data setting'' will be taken
to denote a certain angle and central momentum setting of the HMS.

\subsection{Approaches to obtaining $\sigma$}
There are two approaches to extracting cross sections: the method of
corrections and the ratio method.  In the first approach, one starts
with the measured counts, obtains a charge-normalized yield ($Y_{data}$, corrected for charge and detector
efficiencies) and then applies a series of corrections to extract the
differential cross section:
\begin{equation}
\frac{d \sigma}{d \Omega dE}=\frac{Y_{data} \cdot C_{rad} \cdot
  C_{bin}}{\Delta E'\cdot \Delta \Omega \cdot
  N_{scatterers}}
\label{cs_method_cors}
\end{equation}
where  $C_{rad}$ is a correction due to radiative effects, $C_{bin}$ is
  a bin-centering correction, ($\Delta E'\cdot \Delta \Omega$) is the phase space, and $N_{scatterers}$ is the number of
  scattering centers in the target.  The hard part of this process is obtaining
  the correction factors.

The ratio method requires the same correction factors, but the
idea behind it is different.  The goal is to take a Monte Carlo
yield and apply effects that appear in the data, such as radiative
corrections or spectrometer acceptance.  If the experimental effects
are properly simulated, the yields from the Monte Carlo and the
experiment will be in agreement and the input cross section model will
be a good representation of the measured cross section.  This is
equivalent to:
\begin{equation}
\frac{\sigma ^{born}_{data}}{\sigma
  ^{born}_{model}}=\frac{Y_{data}}{Y_{MC}} 
\label{cs_method_ratio}
\end{equation}
where $Y_{data}$ is the charge normalized data yield integrated over
the acceptance of the spectrometer, $Y_{MC}$ is the Monte
Carlo simulated yield, $\sigma ^{born}_{model}$ is the Born cross
section model (description follows in Sec.~\ref{cs_model}), and $\sigma ^{born}_{data}$ is the
quantity we're interested in.
If the acceptance function, \textit{A}, is known, then the data yield
can by simulated via
\begin{equation}
Y_{MC}=N_{scatterers} \cdot 
\int _V A(V)\: \sigma^{model}R(V)\:C_{det}\:dV, 
\label{y_accp}
\end{equation}
where $A$ is the acceptance function of the
 spectrometer (discussed in Sec.~\ref{accp_section}), $C_{det}$ are
 kinematic-dependent detector efficiencies, $\sigma_{model}$
 is a cross section model, $V$ is the volume of the phase space, and $R$ is radiative effects that are
 also present in the data.
 
 While the first method is more intuitive and easier at the beginning
of the analysis process (before a good model of the cross section is
obtained), the code associated with it can
quickly become extensive, difficult to maintain, and not transparent
to the outside observer.  Also, the first method assumes that all of
the corrections factorize and can be calculated and applied separately. While the cross sections were obtained using
both methods and were found to be in excellent agreement, the ratio
method is more reliable and the final results quoted will be those
obtained with this method.
The subsequent sections will outline the
procedures followed obtain cross sections and the processes
for calculating the various corrections.
%
%
\subsection{Obtaining Yields \label{yields}}
The first step to extracting a cross-section is to obtain a
charge-normalized data yield.
For a given data setting, a list of all the data runs for each target is made and the
ntuples are opened and read in one at a time.  Each ntuple is cycled
through event by event, and electron as well as acceptance cuts are
applied.  If the event passes all the cuts, it then placed into a 1-D
histogram, binned in
$x_{bj}$, which is calculated from $\theta$ and $E'$, which in turn are
calculated from $x'_{tar}$, $y'_{tar}$, and $\delta$.
To obtain a yield for a given
run, $i$, the following is used:
\begin{equation}
Y(i)=\frac{N(i)}{Q(i)*\rm{\textit{Eff}}(i)}
\end{equation}
where $Q(i)$ is the accumulated charge for a given run and \textit{Eff(i)}
is the product of several efficiencies (computer live time, fiducial efficiency,
prescale factor).  One possibility is to store each run in a separate
histogram, which would be combined by way of an
error-weighted mean with
\begin{equation}
Y= \frac{\sum(Y_i/\sigma_i^2)}{\sum1/\sigma_i^2}.
\end{equation}
Poisson statistics are used to determine the error on the yield, which
assumes \mbox{$\sqrt{N} \ll N$}.  However, since statistics are limited in
some bins, we use a different way of
combining runs that is a better application of Poisson statistics when
there are few events:
\begin{equation}
<Y>=\frac{\sum_i N(i)}{\sum_i Q(i)\cdot \rm{\textit{Eff(i)}}}.
\end{equation}
This method acknowledges the fact that the divisions between runs
are arbitrary and treats all the runs at a given setting as one long
data taking run.  As the counts from each run are placed into the histogram, we also keep a running total of the quantity in the
denominator, the efficiency-corrected accumulated charge.  The
efficiencies included in this total are those that do not vary with
kinematics, but are constant for a given run (e.g. dead time, trigger efficiency).

%

\subsection{Electronic Dead Time \label{e_dt}}
Once a trigger is formed in the HMS, the gate is activated, and if
another event arrives, it will be ignored.  This is the main source of
electronic dead time.
The events in the HMS occur randomly in time and obey Poisson statistics.  For a mean particle
rate $R$, the probability for detecting $n$ events in time $t$ is
given by:
\begin{equation}
\label{poisson}
P(n)=\frac{(Rt)^n e^{-Rt}}{n!}
\end{equation}
with the probability distribution for the time between events given
by:

\begin{equation}
P(t)=R \\e^{-Rt}
\end{equation}

If an event produces a trigger and is accepted, no new events can be
accepted for time $\tau$, which is the gate width of the logic
signal.  For small dead times, the fraction of events that will be
detected is the probability that the time between events exceeds
$\tau$, which gives us the live time, $t_{live}$:

\begin{equation}
t_{live}=\frac{N_{triggers}}{N_{total}}=\int^{\infty} _{\tau} Re^{-Rt}\\dt=e^{-R\tau}
\end{equation}
 
For E02-019, the gates in the logic modules were 40ns with the
exception of the hodoscope discriminators, whose width was 50ns.
However,  the hodoscopes continue to accept signals even if their
outputs are active and extend the output signal to 60ns after the most
recent hit.   The rates
for E02-019 ($R\le$ 1MHz) allow for the live time to be  be approximated by the
first few terms of the Taylor expansion ($1-R\tau$), giving $R\tau$
as the dead time.  In this approximation, the dead time is a simple
linear function of the gate width and can be calculated by measuring
triggers of different gate widths and extrapolating to zero dead
time.  This allows us to calculate the number of triggers lost in time
$\tau$, giving the dead time as:
\begin{equation}
t_{dead}=\frac{N_{lost}}{N_{total}}=\frac{1}{N_{total}} \cdot
\frac{N_1-N_2}{\tau _2 -\tau _1} \cdot \tau ,
\end{equation}
where $N_1$ and $N_2$ are the numbers of events corresponding to gate
widths of 100ns and 200ns ($\tau _1$ and $\tau _2$), and $\tau$ is the
hodoscope gate width.  $N_1$ was used in the denominator since it was
measured directly, and for the small dead times of E02-019, this is a
good approximation.
The electronic dead time was calculated for each run and was never
greater than 0.2$\%$.  It was
applied as an efficiency (in the form of 1-$R\tau$) to the yield.
Since the inefficiency was extremely small, no systematic uncertainty
is assigned to it.

\subsection{Computer Dead Time\label{comp_dt}}

Another source of dead time is the data acquisition system, which
cannot accept new events while the previous one is being written.  It
takes about 350$\mu$s to write the data for an event to storage.  When
the computer dead time became significant, the prescale factor was increased,
so that not all good events were recorded.  The computer live time is 

\begin{equation}
t_{c.l.t.}=\frac{N_{triggers}}{N_{pretriggers}},
\end{equation}
where $N_{pretriggers}$ is the number of
good triggers sent to the TS, and $N_{triggers}$ is the number of
triggers processed and recorded.   The prescale factor was adjusted
throughout the data taking to keep the dead time $<$20$\%$.  The
final correction factor for the computer live time includes the
prescale factor.

\subsection{Trigger Efficiency \label{trig_eff_section}}

Not all good electron events form a trigger since the detectors are
not 100$\%$ efficient.  The trigger, discussed in Sec.~\ref{trigger},
is made up of 2 components, which are ORed together. Both components
have particle identification elements in them as well as a scintillator trigger. Since the software cuts in the analysis are
more strict than the PID in the trigger, we correct for the
inefficiency at the software level.  However, there's still an
inefficiency associated with the scintillator trigger, SCIN or '3/4'.

\begin{figure*}[h!]
\center
\includegraphics[height=4in,clip]{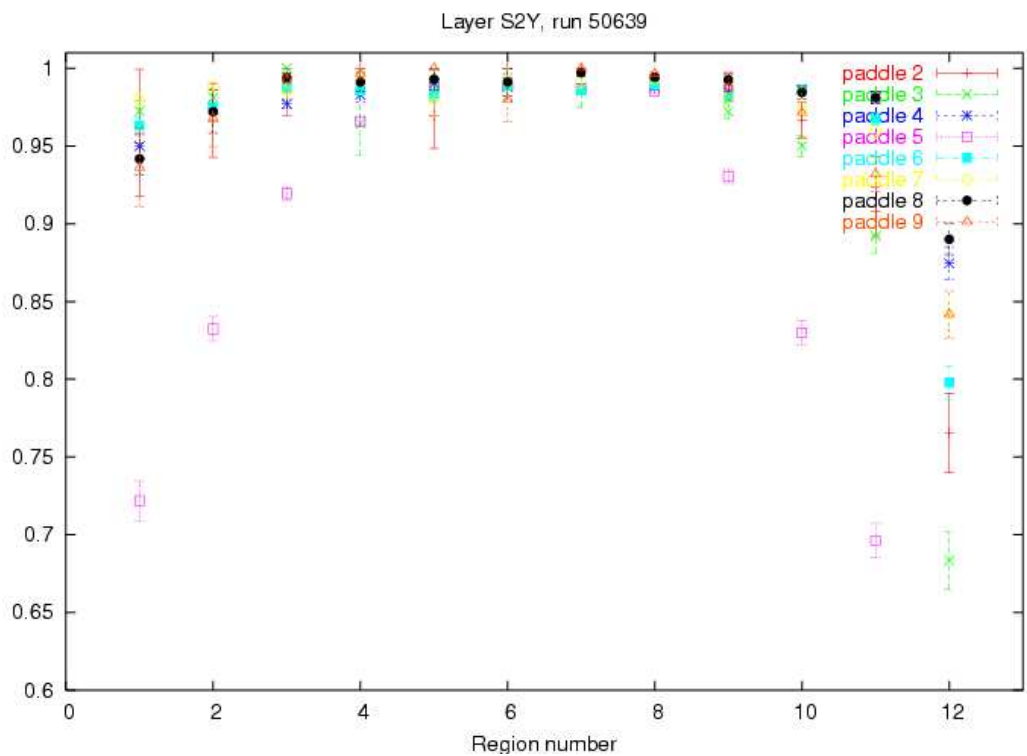}
\caption['SCIN' efficiency in the S2Y HMS plane.]{The fractional
  efficiency, '3/4', for the paddles of the fourth layer of the
  hodoscope detector.  The paddle was divided into 12 equally sized
  regions and the 3/4 efficiency was calculated for each using the
  same method as for the whole detector.  The efficiency shows a
  fall-off at the edges of the paddles.  This effect comes from the attenuation of the
  signals from events at the other end of the scintillator pads, which
  are not always detected by the aging PMTs.}

\label{trigparam}
\end{figure*}
The SCIN trigger requires hits in three scintillator planes for a
track.  The efficiency is calculated separately for each plane as a
ratio between the number of events that fired a given plane and the
number of events that should have fired it.  Each event is examined
and if it produced a hit in the other three scintillator planes, the
running sum
in the denominator is incremented. If that event also fired the plane
being examined, then the total in the numerator is also incremented.

The efficiency
for the entire detector is then determined using the possible permutations
of three planes firing.   This efficiency was found to be constant at
99.3$\pm$0.05$\%$ for runs at x$<$1 kinematics, and decreased for x$>$1 runs.  This trend
was investigated and it was determined to be the effect of aging PMTs, as the
efficiency was higher in previous years.  The efficiency of small
regions of the scintillator paddles was calculated and it was found that
the center regions are almost 100$\%$ efficient and the efficiency
decreases as one moves toward the ends of the paddles (Fig.~\ref{trigparam}).  The PMTs at both ends of a
given scintillator paddle need to fire in order to count as a hit in
that plane.  A signal
from an event at the end of the paddle gets attenuated on the way to the
PMT at the opposite end and the signal may fail to fire the
discriminator.  The runs at x$>$1 kinematics don't
populate the entire focal plane, but rather tend to live near the edges of the
acceptance, which is why they have a decreased SCIN
efficiency. This event distribution also results in a correlated
inefficiency, where a low efficiency from edge events in the 1Y plane
is combined with an even lower efficiency for the same events in the
2Y plane, as they are even closer to the edges of the detector.
However, since this is a position-dependent effect, it is corrected
for with a position-dependent acceptance function described in
Sec.~\ref{accp_section}. 

 This overall trigger efficiency is determined by the ELLO leg of the
 trigger, which has a higher efficiency than the ELHI leg, given that
 STOF (2/4) and SCIN (3/4) are
correlated.   Working out an expression for ELLO using the diagram in
Fig.~\ref{elreal}, we find that the efficiency is given by
ELLO=STOF*PRLO+SCIN*(1-PRLO). The PRLO efficiency was determined to be
99.95$\%$ using elastic scattering data, and calculating the ELLO
efficiency run-by-run, it was found be 99.7
$\pm$0.1$\%$, which some decrease for $x>1$ similar to the behavior of
the SCIN efficiency.  

\subsection{Tracking Efficiency \label{tracking_eff_section}}

A good event must not only fire the electron trigger, but also
reconstruct a good track.  In order to count as good, an event must fire the trigger, have
'forward going' time of flight, and have fewer than 15 hits in one of
the chambers, but at the same time have enough hits to resolve the
left-right ambiguity (not only which wire detected the hit, but which
side the particle passed on).  Events with more than 15 hits are
discounted on the assumption that those hits are  generated by
electrons scraping the dipole exit window and creating a shower of
particles.  The tracking efficiency is
calculated by taking the number of good tracks formed and dividing by
the events with good hits that should have formed tracks.  

\begin{figure*}[htpt]
\center
\includegraphics[angle=270,width=0.7\textwidth]{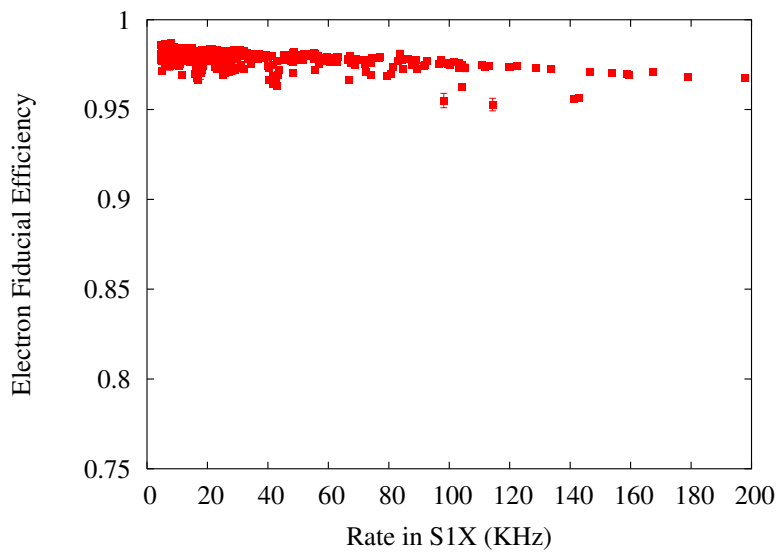}
\caption[HMS Tracking Efficiency.]{HMS Tracking efficiency as a
  function of rate in S1X.  Lower efficiencies were a result of higher rates.}
\label{trackingrate}
\end{figure*}
The tracking efficiency is calculated by selecting events within a small region
around the central ray (fiducial cut) and applying a particle ID
cut.  The fiducial cut selects events from the region where signal to
noise ratio is the largest and the particle ID cut rejects
background events. Pions are able to dissipate energy through elastic collisions or
molecular excitations in addition to ionization, making their
ionization cross section lower than that of electrons. This results in a
lower tracking efficiency for pions.  

  Any events within the fiducial region that  fire the
\v{C}erenkov (0.5 or more photoelectrons) and calorimeter (E$\geq$0.7$\cdot
p_{track}$) should form a track.  The
fraction of those that do form a trigger give the tracking efficiency. 
 The tracking efficiency for E02-019 is shown in
 Fig.~\ref{trackingrate}.  The efficiency is calculated for every run
 and the yields are corrected for it.  A systematic uncertainty of
 0.5$\%$ is assigned to the tracking efficiency based on its variation
 with the event rate.

\subsection{Detector Cut efficiencies \label{detector_eff_section}}                   

The \v{C}erenkov and Calorimeter detectors both have momentum
dependent efficiencies and corrections must be made for these dependences.

\begin{figure*}[h!]
\center
\includegraphics[height=4in,angle=270,clip]{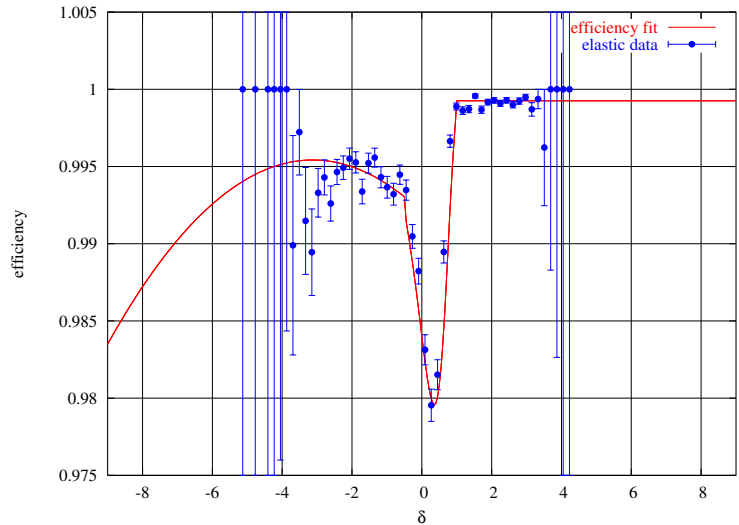}
\caption[HMS \v{C}erenkov efficiency.]{\v{C}erenkov efficiency as a
  function of $\delta$ ($\Delta p/p$) for a
  Carbon run at 4.35 GeV/c and the corresponding parametrization.
  There are three different parametrizations: one for $\delta < $-0.45,
  another for $\delta >$0.99, and a third for -0.45$\leq \delta \leq$0.99.}
\label{cereff}
\end{figure*}

The \v{C}erenkov detector has two elliptical mirrors and the region
where they meet as well as their edges have lower efficiencies than the rest of the detector.
  The \v{C}erenkov was
 run at 0.35 Atm for this experiment to raise the momentum for
which the pions fire it (4.2 GeV/c), and the low pressure  decreased its efficiency.
Data on elastic $ep$ scattering was used to select pure $e^-$ samples to parametrize the \v{C}erenkov efficiency
as a function of $\delta$ ($\%$ offset from central momentum,
correlated with vertical position) as well as the central momentum
setting of the HMS.  This correction was applied to each event in the
analysis.  The shape of the \v{C}erenkov efficiency function is shown
in Fig.~\ref{cereff}.  A systematic uncertainty of 0.2$\%$ was assigned to this efficiency.

The calorimeter cut (E$>$0.7$\times$p the position of the
electron peak) is also not a 100$\%$ efficient and its ability to
reject pions depends on the energy resolution, hence the central
momentum setting of the HMS.  From past experiments ~\cite{john_thesis}, the
resolution is known to be 6.5$\%$/$\sqrt{E}$ for the HMS calorimeter.  The efficiency was found to be fairly
stable for higher momenta, but was fit to a second order polynomial
for lower momenta.  This gave a calorimeter efficiency of
$\approx$99.7$\%$ for the lowest momentum setting (0.86 GeV/c). For momenta above 1.7 GeV/c, a constant
efficiency of 99.895$\%$ was used.  The variation in the calorimeter
efficiency over the kinematic range of the data was minimal and
therefore no systematic uncertainty was assigned.

\subsection{Energy Loss Corrections}

The incoming and scattered electrons can lose energy through
interactions in the target. The energy of the incoming electron can
be lower than the measured beam energy at the vertex, and likewise, the measured
energy of the detected electron can be lower than its energy at the
vertex.  The calculated physics quantities in the analysis engine are
corrected for an average energy loss, while spectrometer-related
quantities are not.  For example, $x$ is calculated using the
corrected incoming and scattered electron energies, but if one
performs the physics analysis in terms of $\delta$ (a common
practice), then any quantities calculated from it (i.e. not using the
ones already in the engine) need to be
corrected for energy loss.

The energy loss is a result of the charged particle interacting with
matter.  The incoming electron loses energy in the cryotarget cell
walls (for example) and in the target material it traverses before
the scattering vertex.  The scattered particle passes through the
remainder of the target material, which for solid targets depends on
the scattering angle, then the target-chamber exit window, the air gap
and finally, the entrance window to the spectrometer.  Since the
energy loss occur after the beam energy measurement and before the
scattered particle is detected, an energy loss correction must be applied.

The energy loss is calculated by using the Bethe-Bloch equation~\cite{Leo:1987a}, and the energy loss is parametrized in terms
  of momentum transfer with the assumption that the scattering event takes place in the center of the target.

One energy loss correction was applied to the beam energy for the entire
experiment since it was found to be fairly constant. That correction
is 1 MeV, giving a beam energy of 5.766 GeV.  The corrections made to
the energy of the scattered electron were of the same order.  The
corrected momentum was calculated since some of the efficiency
corrections are parametrized as functions of momentum.  The effect on the
cross-section was negligible.
\subsection{Monte Carlo Yield \label{accp_section}}
There are two analysis procedures needed to account for the finite acceptance of the HMS:
software cuts and a model of the spectrometer acceptance.  The first method
involves cutting on the spectrometer quantities listed in
Table.~\ref{accpcuts}.  These cuts are selected to reject events that
did not originate in the target and also to select a region where
the reconstruction matrix elements are well known.
\begin{table}[h!]
\begin{center}
\vspace*{0.25in}

\begin{tabular}{|c|c|}
\hline
Variable & Cut \\

\hline\hline
abs($\delta$) & $\leq$ 8 $\%$\\
abs(y'$_{tar}$) &  $\leq$ 0.12 \\
abs(x'$_{tar}$) &  $\leq$ 0.04  \\
\hline
\end{tabular}
\caption{Acceptance Cuts used in the analysis for both data and MC.}
\label{accpcuts}
\end{center}
\end{table}

For a model of spectrometer acceptance, a simulation of electrons
going through the HMS is required.  In the simplest picture, spectrometer acceptance can be defined as the
probability that an electron event within a certain phase space
(defined by $\delta$, $x_{tar}'$ and $y_{tar}'$) will be
accepted.  
%
Since the HMS can only detect events within a limited range  around the
central momentum (ideally, $\pm$15$\%$) and angle setting ($\pm$30~mr), most of the events will be lost at
the edges due to things like scraping the dipole upon exit or hitting
the collimator.  This is modeled with a Monte Carlo, with the acceptance
being defined as the number of particles that are successfully
transported through a model of the spectrometer.

The Hall C single arm Monte Carlo generates events uniformly distributed in
$x,y,z, \delta, \theta$, and $\phi$.  The particles are then
transported through the magnets of the HMS, which are modeled using
the COSY INFINITY program.  Using a list of magnet parameters such as
their positions, dimensions and field maps, COSY generates a forward
matrix that projects rays at the target to the focal
point. The position of each event is calculated for several points:
beginning and end of each magnet as well as position 2/3 of the way
through the first two quadrupole magnets. 
The tracks in the focal
plane are recorded for all particles which make it all the way through
the detector stack.  The focal plane quantities are determined using the target quantities in the following way:
\begin{equation}
a_{fp}=\sum_{i,j,k,l,m} F^x_{ijklm} (x_{tar})^i (x'_{tar})^k
(y_{tar})^j (y'_{tar})^l\delta ^m\;\;\rm{for}\;\;(1 \leq i+j+k+l+m \leq N)
\end{equation}
where N is the order of the expansion (6 for the HMS) and $ F^x_{ijklm}$ is one
column of the forward transport matrix (there are 4 total, one for
each focal plane variable).

In reality, the cross section (and the acceptance function) is a
function of 6 variables: $x$, $y$, $z$, $x'$, $y'$, and $\delta$.
However, it is possible to simplify it by averaging over the behavior
of several variables.  For example, we can average over $x, y,$
(related to the size of the rastered beam) and
$z$ (related to the target length) for a given target.  We simplify further by producing a separate
acceptance ntuple for every setting of the HMS and each target geometry.  Since the central angle of the spectrometer is fixed
for a given setting, $x'$ and $y'$ can be converted to the lab angles,
and integrated over $\phi$,
so that finally, A=A($\delta$, $\theta$).  In order to reproduce the
measured yield, it's necessary to calculate the number of scattering
centers for a given target, $N_{scatterers}$, and the acceptance integral must be weighted by the cross
section and detector inefficiencies, giving:
\begin{equation}
Y_{MC}=N_{scatterers}\int {A(\delta, \theta) \: \sigma^{rad}_{model}(\delta, \theta)\:C_{det}\:p_{HMS}\:d \delta d\Omega},
\label{y_accp_2}
\end{equation}
where the subscript \textit{model} and superscript \textit{rad} denote the radiated model cross section
evaluated at a given point, $p_{HMS}d\delta$ is equivalent to $dE'$, and $C_{det}$ is the combined
kinematic-dependent detector efficiency, which includes calorimeter
and \v{C}erenkov efficiencies.

\begin{figure*}[h!]
\center
\includegraphics[height=4in]{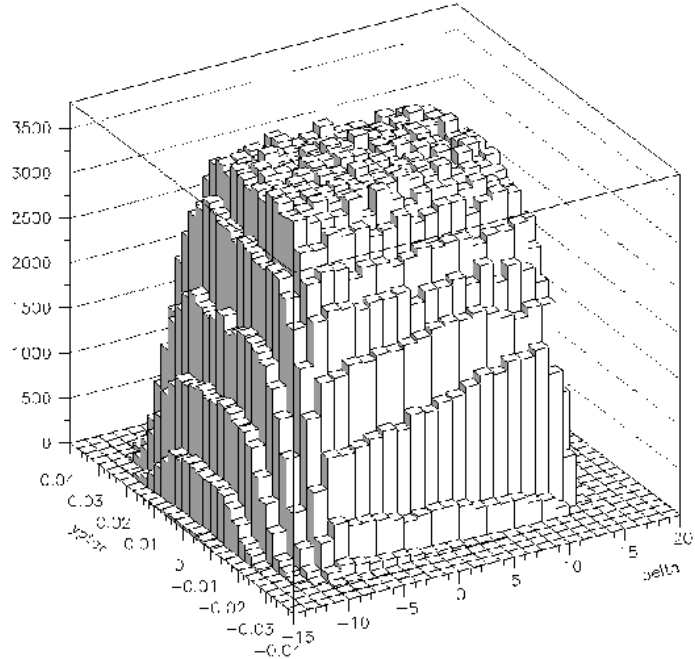}
\caption[HMS Acceptance.]{Shape of the acceptance of the HMS from Monte Carlo in $\delta$ and
  $y' _{tar}$ (unweighted MC counts are shown on the vertical axis).  The variable $y' _{tar}$ is often used in place of
  $\theta_{lab}$ as it is the slope of the
  track in the $y$-direction (horizontal) and is therefore the arctan of the
  angle the track makes with the central angle of the HMS.  This is
  not a good practice for small angles, where the contribution from
  $x' _{tar}$ to the correct calculated scattering angle is non-negligible.}
\label{accept}
\end{figure*}

The shape of the acceptance function is shown in Fig.~\ref{accept} as
a function of $\delta$ and $y' _{tar}$.  The variable $y' _{tar}$ is often used in place of
  $\theta_{lab}$, since it's the arctan of the angle the electron
track makes with the central ray of the HMS in the horizontal plane.  The complete definition of $\theta _{lab}$,
or just $\theta$, is:
\begin{equation}
\theta _{lab} =\frac{\cos\: (\theta _{HMS})+y' _{tar}\:sin\:(\theta
  _{HMS})}{\sqrt{1+x'^2 _{tar}+y'^2 _{tar}}}
\label{theta_lab}
\end{equation}
where $\theta _{HMS}$ is the central angle of the spectrometer.  The
complete definition of $\theta$ was used in this analysis, since at
small angles, the contribution from $x' _{tar}$ cannot be ignored.

Binning the unweighted MC counts in the exact same manner as the data, automatically
generates the acceptance function.
The weighted, simulated yield in a particular bin is given by:
\begin{equation}
Y_{MC}=N_{scatterers}\sum_{events}{ \sigma^{rad}_{model}(\delta, \theta)\:C_{det}\:p_{HMS}\:
  (\Delta \delta \Delta \Omega)_{bin}},
\label{y_accp_2}
\end{equation}
where $\Delta \delta \Delta\Omega_{bin}=(\Delta \delta \Delta
  \Omega)^{gen}_{bin}/(N^{bin}_{gen})$ represents the relative phase space for a
  given event.  The generated solid angle depends on the generation
  limits in $x' _{tar}$ and
$y' _{tar}$, the in- and out-of-plane angles.  For this analysis, 5x10$^6$
  events were generated for each acceptance ntuple, with the
  generation limits of $\delta\pm$15$\%$, $x' _{tar}\pm100$~mr, and $y'
  _{tar}\pm50$~mr.

\begin{figure*}[h!]
\center
\includegraphics[width=\textwidth,clip]{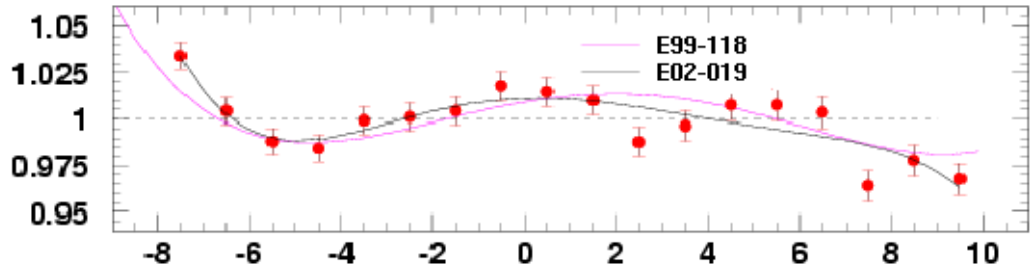}
\caption[Data/MC vs. $\delta$.]{Ratio of Data to Monte Carlo vs. $\delta$.  This is a
  residual acceptance effect, independent of momentum or angle with 2
  fits, one from our experiment, and the other from a different
  analysis~\cite{vladas_thesis}.}
\label{vladas}
\end{figure*}

It was discovered, when combining data from adjacent HMS momentum
settings, that the acceptance function does not
perfectly model the acceptance of the detector and there's an
additional $\delta$-dependent effect.  A polynomial fit from a
previous experiment ~\cite{vladas_thesis} was used since it described the data well
(Fig.~\ref{vladas}). This correction was applied on an event-by-event
basis in the Monte Carlo.  A 1$\%$ scale and a 1$\%$ relative
uncertainty were applied to the acceptance based on
experience from previous experiments in the Hall. 

\subsection{Bin-Centering\label{bc_section}}
The data in this experiment were taken at several angles. Because the HMS
has a finite acceptance
around the central angle setting, the counts have some distribution in
$\theta_{lab}$ for a given $x$  
bin.  This means that the measured data yield in any given
($x,\theta$) bin is an average value rather than the yield at the
center of the bin.  These are not the same unless the cross section has
a linear dependence on $\theta$.  Instead of
correcting the data and moving each event to the center of its bin
using a cross section model, the correction was applied to the MC.  By
putting in a cross section model with the same angular dependence as
the data into Eq.~\ref{y_accp_2}, the simulated yield is averaged
over the acceptance in the same manner as the data.

\begin{figure*}[htbp]
\centerline{\includegraphics[angle=270,width=\textwidth]{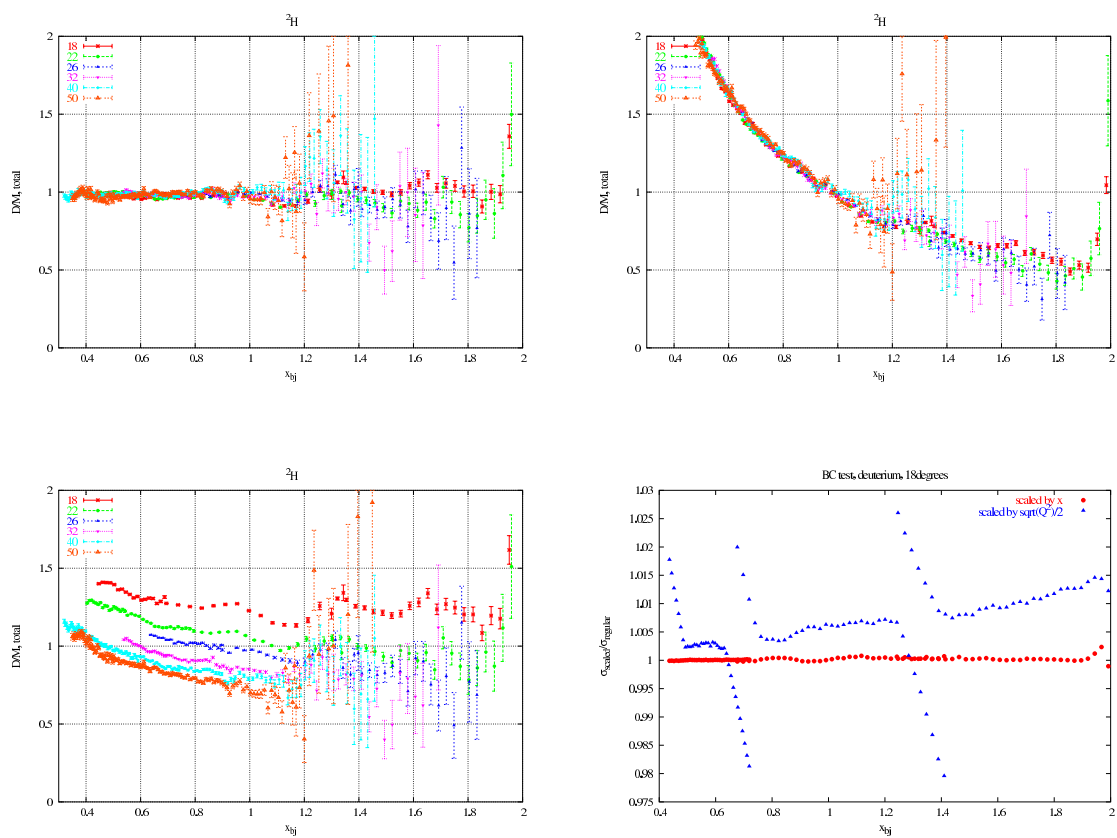}}
\caption[Systematic uncertainty due to BC model.]{The top left plot
  shows the data to model ratio before the bin-centering test.  The
  top right and bottom left plots show the effect on that ratio if the
  model is multiplied by $x$ in one case and $\sqrt{Q^2}/2$ in the other
  case.  The scale factor is a gross overestimate of the disagreement
  between data and optimal model.  The bottom right plot shows ratio of the
  experimental cross sections extracted with the altered model to those
  extracted with the optimal model.  There are three HMS settings in
  the 18$^{\circ}$ data set, and the features at the edges of the
  settings for the  $\sqrt{Q^2}/2$ test are a result of the data being
  binned in $x$, where the edge bins do not include events from the
  whole acceptance.}
\label{bc_test}
\end{figure*}

To determine the systematic uncertainty associated with the choice of
the bin-centering model, it was varied and the experimental cross
sections were recalculated.  The input model was altered in two ways
for this test: in the first, the calculated cross section at each point was multiplied by $x$; in the
second, it was multiplied by $\sqrt{Q^2}/2$.  Fig.~\ref{bc_test} shows the
result for 18$^{\circ}$ deuterium data.  Since the input model was
scaled by more than the disagreement between the original cross
section model and the data, this variation in the extracted cross
section is an overestimate. A conservative uncertainty of 0.5$\%$ in
the bin-centering model was assigned.

\subsection{Radiative Corrections \label{rc_section}}

\begin{figure*}[htbp]
\centerline{\includegraphics[height=1.5in,clip]{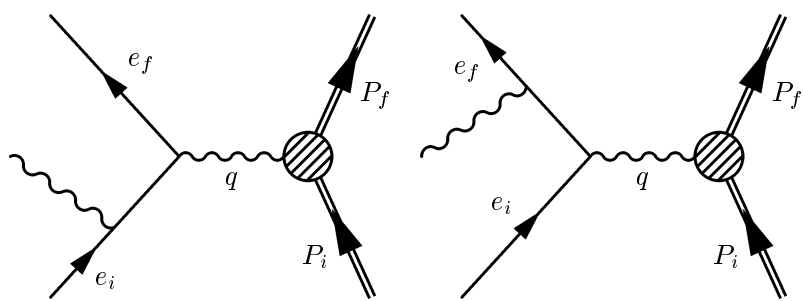}}
\caption[Radiative Processes.]{The incoming and/or scattered electron can lose energy through
  bremstrhalung.  The radiative corrections procedure restores events
  that lose energy in this way to the correct data bin since the event
  may have ended up in a higher or lower energy bin, depending on when
  it underwent bremstrhalung.}
\label{rcdiagram}
\end{figure*}

 The measured cross sections also need to be corrected for effects
 from internal and external radiative processes.  Internal radiative processes
 include vacuum polarization, vertex corrections, and internal
 bremsstrahlung.  An electron can also lose energy through
 bremsstrahlung while passing through the target, which is an external
 effect. Fig.~\ref{rcdiagram} illustrates external bremsstrahlung.  No correction was made for the nuclear elastic contribution.

The  challenge that radiative processes present can be understood by
examining Fig.~\ref{rcdiagram}.  In order to calculate the
contribution of, for example, the process depicted on the left hand
side in the figure, the cross section must be known for beam energies
that are lower than that of the experiment. Similarly, for the
figure on the right hand side, the cross section must be known for
scattered energies greater than those of the experiment.  This
requires a realistic cross section model, which was developed for this
analysis and is described in Sec.~\ref{cs_model}.  

In this analysis, the radiative correction was originally calculated with the method described by Stein
in~\cite{Stein:1975yy} using the peaking approximation method of
Mo and Tsai ~\cite{Mo:1968cg}.  However, this approximation does not
do a good job for thick targets at low $x$ and the full 2-D integral
needs to be calculated.  

The approach adopted was one described in detail by Dasu~\cite{dasu_thesis} and used to do radiative corrections for SLAC
experiments.  In this approach, a complete calculation of Mo and
Tsai's formula for external effects is done.  An equivalent radiator is used to
take into account materials before and after the target, such as air,
aluminum target exit window, mylar and kevlar entrance and exit
windows for the magnets.  Unlike in Dasu's analysis, where the Bardin
prescription ~\cite{Akhundov:1977bh} is used for the internal
corrections, Mo and Tsai's approach is used in this analysis.

The radiated and Born versions of the model cross section
are used to form the correction factor:
\begin{equation}
C_{rad}=\frac{\sigma^{rad}_{model}(x,\theta)}{\sigma^{born}_{model}(x,\theta)}=\frac{\sigma_{MT}^{i+e}(x,\theta)}{\sigma^{born}_{model}(x,\theta)}
\end{equation}
The subscript \textit{MT} denotes Mo and Tsai's equivalent radiator
calculation, and the superscripts \textit{i} and \textit{e} refer to
the internal and external radiative corrections, respectively.
\begin{figure*}[htbp]
\centerline{\includegraphics[angle=270,width=.9\textwidth]{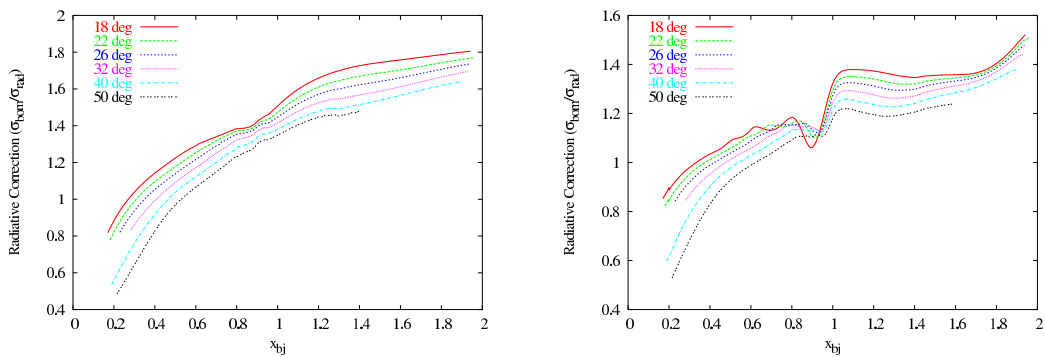}}
\caption[Radiative Corrections for $^2$H and $^{197}$.]{Radiative Corrections for
  $^2$H (right) and $^{197}Au$ (left) at all angle settings.  The y-axis shows the size of the
  radiative correction, which is given by the ratio $\sigma
  _{born}/\sigma _{rad}$.}
\label{rcsize}
\end{figure*}
A cross section model is used to generate a list of Born and radiated
cross section values in increments
10 MeV for each central angle setting of the HMS. 
The correction for each data point is interpolated from the two
closest points and is applied as a correction to the binned, radiated
cross sections.

The size of the correction can be seen if Fig.~\ref{rcsize}.
where the ratio of the extracted cross sections with and without
radiative corrections is shown.

To determine the systematic uncertainty from the choice of the
cross section model used to calculate the radiative correction, the
model was scaled by $x$ in one case, $\sqrt{(Q^2)}/2$ in another case,
(as was done for the bin-centering uncertainty)
and the correction recalculated.  At low values of $x$, the radiative
correction is progressively more sensitive to the quasielastic tail
with increasing values of Q$^2$.  The cross section model was
calculated for the kinematics of previous experiments using the
quasielastic archive~\cite{Benhar:2006er} and found to be in good
agreement in the region around the quasielastic peak down to Q$^2$
values of $\approx 0.5$ GeV$^2$. To determine the uncertainty associated with the quasielastic tail,
the quasielastic contribution alone was scaled by 10$\%$ (somewhat
higher than the level of disagreement between our cross section and
the quasielastic archive) and the correction
factor recalculated.

A systematic uncertainty of 1$\%$ due to the radiative correction
model is applied for the entire kinematic range.  At low $x$, this is
the combined effect of rescaling the entire model and the
quasielastic tail alone, and for $x>1$, the uncertainty is the result of the
model rescaling alone.  An additional uncertainty of 1$\%$ is included
due to the limitations inherent in the calculation, for a total of
1.4$\%$ systematic uncertainty.
\subsection{Corrected Cross Section}
\begin{figure*}[htbp]
\centerline{\includegraphics[height=3.5in,clip]{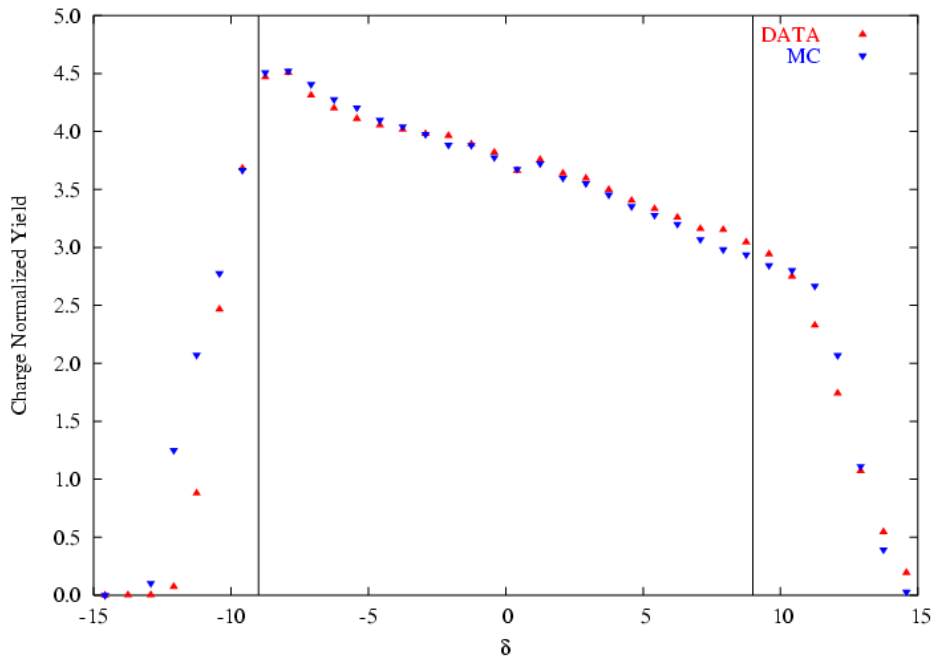}}
\caption[Data and MC yields for Carbon, 40$^{\circ}$, p$_{HMS}$=1.14
GeV/c.]{Data and MC yields for Carbon, 40$^{\circ}$, p$_{HMS}$=1.14
GeV/c, arbitrary units.  The vertical lines at $\pm$9 represent the
software cut applied to the data and MC.  The discontinuity at $\delta \approx$0
is the result of a region of decreased efficiency in the
\v{C}erenkov. This region corresponds to the overlap between the two
mirrors and is well reproduced in the MC.}
\label{data_mc_yields}
\end{figure*}

%
%
Once the measured data yield and the simulated MC yield have been
obtained, the experimental Born cross section can be extracted through:
\begin{equation}
  \sigma(x, \theta _c)=\frac{Y_{data}}{N_{scatterers}\int{\sigma
  ^{rad}_{model}(x,\theta)\:C_{det}\:A(x,\theta) d\Omega}} \cdot
  \sigma ^{born}_{model} (x,\theta _c)
\label{final_answer}
\end{equation}
where the denominator is the Monte Carlo yield, previously described
in Sec.~\ref{accp_section}, and the subscript \textit{c} denotes the central
angle setting of the HMS.  If the Monte Carlo perfectly describes
the spectrometer acceptance and the cross section model used as a
weight for the simulated yield agrees well with the data, the ratio in
Eq.~\ref{final_answer} should be about 1.  An example of data and MC yields is shown
in Fig. ~\ref{data_mc_yields}, where the agreement between the two
is excellent in the $\delta$-region to which this analysis is confined.

\subsection{Coulomb Corrections \label{cc_section}}

It is also important to correctly treat the Coulomb distortion of the
electron wave function by the electrostatic field of the nucleus.  The
nucleus has two effects on the electron: it is accelerated due to
the attractive force when in close proximity to the nucleus,
increasing the electron's momentum and also, the attractive force focuses the electron wave function.

A distorted wave Born approximation (DWBA) calculation using Dirac wave
functions is the correct approach ~\cite{Kim:1996ua} to the problem but it's not a practical solution given the computational load
involved.  Instead, a local Effective Momentum Approximation (EMA) is used ~\cite{Aste:2005wc}.

The first effect is quantified by calculating the shift in the
momentum due to the acceleration by the nucleus.  The enhancement of the momenta is given by: $k' _f=k_f+\Delta
k$ and  $k' _i=k_i+\Delta k$, with $\Delta k=-.775V_0/c$, where
$V_0$ is the potential energy of the electron in the center of the
nucleus.  $V_0$ is the lowest order of the electrostatic potential inside a charged
sphere and is given by 
\begin{equation}
V_0=\frac{-3\alpha(Z-1)}{2R}, \nonumber
\end{equation}
 where $R$ is the
radius of the nucleus. Note that $Z-1$ is used instead of the $Z$ in
the given reference, since we're calculating acceleration due to the
A-1 nucleus.  The nuclear radii used for the different targets are
given in Table~\ref{coulomb_radii}.  The extra factor of 0.775
~\cite{Aste:2005wc} comes
from the fact that the scattering is distributed over the volume
of the nucleus and the potential needs to be modified to reflect the
average potential inside a homogeneously charged sphere.

\begin{table}[h!]
\begin{center}
\vspace*{0.25in}

\begin{tabular}{|c|c|}
\hline
Target & Radius (fm) \\
\hline
$^3$He & 1.96\\
$^4$He & 1.67\\
$^9$Be & 2.70\\
$^{12}$C & 2.89\\
$^{63}$Cu & 4.60\\
$^{197}$Au & 6.55\\
\hline
\end{tabular}
\caption{RMS charge radii.  For heavy targets, the radius is given by
  $R=1.1A^{1/3}+0.86A^{-1/3}$ and previously determined values are used for helium ~\cite{morton:034502}.}
\label{coulomb_radii}
\end{center}
\end{table}

\begin{figure*}[htbp]
\centerline{\includegraphics[angle=270, width=.8\textwidth]{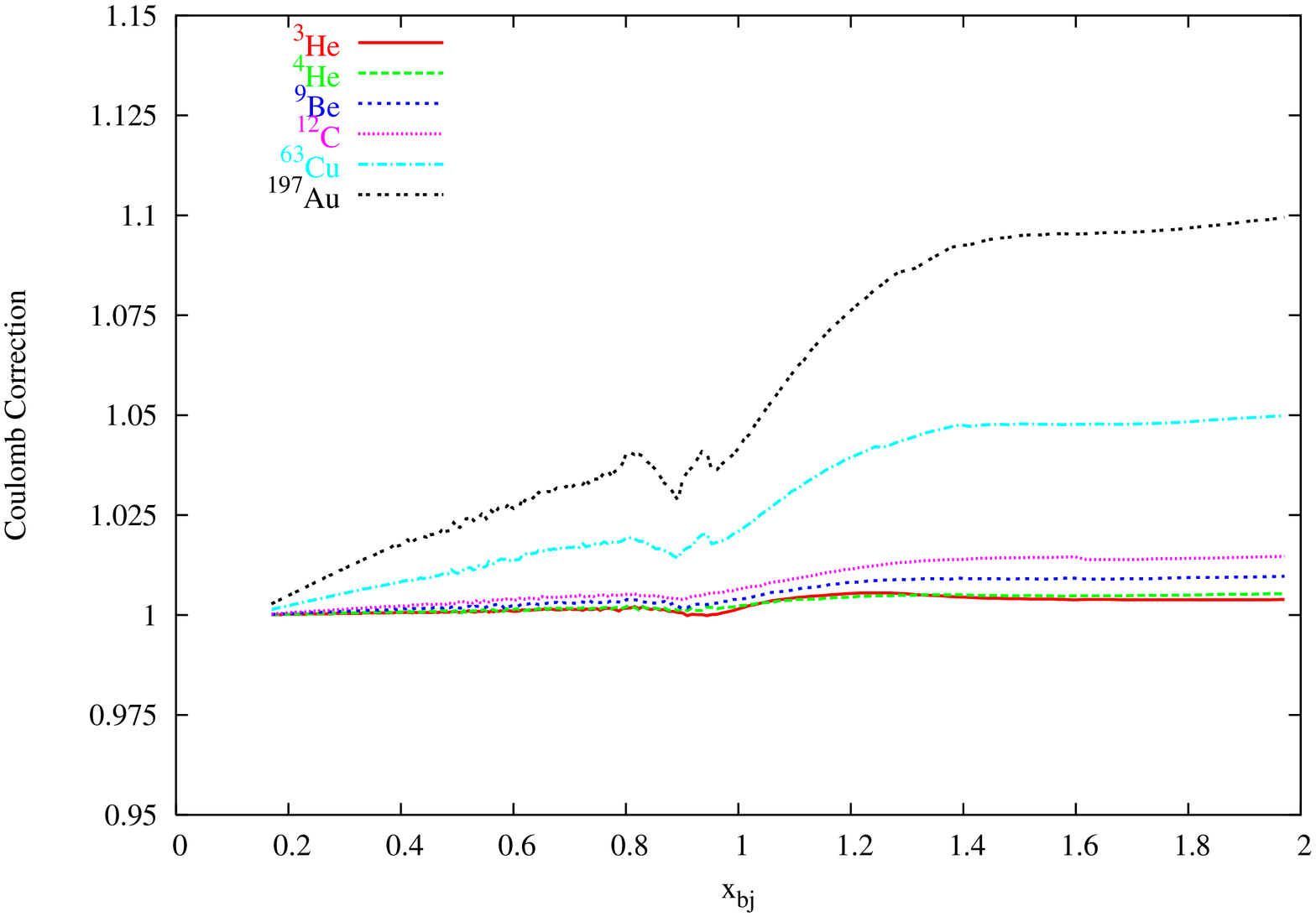}}
\caption[Coulomb Corrections at 18$^{\circ}$.]{Coulomb correction factor for all targets at 18 degrees.  The
dip around $x_{bj}=0.9$ is a result of a transition between two
different models for the inelastic cross section. }
\label{coulcor}
\end{figure*}

The second effect is the focusing of the electron wave function by
the attractive potential of the nucleus and is accounted for with a
``focusing'' factor, which enters into the cross section quadratically.
The wave functions of incoming and outgoing electrons are both
enhanced, but the artificial enhancement of the phase space in
the shifted cross section cancels the focusing factor of the outgoing electron~\cite{Aste:dearjohn}.
 The remaining focusing factor is given by $(k'
_i/k_i)^2$.  

The complete form of the Coulomb correction is then:
\begin{equation}
C_{coulomb}=\frac{\sigma_{born}(k_i,k_f)}{\sigma_{born}(k' _i, k' _f)} \frac{1}{(k'_i/k_i)^2}
\end{equation}
 The size of the correction is shown in Fig.~\ref{coulcor} for all
 targets at 18$^{\circ}$.  The correction is largest for the heaviest target, which for E02-019
is $^{197}$Au, where it reaches 10$\%$ for 18$^{\circ}$, and 20$\%$
 for 50$^{\circ}$.

There is a 10$\%$ uncertainty associated with the Coulomb potential,
which needs to be included in the measured cross section.  To
determine the contribution from this uncertainty to the cross section,
the coulomb corrections were evaluated for each target and kinematic
setting with a 10$\%$ shift applied to the potential.  The ratio of
this shifted correction to the normal correction was parametrized as a
function of $x$ for each $\theta _{HMS}$ setting and for each target.
Evaluation of this parametrization at any data point gives the
systematic uncertainty associated with this correction.  The largest
corrections and therefore, the largest uncertainties are for
$^{197}$Au, where the uncertainty at $x>$1 reaches 0.5$\%$ and 2$\%$
for 18$^{\circ}$ and 22$^{\circ}$, respectively.

\subsection{Cross Section Model \label{cs_model}}

The three sets of corrections described above all require an input
model of  the cross section.  For E02-019, the cross-section model
consists of 2 parts: a quasi-elastic
contribution and a deep inelastic contribution. 

The quasi-elastic contribution is calculated from a scaling function,
$F(y)$, the off-shell electron-nucleon cross-section, and a kinematic
factor, $K$.
\begin{equation}
\frac{d\sigma}{d\Omega
  d\nu}=F(y)\cdot(Z\cdot\sigma_p+N\cdot\sigma_n)\cdot K
\end{equation}
The scaling function $F(y)$ used for $^2$H, a variation of the form
used in~\cite{CiofidegliAtti:1997km}, is given by:
\begin{equation}
F(y)=(f_0-B)\cdot\frac{\alpha ^2 e^{-(ay)^2}}{\alpha ^2+y^2}+B e^{-b|y|}
\label{scaling}
\end{equation}
This form was modified for heavier targets to get:
\begin{equation}
F(y)=(f_0-B)\cdot\frac{\alpha ^2 e^{-(ay)^2}}{\alpha ^2+y^2}+B e^{-(by)^2}
\label{nuc_scaling}
\end{equation}
where the parameters $a$, $b$, $f_0$, $B$, and $\alpha$ are fit to the
$F(y)$ extracted from the data for each target (see Table~\ref{fit_params}).  This is done by
taking the data cross section, subtracting the inelastic contribution
(calculated using the inelastic part of the model), and dividing out
the kinematic factor and the electron-nucleon cross section, which is
described in Sec.~\ref{yscaling}.  After
the fit, the iterated model was used as the input to the cross section
extraction and the process was repeated until good agreement between
the data and the model was achieved for all settings.  
\begin{table}[h!]
\begin{center}
\vspace*{0.25in}

\begin{tabular}{|c|c|c|c|c|c|c|c|}
\hline
 & $^2$H & $^3$He & $^4$He & $^9$Be &$^{12}$C & $^{63}$Cu & $^{197}$Au \\
\hline
f$_0$ &8.742e-03&5.309e-03&4.020e-03&3.481e-03&3.182e-03&2.874e-03&2.642e-03\\
\hline    
B &8.239e-04&2.184e-03&1.345e-03&1.161e-03&1.359e-03&8.866e-04&7.632e-04\\
\hline    
a &7.727e-03&2.886e-03&2.699e-03&3.120e-03&3.027e-03&3.096e-03&3.065e-03\\
\hline    
b &9.394e-03&1.035e-02&7.494e-03&7.840e-03&7.050e-03&7.094e-03&6.768e-03\\
\hline    
$\alpha$ & 45.3& 64.2&100.2&110.9&137.2&132.4&132.4\\
\hline
\end{tabular}
\caption{F(y) fit parameters as determined by fitting
  the experimental scaling function to the form of Eqs.~\ref{scaling}
  and~\ref{nuc_scaling}.}
\label{fit_params}
\end{center}
\end{table}

An additional correction to the quasi-elastic model was introduced.
Since the quasi-elastic model is based on $y$-scaling, it doesn't work
perfectly for the largest negative $y$ values, where the low $Q^2$
data is affected by final state interactions (FSIs) and doesn't
scale.  A polynomial fit for each target was done for the
ratio of the data to the model at large negative $y$'s ($x_{bj}>1.4$) to account for this effect.

The inelastic cross section is calculated slightly differently for
over the range of $x$.  For $x<0.8$, parametrizations~\cite{Bosted:f2stuff} of the proton and neutron structure functions
($F^{n}_2$ and $F^{p}_2$) are used and they're smeared using the
momentum distribution $n(k)$ which is determined from the derivative
of $F(y)$ used in the quasi-elastic model cross section via Eq.~\ref{fy_nk_simple}.   The
inelastic cross section is obtained using the resulting nuclear
structure functions through Eq.~\ref{inelastic_sigma_simple}. A polynomial
correction function is then applied to the result to reproduce the
shape of the data. This function is different for every target.
\begin{figure*}[htpt]
\center
\includegraphics[width=\textwidth,clip]{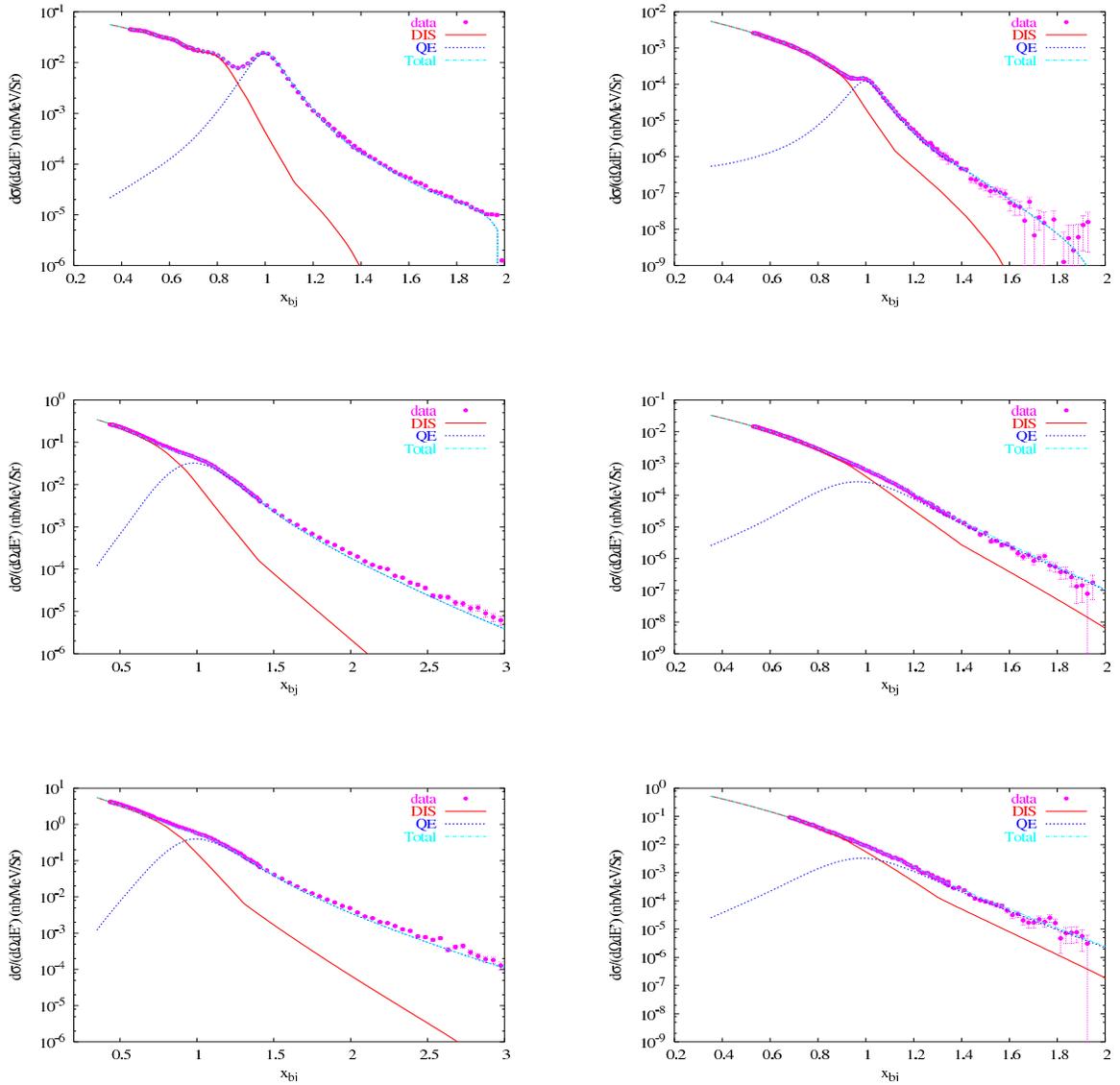}
\caption[Data and model cross sections for $^2$H, $^{12}$C, and
  $^{197}$Au at  18$^{\circ}$ and
  32$^{\circ}$.]{Data and model cross section for $^2$H (top),
  $^{12}$C (middle), and
  $^{197}$Au (bottom) at
  18$^{\circ}$ (left) and 32$^{\circ}$ (right) .  The model cross
  section is separated into the inelastic and quasi-elastic pieces to
  show their relative contributions.  The rise of the data cross section
  around $x\approx$ 2 for $^2$H is the result of elastic electron-deuteron
  scattering contribution.}
\label{ld2_data_model}
\end{figure*}
In the $x_{bj}>0.9$ region, only the smearing prescription is used with the
corresponding $n(k)$ for a given target.   For the region where $0.8 < x_{bj} < 0.9$, a weighted average of the
two prescriptions is used so that the transition between the two regions
is continuous.  Once a good model was obtained
for the quasielastic region, it was subtracted from the measured data
cross section and the resulting inelastic data was compared to the
inelastic model.  A residual slope in the inelastic data/model ratio was found for
$x>0.9$ for all angles, and it was corrected for with a straight line
fit.  

The data as well as model cross sections, including quasielastic and
inelastic contributions, are shown in Fig.~\ref{ld2_data_model} for several targets at 18$^{\circ}$
and 32$^{\circ}$ .


\chapter{Results}
In this chapter, the results will be presented, starting with cross
sections and following up with the extracted scaling and structure
functions.  The sources of systematic uncertainties will be discussed
in Sec.~\ref{sys_err_section}. Throughout this chapter, the results
shown will include only the statistical error bars.
\section{Cross Sections}
The measured cross sections are shown in
Figs. \ref{ld2_he3_sigma_model}-\ref{cu_au_sigma_model}, excluding the
Beryllium and Copper targets for space reasons alone.  The six sets of data in each plot correspond to the
six central angle settings of the HMS: 18$^{\circ}$, 22$^{\circ}$,
26$^{\circ}$, 32$^{\circ}$, 40$^{\circ}$, and 50$^{\circ}$.  The
Q$^2$ and $E'$ values associated with the center of the quasielastic
peak are shown in Table~\ref{q2_values} for each central angle
setting. The data
are plotted as a function of energy loss, $\nu$, and the characteristic
quasielastic peak can be seen especially well in the deuterium data
and its evolution as the relative contributions from the quasielastic
and inelastic reactions change can also be seen. 
\begin{table}[h!]
\begin{center}
\caption{Central Angle settings and corresponding Q$^2$ values at the
  quasielastic peak ($x$=1)}
\vspace*{0.25in}

\begin{tabular}{|c|c|c|}
\hline
$\theta_{HMS}$ &
E' (GeV) &
Q$^2$  (GeV$^2$) \\
\hline\hline
18 & 4.43 & 2.5\\
22 & 3.98 & 3.3\\
26 & 3.55 & 4.1\\
32 & 2.98 & 5.2\\
40 & 2.365 & 6.4\\
50 & 1.805 & 7.4\\
\hline
\end{tabular}
\label{q2_values}
\end{center}
\end{table}

\begin{figure*}[h!]
\center
\includegraphics[angle=270,width=\textwidth]{ld2_he3_data_model_thesis_2.epsi}
\caption[$^2$H and $^3$He cross sections: experimental results and
  Born model.]{$^2$H (left) and $^3$He (right):  experimental cross section and Born cross section
  model.  Errors shown are statistical only.}
\label{ld2_he3_sigma_model}
\end{figure*}
\begin{figure*}[h!]
\center
\includegraphics[angle=270,width=\textwidth]{he4_carbon_data_model_thesis_2.epsi}
\caption[$^4$He and $^{12}$C cross sections: experimental results and
  Born model.]{$^4$He (left) and $^{12}$C (right):  experimental cross section and Born cross section
  model.  Errors shown are statistical only.}
\label{he4_carbon_sigma_model}
\end{figure*}
\begin{figure*}[h!]
\center
\includegraphics[angle=270,width=\textwidth]{cu_au_data_model_thesis_2.epsi}
\caption[$^{63}$Cu and $^{197}$Au cross sections: experimental results and
  Born model.]{$^{63}$Cu (left) and $^{197}$Au (right):  experimental cross section and Born cross section
  model.  Errors shown are statistical only.}
\label{cu_au_sigma_model}
\end{figure*}
\begin{figure*}[h!]
\center
\includegraphics[angle=270,width=0.9\textwidth]{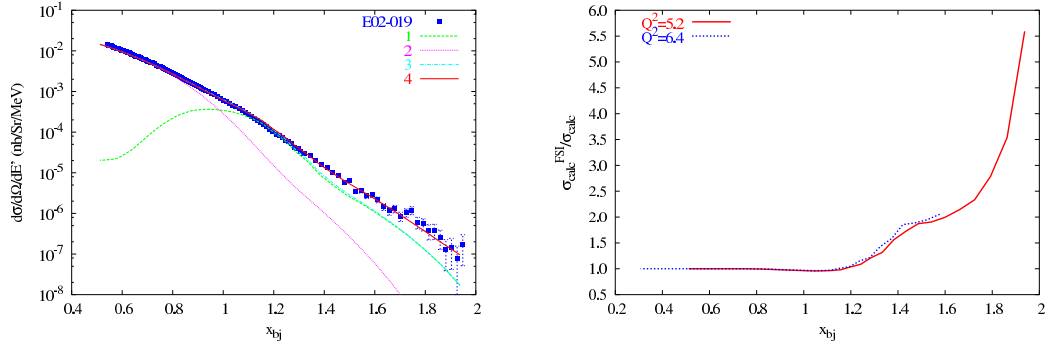}
\caption[Carbon cross section at $\theta _{HMS}=32^{\circ}$, data and
  theory calculation.]{Carbon cross section at $\theta _{HMS}=32^{\circ}$, data and
  theory calculation (left).  The contributions in the theoretical
  calculation are as follows: (1) quasielastic, (2) inelastic,(3) sum
  of (1) and (2),(4)
  total cross section, including FSIs.  Calculation provided by
  O.Benhar ~\cite{Benhar:2007dd}.  The right plot shows the relative
  contribution from FSIs alone for 32$^{\circ}$ and 40$^{\circ}$, obtained by comparing the final folded
  cross section to the sum of the quasielastic and inelastic pieces.
  Note that the contribution from FSis is very similar at these values
  of Q$^2$.}
\label{benhar_plot}
\end{figure*}

Fig.~\ref{benhar_plot} shows the cross section for Carbon at
32$^{\circ}$ along with a theoretical calculation by O. Benhar~\cite{Benhar:2007dd}.  The
experimental cross section is very well reproduced by the calculation,
which includes the contributions from Final State Interactions.  The
FSIs can be seen to be a significant effect at low energy transfers,
where they
double the total calculated cross section by $x\approx 1.4$ and the
contribution continues to grow with $x$. 

 In the work of Benhar et al., the theoretical cross
section is calculated using a convolution integral of the form:
\begin{equation}
\frac{d\sigma}{d\Omega d\nu}=\int{d\nu '\:
  \left(\frac{d\sigma}{d\Omega d\nu '}\right)_{IA}\:f_{\textbf{q}}(\nu-\nu')},
\end{equation}
where $f_{\textbf{q}}(\nu)$ is a folding function that accounts for FSI
effects, and $\left(\frac{d\sigma}{d\Omega d\nu}\right)_{IA}$ is the inclusive cross
section in the Impulse Approximation obtained from a nuclear matter
spectral function.  The spectral function is constructed from two
pieces~\cite{Benhar:1991af}: one coming from the shell model picture and the second from
correlations.  The former is calculated using a momentum-space wave
function for a single particle in a given shell state and summing over
all the occupied states.  The latter piece is calculated by integrating the
correlation contribution from the nuclear matter spectral function~\cite{Benhar:1994hw}
with the nuclear density distribution to give the contribution for a finite
nucleus.  

 The FSIs were treated with the Correlated Glauber
Approximation~\cite{Benhar:1991af}, which employs the eikonal and
frozen approximations.  The former assumes that the struck
particle moves in a straight line and the latter assumes that the
spectator nucleons are fixed.  The folding function includes two FSI
effects:  an energy shift in the cross section and a redistribution of
the cross section strength from the quasielastic peak to the region of
low energy transfer.  The cross sections in this region are very small
and decreasing rapidly, so even a small contribution from FSIs results
in a significant enhancement.  In Benhar's calculation, the FSIs do
not disappear at high Q$^2$, but rather become constant as the width
of the quasielastic peak in $\nu$ becomes constant with increasing
Q$^2$ and the folding function contains no Q$^2$ dependence.  This seems
to have already happened by the kinematics of 32$^{\circ}$ and
40$^{\circ}$, as can be seen in the second plot of Fig. \ref{benhar_plot}.

\section{F(y,$|$\textbf{q}$|$) Scaling Function}
\begin{figure*}[h!]
\center
\includegraphics[width=0.6\textwidth,angle=270,clip]{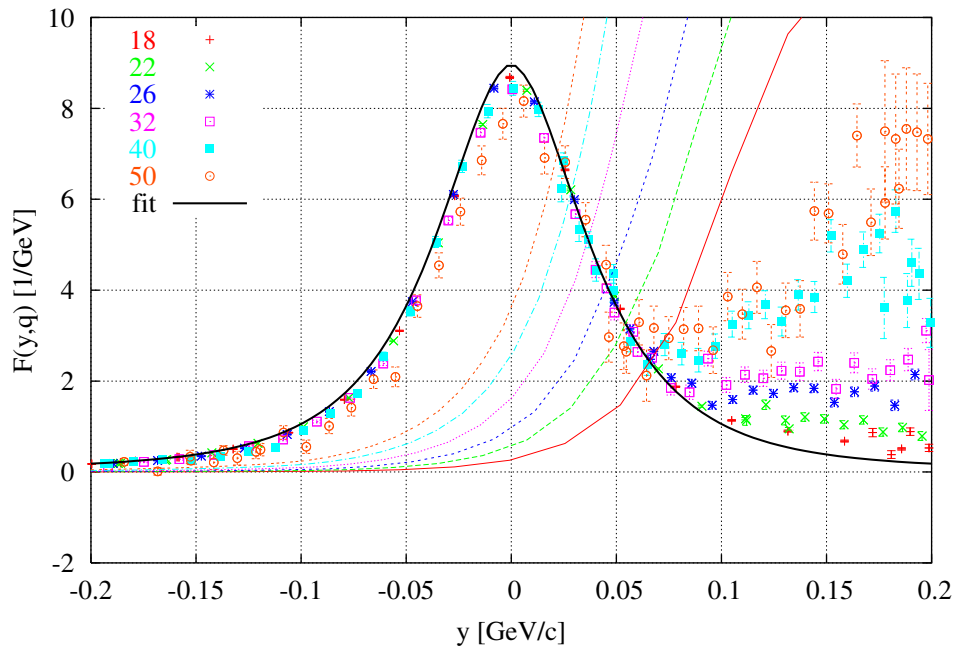}
\caption[F(y,q) for deuterium, model and experiment.]{Scaling function
F(y,q) for deuterium, as extracted from experimental data.  The inelastic cross
section (solid color lines) was calculated using a model described in Sec.~\ref{cs_model} and
subtracted from the experimental cross section.  The solid black line
is the fit to the experimental scaling function that was then used in the
cross section model.}
\label{ld2_fy_pieces}
\end{figure*}

\begin{figure*}[h!]
\center
\includegraphics[angle=270,width=\textwidth]{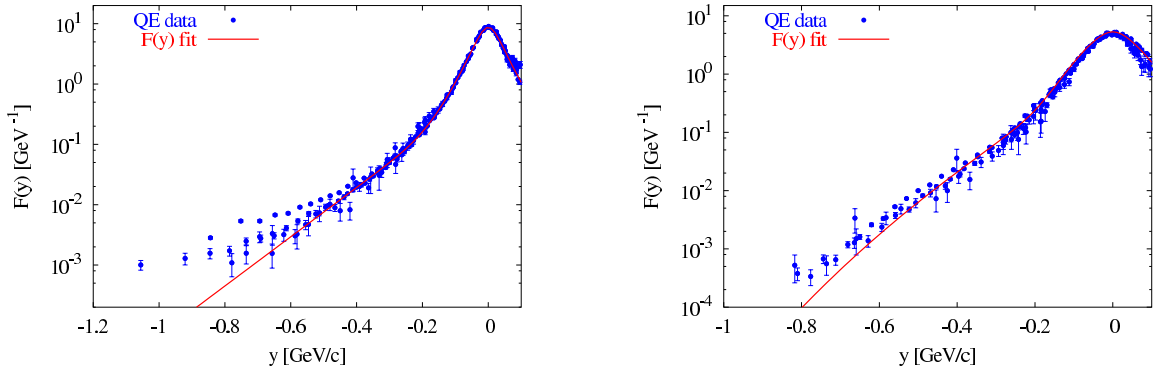}
\caption[$^{2}$H and $^{3}$He scaling functions: experimental result
  and fit.]{$^{2}$H (left) and $^{3}$He (right): experimental scaling function, F(y,$|$\textbf{q}$|$),
  and the fit to it. The fit was used to calculate the quasielastic
  contribution to the model cross section.  Errors shown are
  statistical only.  For $^2$H, the points that lie above the fit
  correspond to lower Q$^2$ values, and have the highest contributions
  from FSIs.  The inelastic contribution was subtracted using the
  cross section model described in Sec.~\ref{cs_model}.}
\label{ld2_he3_fy_final}
\end{figure*}

The scaling function F(y,$|$\textbf{q}$|$) was extracted for each target via
the method described in Sec.~\ref{extract_fy_section}.   The inelastic
contribution has been subtracted from the measured cross section using the model
described in Sec.~\ref{cs_model}.  Fig.~\ref{ld2_fy_pieces} shows the
region around the quasielastic peak for deuterium and the size of
the inelastic contribution can be seen.  The inelastic contribution
is the largest for the setting with the
largest Q$^2$, 50$^{\circ}$ for E02-019, with
$\approx$ 30$\%$ of the total strength at the quasielastic peak. Experimental
scaling functions are shown in
Figs. \ref{ld2_he3_fy_final}-\ref{cu_au_fy_final}, along with the fit to the
data that was used as an input to the cross section model.

\begin{figure*}[h!]
\center
\includegraphics[angle=270,width=\textwidth]{fy_he4_carbon_fit_thesis_2.epsi}
\caption[$^{4}$He and $^{12}$C scaling functions: experimental result
  and fit.]{$^{4}$He (left) and $^{12}$C (right): experimental scaling function, F(y,$|$\textbf{q}$|$),
  and the fit to it. The fit was used to calculate the quasielastic
  contribution to the model cross section.  The inelastic contribution was subtracted using the
  cross section model described in Sec.~\ref{cs_model}. Errors shown are statistical only.}
\label{he4_carbon_fy_final}
\end{figure*}
\begin{figure*}[h!]
\center
\includegraphics[angle=270,width=\textwidth]{fy_cu_au_fit_thesis_2.epsi}
\caption[$^{63}$Cu and $^{197}$Au scaling functions: experimental result
  and fit.]{$^{63}$Cu (left) and $^{197}$Au (right): experimental scaling function, F(y,$|$\textbf{q}$|$),
  and the fit to it. The fit was used to calculate the quasielastic
  contribution to the model cross section. The inelastic contribution was subtracted using the
  cross section model described in Sec.~\ref{cs_model}.  Errors shown are statistical only.}
\label{cu_au_fy_final}
\end{figure*}

PWIA predicts that scaling will be approached from below with
increasing Q$^2$, as more
strength from the spectral function is integrated over.  In reality,
scaling appears to be approached from above, as is very well seen in
deuterium data (Fig.~\ref{ld2_he3_fy_final}).  This is because much of
the strength in the tail comes from FSIs, which are not included in
the PWIA, but their contribution is large enough to change the
approach to scaling.  
%

\begin{figure*}[h!]
\center
\includegraphics[width=0.7\textwidth,angle=270,clip]{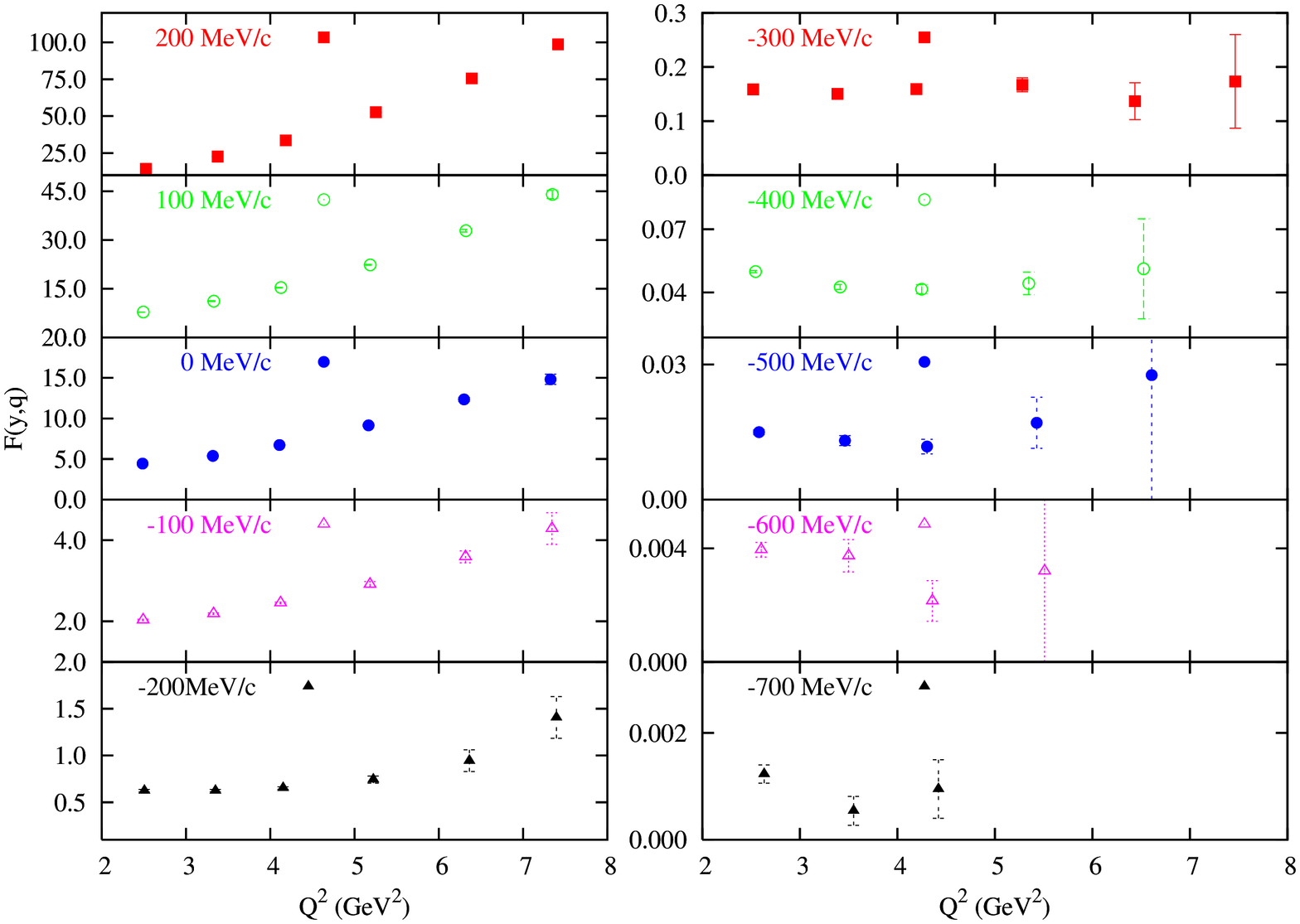}
\caption[Approach to scaling for $^{12}$C as a function of Q$^2$.]{$^{12}$C scaling function, shown for fixed values of $y$,
  as a function of Q$^2$.  The inelastic contribution has not been
  subtracted and it is the cause of scaling violations for positive
  and small negative values of $y$ at high Q$^2$.  At large negative
  values of $y$, scaling is broken at low Q$^2$ due to contributions
  from FSIs which decrease with increasing Q$^2$.}
\label{scaling_approach_carbon}
\end{figure*}

\begin{figure*}[h!]
\center
\includegraphics[width=0.7\textwidth,angle=270,clip]{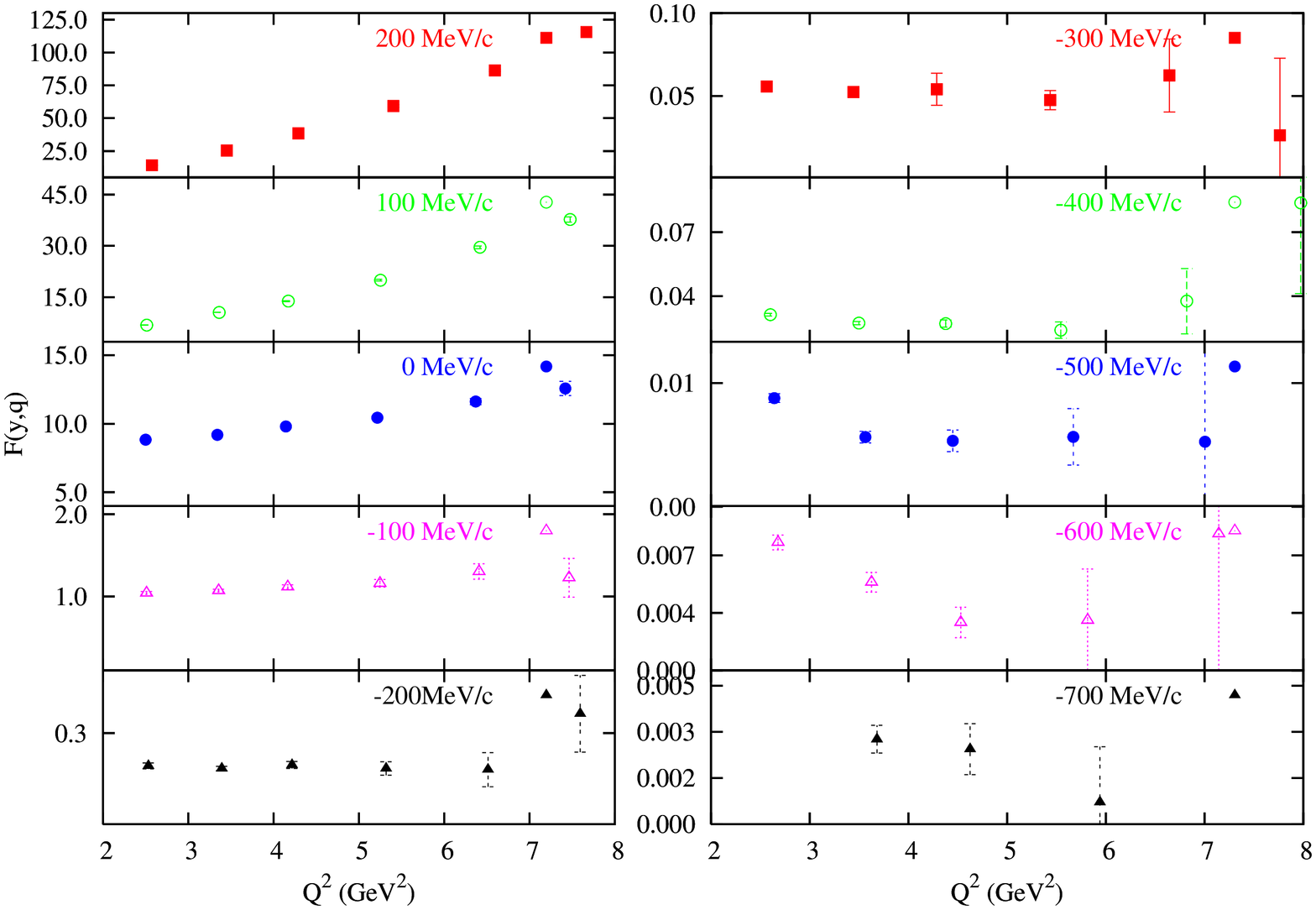}
\caption[Approach to scaling for $^{2}$H as a function of Q$^2$.]{$^{2}$H scaling function, shown for fixed values of $y$,
  as a function of Q$^2$. The inelastic contribution has not been
  subtracted and it is the cause of scaling violations for positive
  and small negative values of $y$ at high Q$^2$.  At large negative
  values of $y$, scaling is broken at low Q$^2$ due to contributions
  from FSIs which decrease with increasing Q$^2$. }
\label{scaling_approach_ld2}
\end{figure*}

\begin{figure*}[h!]
\center
\includegraphics[width=0.7\textwidth,angle=270,clip]{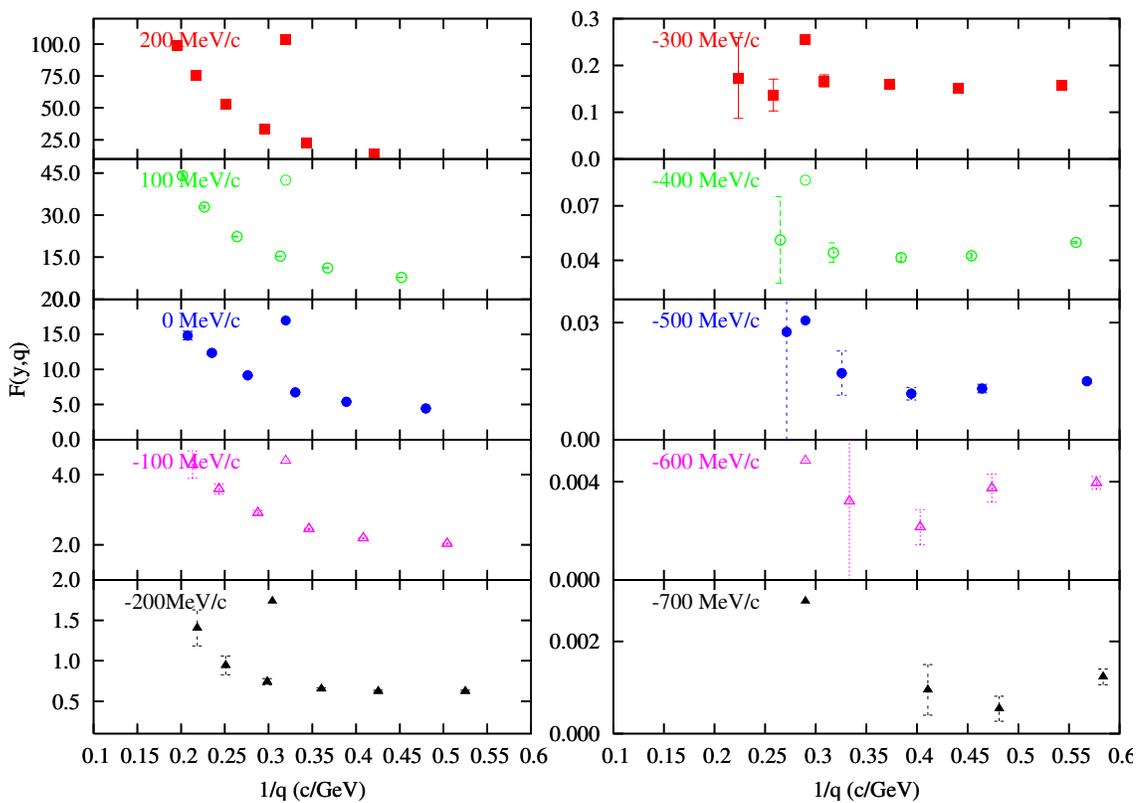}
\caption[Approach to scaling for $^{12}$C as a function of $1/q$.]{$^{12}$C scaling function, shown for fixed values of $y$,
  as a function of $1/q$. Following the approach of ~\cite{CiofidegliAtti:1989qs} described
  in Sec.~\ref{yscaling}, it maybe possible to extract the asymptotic limit of
  the scaling function, provided that the inelastic contribution is
  subtracted at $1/q<<$1.}
\label{scaling_approach_carbon_1q}
\end{figure*}

This approach can be seen when the scaling function is
plotted for fixed values of $y$, as a function of Q$^2$. Once the
points lie on a flat line - there's no Q$^2$-dependence and the
scaling regime is reached.  This can be seen in 
Figs. \ref{scaling_approach_carbon} and \ref{scaling_approach_ld2} for
$^2$H and $^{12}$C, respectively.  Note that in Figs.~\ref{ld2_he3_fy_final}-\ref{cu_au_fy_final}, the
inelastic contribution was subtracted before the extraction of
F(y,$|$\textbf{q}$|$).  That is not the case in
Figs. \ref{scaling_approach_carbon} and \ref{scaling_approach_ld2},
where the scaling function is determined using the total measured
cross section.  This way, the decrease
in the inelastic contribution can be seen.  In the plots, $y$ values
between 200 MeV/c and -700 MeV/c are shown, even though $y$-scaling is not
expected for positive $y$, as those data correspond to the high
energy loss side of the quasielastic peak.  The inelastic contribution breaks the scaling for low values
of $y$ (positive and negative).  As $y$ becomes more negative, the
onset of this scale-breaking moves to larger values of Q$^2$, where
the inelastic contribution is largest.  At large negative values of $y$, the scaling is broken
for low values of Q$^2$, which are the data with the largest
contributions from FSIs. The $^{12}$C scaling function is also plotted for
constant $y$ values as a function of $1/q$ in
Fig.~\ref{scaling_approach_carbon_1q}.  In order to determine the
asymptotic limit, the inelastic contribution would need to be
subtracted and the $1/q^2$ term must be shown to be negligible
(i.e. $1/q<<1$), which is not the case for all the data points.  If
one wishes to extract the asymptotic limit of the scaling function, it
might be best to combine this data set with other available data~\cite{Benhar:2006er}.  
\clearpage
\section{Superscaling}
\begin{figure*}[h!]
\center
\includegraphics[width=0.5\textwidth,angle=270,clip]{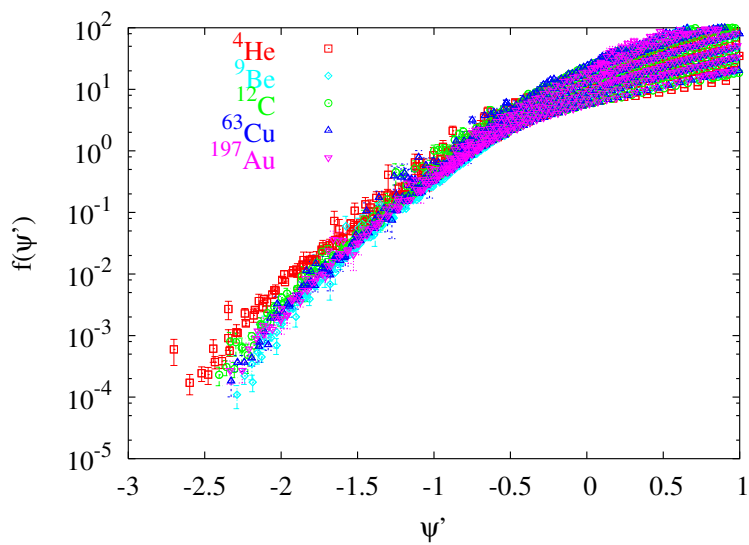}
\caption[Superscaling function $f(\psi')$ for A$>$3.]{Superscaling function $f(\psi')$, extracted from the
  E02-019 cross sections for all available kinematics, A$>$3.}
\label{superscaling_all}
\end{figure*}

Superscaling function $f(\psi ')$ was extracted from the measured
cross sections with the method described in ~\cite{Maieron:2001it}.  In this
reference, the lightest nucleus that was included in the analysis was
$^6$Li and superscaling was observed for all nuclei studied.  For
E02-019, the lightest nucleus examined as a function of $f(\psi ')$ is
$^4$He, which was included in the superscaling analysis
of~\cite{Donnelly:1999sw} and was found to superscale, except in the
region of the quasielastic peak.  As in the
references, Fig.~\ref{superscaling_all} shows the superscaling function
$f(\psi ')$ for the selected targets for all available kinematics.
In E02-019 data, we see that $^4$He deviates noticeably from the
superscaling curve described by the heavier targets.  Deviations around the quasielastic peak can be
explained by the $^4$He momentum distribution, which has a
well-defined peak.  However, this explanation doesn't extend into the
quasielastic tail.

\begin{figure*}[h]
\center
\includegraphics[angle=270, width=\textwidth]{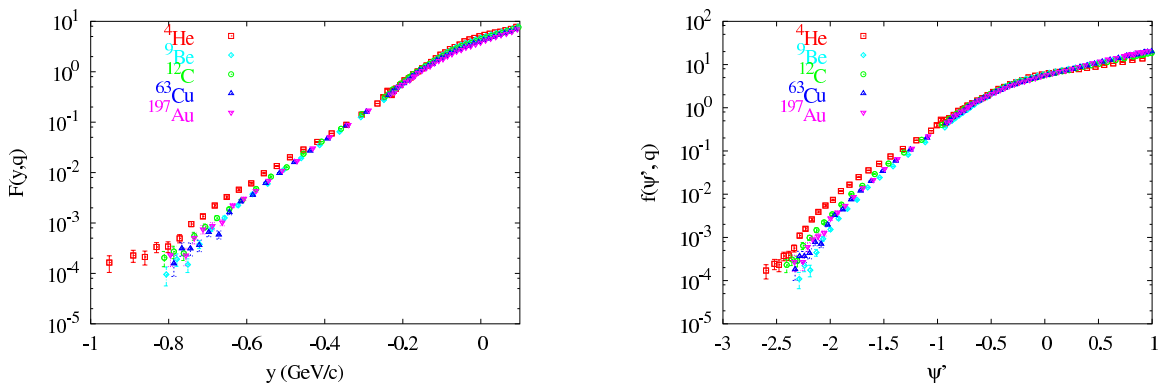}
\caption[F($|\textbf{q}|$,y) and $f(\psi ')$ for A$>$3 at 18$^{\circ}$]{$F(|\textbf{q}|,y)$ and $f(\psi ')$, shown for A$>$3,
  for the central angle setting, 18$^{\circ}$.}
\label{scaling_reg_super_18}
\end{figure*}

\begin{figure*}[h]
\center
\includegraphics[angle=270,width=\textwidth]{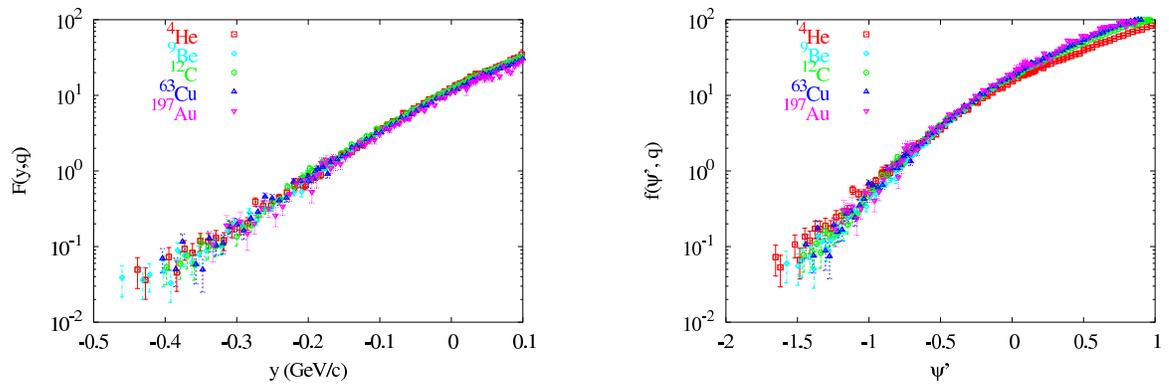}
\caption[$F(|\textbf{q}|,y)$ and $f(\psi ')$ for A$>$3 at 40$^{\circ}$]{$F(|\textbf{q}|,y)$ and $f(\psi ')$, shown for A$>$3,
  for the central angle setting, 40$^{\circ}$.}
\label{scaling_reg_super_40}
\end{figure*}

 When we examine $f(\psi)'$ for just one kinematic
setting, the deviation of $^4$He can be more easily seen.  In addition,
the heavy targets do not appear to form one scaling curve, especially
for large negative values of $\psi '$.
However, when a similar plot is made using the conventional variable,
$y$, and the corresponding scaling function F($|\textbf{q}|$,y), we
observe similar behavior (see Figs.~\ref{scaling_reg_super_18}
and \ref{scaling_reg_super_40}).  While superscaling gives better scaling
behavior at the peak for low values of Q$^2$, as is seen in
Fig. \ref{scaling_reg_super_18}, F($|\textbf{q}|$,y) does a better job
for the high momentum tail and for higher Q$^2$ values (see
Fig.~\ref{scaling_reg_super_40}).

\clearpage
\section{Other Scaling Variables - $y_{cw}$}
\begin{figure*}[h!]
\center
\includegraphics[angle=270,width=\textwidth]{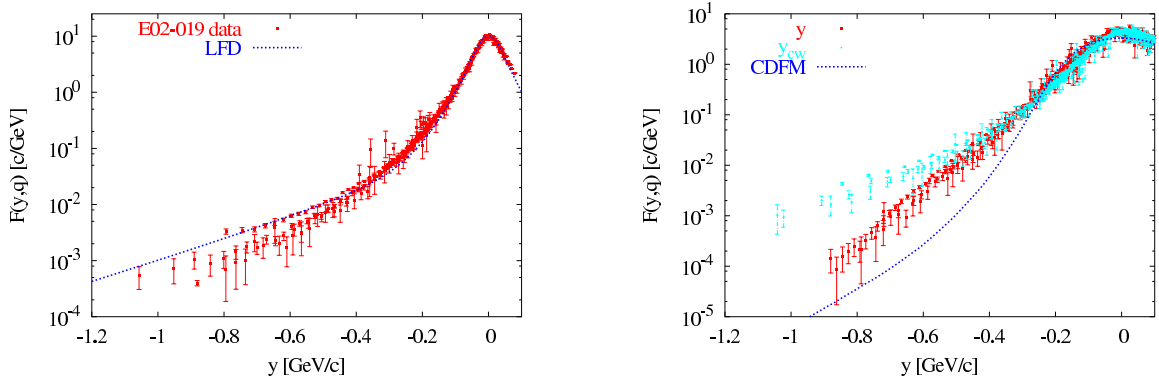}
\caption[F(y,$|$\textbf{q}$|$) for $^2$H and  $^4$He as a function of  $y$, $y_{cw}$:
  data and a theoretical calculation.]{Experimental scaling function,
  F(y,$|$\textbf{q}$|$) for $^2$H and $^4$He, shown as a function of conventional $y$, as well as $y_{cw}$, along
  with a theoretical calculation by the group using the LFD approach.
  For $^2$H, $y$=$y_{cw}$.  The inelastic contribution has been
  subtracted from the data.}
\label{ycw_cryo}
\end{figure*}

\begin{figure*}[h!]
\center
\includegraphics[angle=270,width=\textwidth]{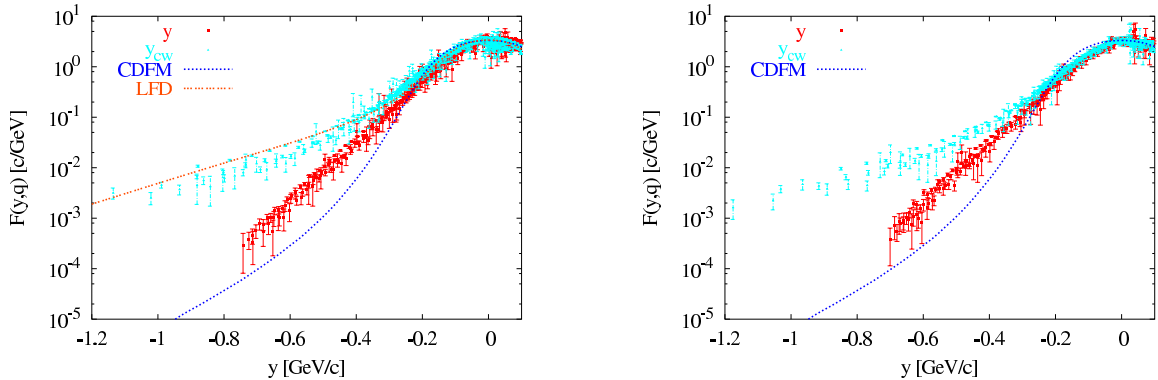}
\caption[F(y,$|$\textbf{q}$|$) for $^{12}$C as a function of  $y$, $y_{cw}$:
  data and a theoretical calculation.]{Experimental scaling function, F(y,$|$\textbf{q}$|$) for
  $^{12}$C and $^{197}$Au, shown as a function of conventional $y$, as well as $y_{cw}$, along
  with a theoretical calculation by the BLANK group using the LFD
  ($^{12}$C only) and CDFM approaches. The inelastic contribution has been
  subtracted from the data.  Remarkable agreement is seen
  for $^{12}$C between the LFD calculation and the data analyzed in $y_{cw}$.}
\label{ycw_solid}
\end{figure*}

The cross section data were also analyzed in terms of the modified
scaling variable $y_{cw}$, described in Sec.~\ref{ycw_section}.  The
results of this analysis can be seen in
Figs. \ref{ycw_cryo} and \ref{ycw_solid}, where the reduced cross section,
F(y,$|$\textbf{q}$|$) is shown as a function of both $y$, and $y_{cw}$.
Also plotted are the results of theoretical calculations of momentum
distributions, $n(k)$ for several nuclei using Light Front Dynamics
and Coherent Density Fluctuation Model ~\cite{Antonov:2001mk,Antonov:email}, converted to
F(y,$|$\textbf{q}$|$) through Eq.~\ref{fy_nk_simple}.  The latter model, CDFM, is a natural extension of
the Relativistic Fermi Gas model, that uses the density and momentum
distributions to explain the observations of both scaling and superscaling.  Both calculations
are only available for Carbon, where the LFD curve shows excellent
agreement with experimental data examined as a function of $y_{cw}$.
This could suggest that $y_{cw}$ is the appropriate variable for this
analysis.  However, the current definition needs to be replaced with a
relativistic version, since it breaks down for large negative values
of momentum, where the quantity under the square root in Eq.~\ref{ycw} becomes undefined.
For deuterium of course, the two variables are the same.

\clearpage
\section{Short Range Correlations - Ratios}
%
%
In order to study $2N$ and $3N$ correlations in nuclei, cross section
ratios were obtained of nuclei with $A>3$ to those of $^2$H and $^3$He.  If
these ratios show plateaus, they are believed to be a signature of
$2N$ or $3N$ correlations, depending on the $x$-range.  At $x>1$, the
electron can scatter from a single nucleon of high momentum, or from a
correlated pair or trio of nucleons.  Different regions in $x$ are
sensitive to these different reactions.  For example for $1<x<1.4$, the scattering is believed to be
from a single moving nucleon, and scattering from a correlated pair
takes place for $x\approx1.4<x<2$.  Similarly, scattering from a $3N$
correlated cluster is believed to be taking place for  $2.5<x<3.$
Previous measurements~\cite{Egiyan:2003vg} of the ratios have shown no evidence of
correlations for Q$^2<1.4\: GeV^2$, which has since then been used as
the threshold for this kind of scaling.  The E02-019 data exceed
this threshold.

To estimate the relative probabilities of SRCs in heavier nuclei
(A$_1$) to those in lighter nuclei like $^2$H or $^3$He (A$_2$), a
ratio of the experimentally measured cross sections,
$R(A_1,A_2)=\sigma _{A_1}/\sigma _{A_2}$, is formed and
an isoscalar correction~\cite{Egiyan:2003vg} is applied to give:
\begin{equation}
r(A_1,A_2)=R(A_1,A_2)\times \frac{A_1(Z_2\sigma _p+N_2\sigma _n)}{A_2(Z_1\sigma _p+N_1\sigma _n)},
\label{src_ratio_eqn}
\end{equation}
where $\sigma _n$ and $\sigma _p$ are the elastic electron-nucleon
cross sections.  The isoscalar correction factor shows little Q$^2$
dependence and varies very slowly with $x$.  For example, the
correction factor for the $^{12}$C/$^{3}$He ratio varies between
$\approx$ 1.125 at 50$^{\circ}$
and $\approx$1.138 at 18$^{\circ}$ at $x$=1.

\begin{figure*}[h!]
\center
\includegraphics[angle=270,width=0.9\textwidth]{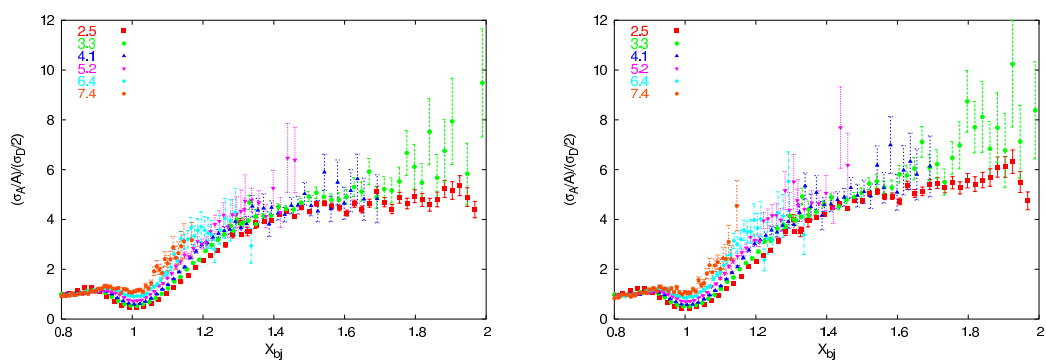}
\caption[Cross section ratios of $^{12}$C (right) and $^{63}$Cu (left)
  to $^2$H.]{Cross section ratios of $^{12}$C and $^{63}$Cu to $^2$H.
  The expected 2N plateau for $x>1.4$ is not flat, since the $^2$H
  cross sections in that region are limited by the subtraction of the
  aluminum cell yield, which dominates the cross section.  The Q$^2$
  values quoted in the figure are calculated at the quasielastic peak.}
\label{2n_cor_result}
\end{figure*}

\begin{figure*}[h!]
\center
\includegraphics[angle=270,width=0.9\textwidth]{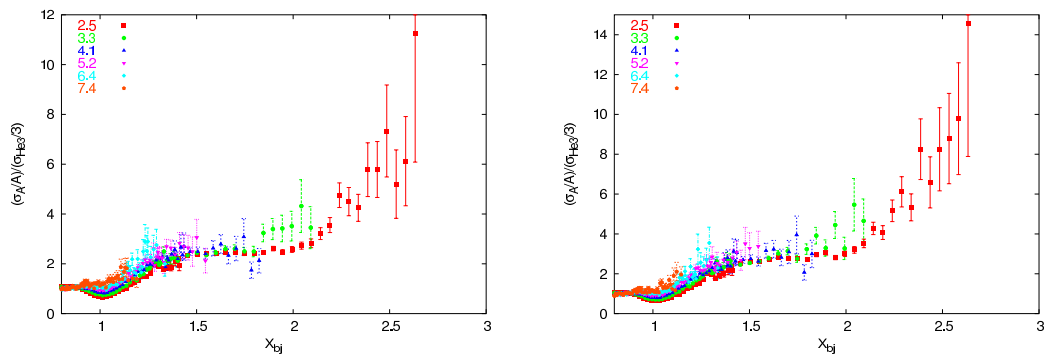}
\caption[Cross section ratios of $^{12}$C (right) and $^{63}$Cu (left)
  to $^3$He.]{Cross section ratios of $^{12}$C and $^{63}$Cu to $^3$He.
  The expected 2N plateau for $x>1.4$ is better visible than in the
  ratios to $^2$H.  However, the expected plateaus at $x>2.5$ suffer
  from poor statistics and the large contribution from the aluminum
  cell to the helium cross section. The Q$^2$
  values quoted in the figure are calculated at the quasielastic peak.}
\label{3n_cor_result}
\end{figure*}

Fig.~\ref{2n_cor_result} shows the cross section ratios of  $^{12}$C
and $^{63}$Cu to $^2$H.  The data at 18$^{\circ}$ have the best
coverage and statistics, and show a possible plateau for both targets
for $x>1.4$.  The ratios are limited by the deuterium data, which has
large contributions for the aluminum container cell that grow with
$x$, reaching 100$\%$ at $x\approx$2. The $2N$ correlation plateau is more clearly
seen in the ratios of the same cross sections to $^3$He (see Fig.~\ref{3n_cor_result}), where the
same end-cap subtraction constraints don't dominate until much larger $x$ values.

It's been suggested that while the ratios are a signature of SRCs,
they cannot be used to provide a quantitative measurement since different targets will have
different FSIs.  However, a more widely accepted idea~\cite{Frankfurt:1981mk} is that the FSIs in the correlations region are
localized to the correlations themselves since they're dominated by
the rescattering of the knocked-out nucleon by the correlated
nucleon(s).  Therefore, the FSIs do not contribute to the
ratios, but rather cancel out.

\begin{figure*}[h!]
\center
\includegraphics[angle=270,width=0.9\textwidth]{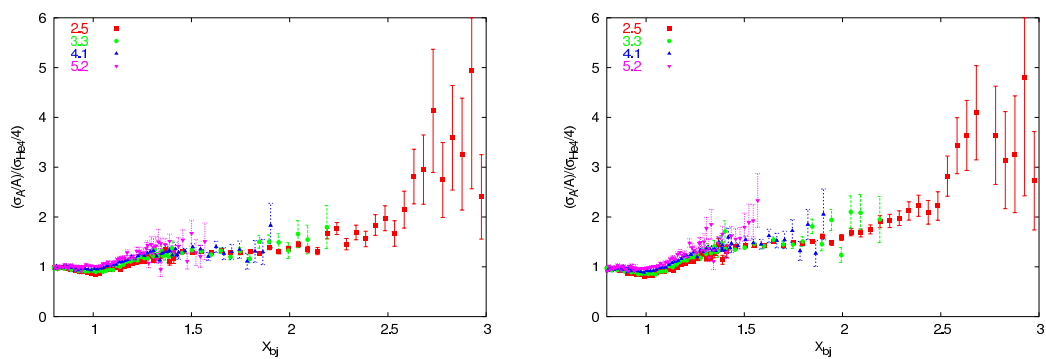}
\caption[Cross section ratios of $^{12}$C (right) and $^{63}$Cu (left)
  to $^4$He.]{Cross section ratios of $^{12}$C and $^{63}$Cu to
  $^4$He.  The $3N$ plateau is seen more clearly in this case than in
  the ratio to $^3$He. The Q$^2$
  values quoted in the figure are calculated at the quasielastic peak.}
\label{3n_cor_he4_result}
\end{figure*}

$3N$ correlations can also be seen in ratios to $^4$He, which is less
affected by the target cell contribution in the 2.5$<x<$3 range than
$^3$He.  This ratio is shown in Fig.~\ref{3n_cor_he4_result} and
suggests a plateau in the $x>$2.5 region, where 3N correlations should
be visible if present.
It appears that the ratios shown in Fig.~\ref{3n_cor_result} are not
independent of Q$^2$.  Therefore, when $r(A_1,A_2)$ was extracted from
the data, it was done separately for each Q$^2$ setting.  Table
\ref{src_2_values} lists the ratios, $r(A_1,A_2)$, of A$_1$ to $^3$He,
in the 2N correlation region (1.5$<x<$2) for the three lowest Q$^2$
settings.  For targets in common with~\cite{Egiyan:2003vg}, the ratios
for this experiment are lower.  However, the two data sets are consistent if
systematic as well as statistical errors are taken into account.  The
systematic uncertainties listed in Table~\ref{sys_error_table} apply
to the cross sections, and not all of them are included when ratios are taken
since many of the corrections (e.g. detector-related quantities) will cancel.  Also, some uncertainties such as
those coming from the radiative corrections, have a contribution associated
with the method and another contribution that's target-dependent. In
such cases, only the target-dependent errors were propagated.  For the
uncertainty on the end-cap subtraction for $^3$He, a 1$\%$ uncertainty
was used (3$\%$ of the $\approx$35$\%$ contribution at 1.5$<x<$2).
\clearpage
\begin{table}[h!]
\begin{center}
\caption{$r(A_1,^3He)$ in the 2N correlation region (1.5 $<x<$ 2) for
  A$>$3 at each kinematic setting with data in the relevant $x$
  region. Q$^2$ values are quoted at $x$=1.5. Errors quoted are
  statistical and systematic.  See text for discussion of
  systematic errors.}
\vspace*{0.25in}

\begin{tabular}{|c|c|c|c|}
\hline
A$_1$ &
Q$^2$ =2.71 (GeV$^2$) &
Q$^2$ =3.73 (GeV$^2$) &
Q$^2$ =4.76 (GeV$^2$) \\
\hline\hline
$^{4}$He & 1.86 $\pm$ 0.009 $\pm$ 0.073& 1.89 $\pm$ 0.031 $\pm$ 0.074& 1.83 $\pm$ 0.079 $\pm$ 0.071 \\ 
$^{9}$Be & 2.09 $\pm$ 0.011 $\pm$ 0.071& 2.06 $\pm$ 0.030 $\pm$ 0.070& 2.10 $\pm$ 0.086 $\pm$ 0.072 \\ 
$^{12}$C & 2.42 $\pm$ 0.013 $\pm$ 0.083& 2.47 $\pm$ 0.038 $\pm$ 0.084& 2.42 $\pm$ 0.099 $\pm$ 0.083 \\ 
$^{63}$Cu & 2.70 $\pm$ 0.015 $\pm$ 0.095& 2.72 $\pm$ 0.041 $\pm$ 0.096& 2.76 $\pm$ 0.114 $\pm$ 0.097 \\ 
$^{197}$Au & 2.83 $\pm$ 0.016 $\pm$ 0.111& 2.81 $\pm$ 0.043 $\pm$ 0.110& 2.97 $\pm$ 0.122 $\pm$ 0.116 \\ 

\hline
\end{tabular}
\label{src_2_values}
\end{center}
\end{table}

%
%
%
\section{Inelastic Response\label{inel_scaling_results}}

The inelastic response is best examined by looking at the F$_2$
structure function.  It is extracted from the measured cross section
with the method described in Sec.~\ref{inel_theory_section} and has an additional systematic
  uncertainty associated with $R$.  

\begin{figure*}[h!]
\center
\includegraphics[angle=270,width=\textwidth]{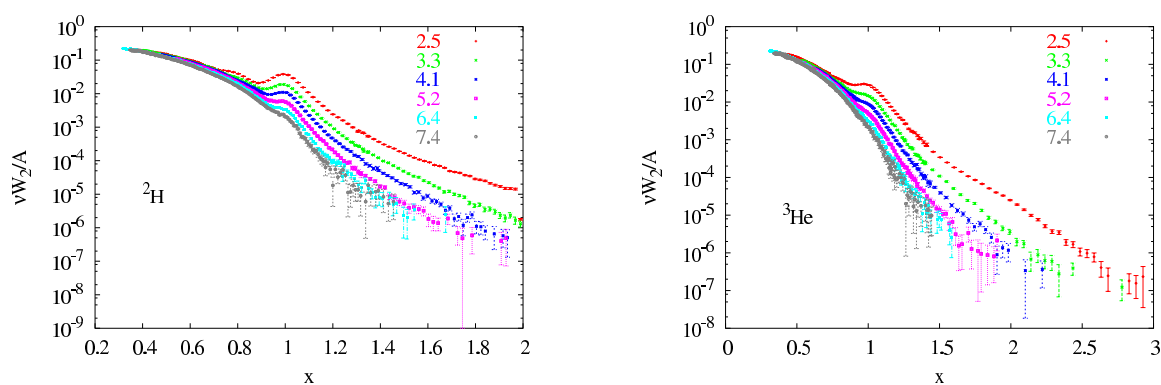}
\caption[F$_2$ structure function for $^2$H and $^3$He as a function
  of $x$.]{F$_2$ structure function for $^2$H (left) and $^3$He (right) as a function
  of $x$, all HMS angle settings. Scaling is seen for the lowest
  values of $x$ only.}
\label{x_scaling_cryo_result}
\end{figure*}

\begin{figure*}[h!]
\center
\includegraphics[angle=270,width=\textwidth]{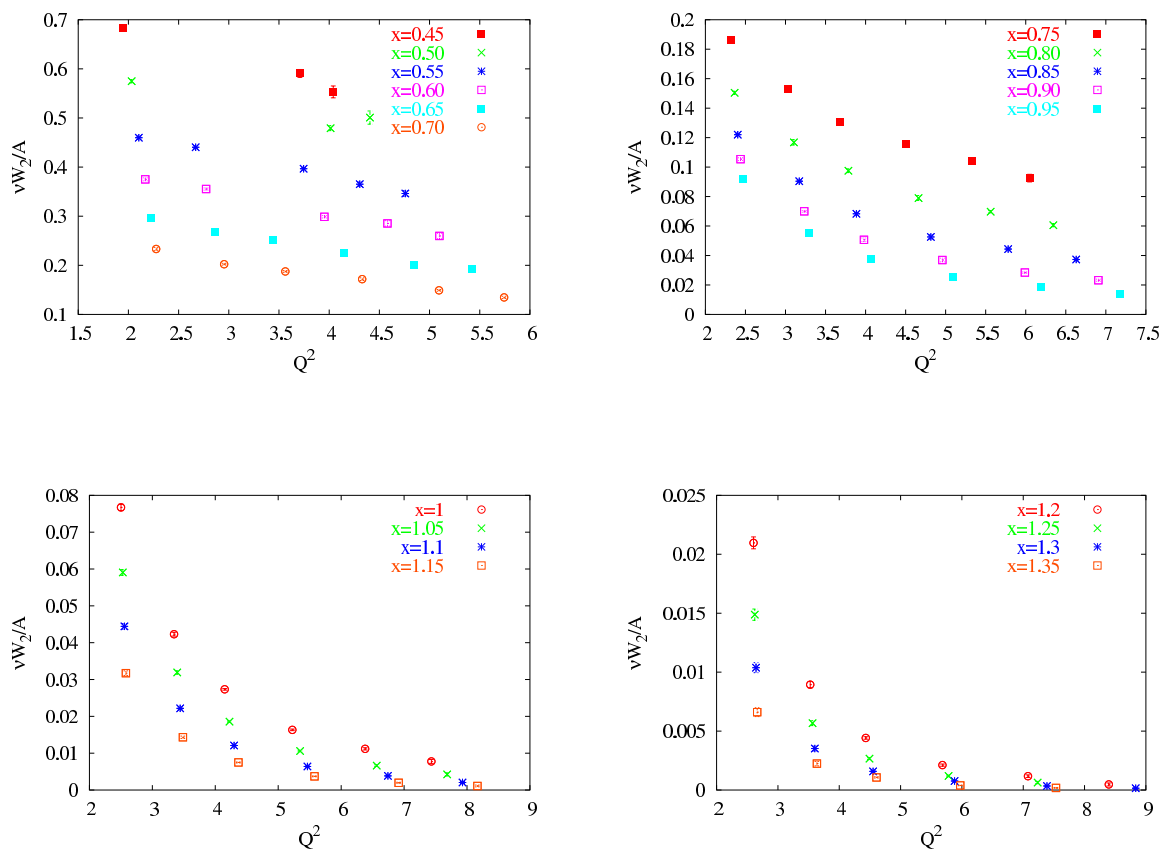}
\caption[Q$^2$ behavior of the F$_2$ structure function for
  $^{4}$He at fixed $x$.]{F$_2$ structure function for $^{4}$He as a function of
  Q$^2$ for fixed values of $x$.  Even at the lowest values of $x$,
  the structure function shows a large decrease over the Q$^2$ range of E02-019.}
\label{x_scaling_he4}
\end{figure*}

\begin{figure*}[h!]
\center
\includegraphics[angle=270,width=.8\textwidth]{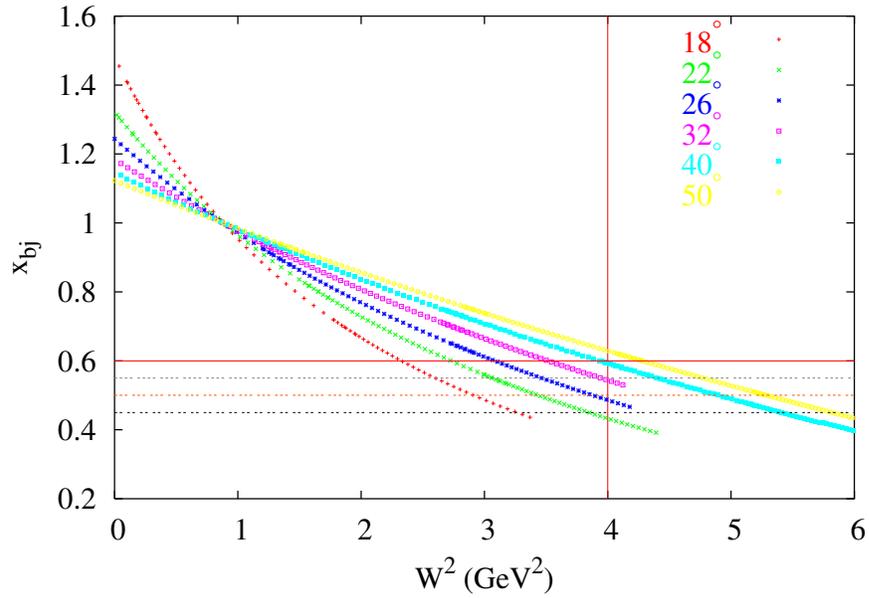}
\caption[Kinematic coverage in W$^2$.]{Kinematic coverage in $x$ and
  W$^2$.  Note that the DIS regime (W$^2>$4) only has data at two
  $\theta _{HMS}$ settings over a large range in $x$.}
\label{w_coverage}
\end{figure*}

As was discussed in Sec.~\ref{qe_intro}, scaling of the nuclear
structure function
has been previously observed in $x$ in the deep inelastic region.
Similar behavior is seen in E02-019 results, shown in
Figs. \ref{x_scaling_cryo_result} and \ref{x_scaling_solid_result},
for low values of $x$.  To examine the Q$^2$-dependence of the
structure function, it is useful to plot it for constant values of
$x$ as is done in Figs. \ref{x_scaling_he4} and
\ref{x_scaling_carbon} for the Helium-4 and Carbon targets.  Even at
low values $x$ ($<$0.65), there is a significant fall-off in F$_2$
over the Q$^2$ range of E02-019.  Data for $x<$0.45 are not shown,
since data at only two settings of $\theta _{HMS}$ reach that kinematic range.  Of the low-$x$
data that are shown, only the two highest Q$^2$ settings reach the DIS
regime (W$^2>$4, shown in Fig.~\ref{w_coverage}), so it's not surprising that scaling is not observed.
\begin{figure*}[h!]
\center
\includegraphics[angle=270,width=\textwidth]{x_scaling_carbon_au_thesis_2.epsi}
\caption[F$_2$ structure function for $^{12}$C and $^{197}$Au as a function
  of $x$.]{F$_2$ structure function for $^{12}$C (left) and $^{197}$Au (right) as a function
  of $x$, all HMS angle settings. Scaling is seen for the lowest
  values of $x$ only.}
\label{x_scaling_solid_result}
\end{figure*}
\begin{figure*}[h!]
\center
\includegraphics[angle=270,width=\textwidth]{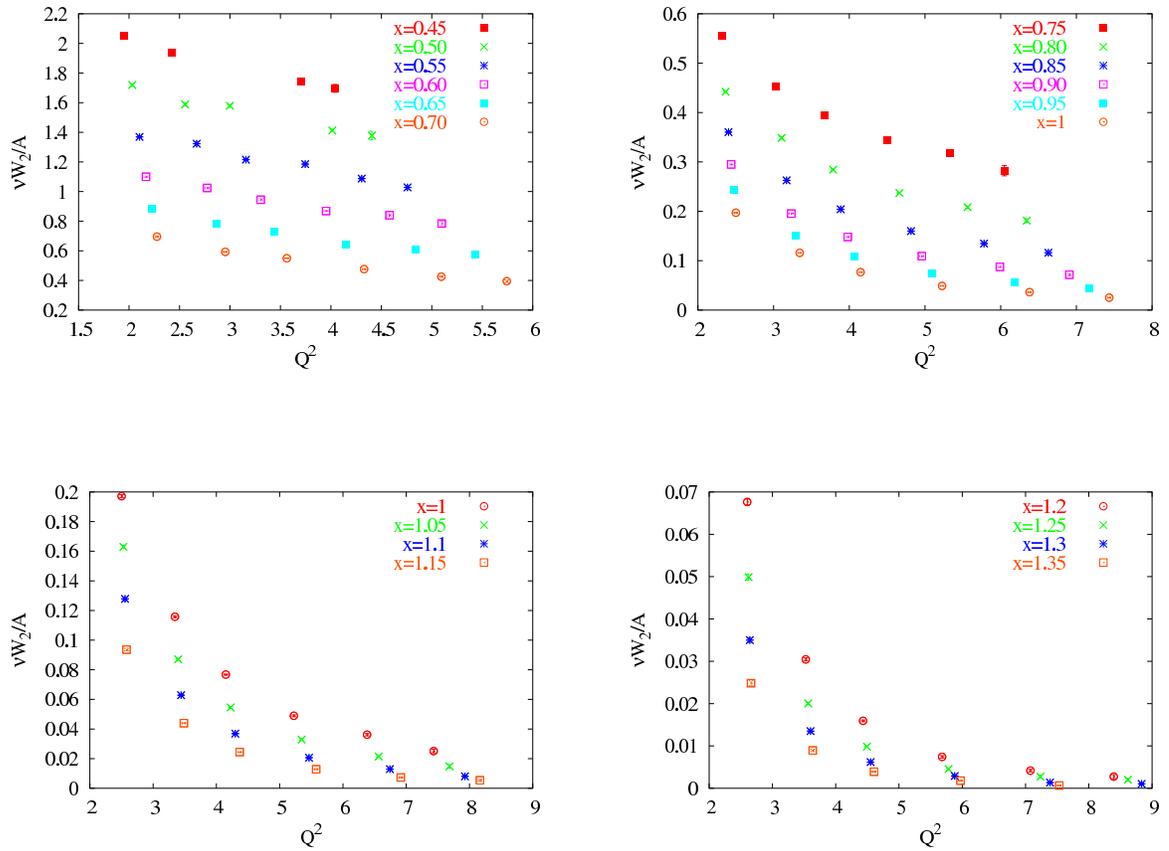}
\caption[Q$^2$ behavior of the F$_2$ structure function for
  $^{12}$C at fixed $x$.]{F$_2$ structure function for $^{12}$C as a function of
  Q$^2$ for fixed values of $x$.  No scaling is observed in $x$.  The
  scaling that was seen for low values of $x$ in
  Fig.~\ref{x_scaling_solid_result}, is for the data at the two
  largest angles (40$^{\circ}$ and 50$^{\circ}$) where the coverage
  extends below $x<0.4$ into the DIS
  region where scaling is expected.}
\label{x_scaling_carbon}
\end{figure*}
\begin{table}[h!]
\begin{center}
\caption{Location of the quasielastic peak in $\xi$ ($x$=1) for the $\theta
  _{HMS}$ settings of E02-019 and the corresponding Q$^2$ values.}
\vspace*{0.25in}
\begin{tabular}{|c|c|c|}
\hline
$\theta_{HMS}$ &
Q$^2$  (GeV$^2$) &
$\xi$ ($x$=1)\\
\hline\hline
18 &  2.5 & 0.784\\
22 &  3.3 & 0.82\\
26 &  4.1 & 0.846\\
32 &  5.2 & 0.871\\
40 &  6.4 & 0.891\\
50 &  7.4 & 0.903\\
\hline
\end{tabular}
\label{xsi_peak_values}
\end{center}
\end{table}

\begin{figure*}[h!]
\center
\includegraphics[angle=270,width=\textwidth]{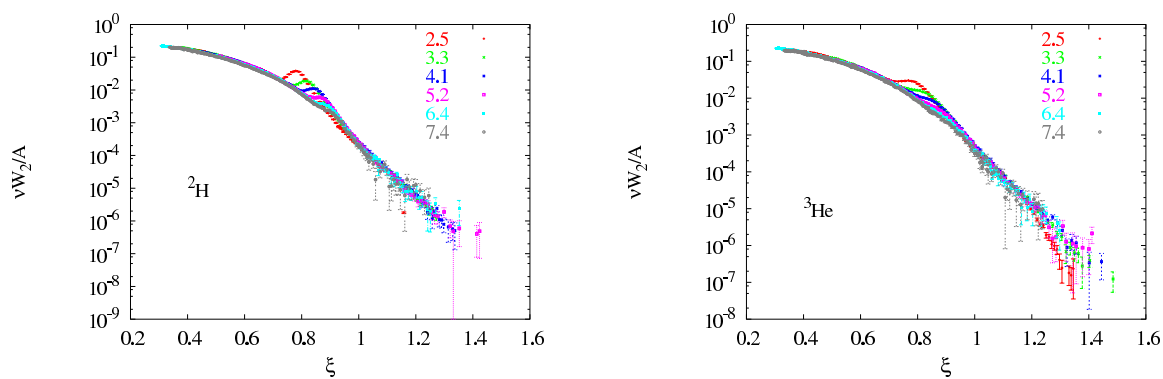}
\caption[F$_2$ structure function for $^2$H and $^3$He as a function
  of $\xi$.]{F$_2$ structure function for $^2$H (left) and $^3$He (right) as a function
  of $\xi$, all HMS angle settings.  The largest scaling violations
  are seen around the quasielastic peak, which is very well defined in
  these light targets.}
\label{xsi_scaling_cryo_result}
\end{figure*}

\begin{figure*}[h!]
\center
\includegraphics[angle=270,width=\textwidth]{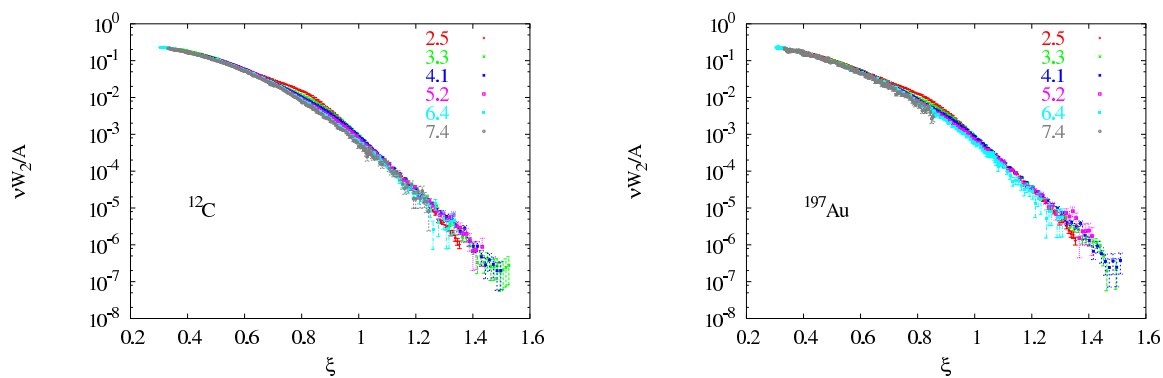}
\caption[F$_2$ structure function for $^{12}$C and $^{197}$Au as a function
  of $\xi$.]{F$_2$ structure function for $^{12}$C (left) and $^{197}$Au (right) as a function
  of $\xi$, all HMS angle settings. While the quasielastic peak is not
  as well defined in the heavier targets as it is in the light ones,
  scaling violations are still seen around the peak.}
\label{xsi_scaling_solid_result}
\end{figure*}

\begin{figure*}[h!]
\center
\includegraphics[angle=270,width=\textwidth]{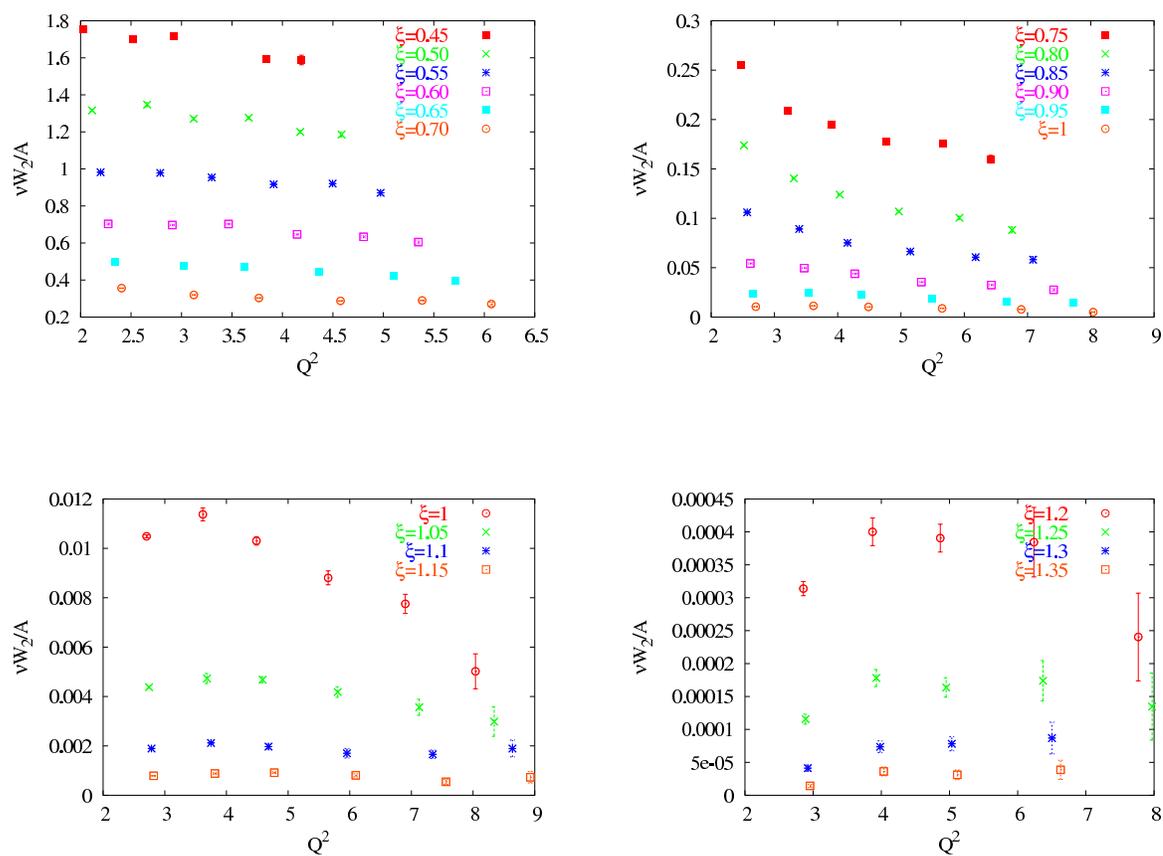}
\caption[Q$^2$ behavior of the F$_2$ structure function for
  $^{12}$C at fixed $\xi$.]{F$_2$ structure function for $^{12}$C as a function of
  Q$^2$ for fixed values of $\xi$. Scaling behavior is best for low
  $\xi$, but scaling extends into the quasielastic region, with some
  violations at low Q$^2$ values.}
\label{xsi_scaling_carbon}
\end{figure*}

Since scaling is only seen in the data over a small range
in $x$, it is more useful to instead look at F$_2$ as a function of
$\xi$, which includes first order target mass effects and extends the
scaling to lower values of Q$^2$.
The general scaling behavior for several nuclei can be seen in
Figs. \ref{xsi_scaling_cryo_result} and \ref{xsi_scaling_solid_result}.
Scaling violations are obvious for the lighter targets, which have a
well defined quasielastic peak appearing at different $\xi$ values for
different Q$^2$ values, rather than at a fixed point, as it does in
$x$.  The locations of the quasielastic peak in $\xi$ for each Q$^2$ are given in
Table \ref{xsi_peak_values}.  The scaling is best for lowest and
highest values of $\xi$, with scaling violations around the
quasielastic peak.  The approach to scaling as a function of Q$^2$ is
examined for constant values of $\xi$ for Helium-4 in
Fig.~\ref{xsi_scaling_he4} and Carbon in
Fig.~\ref{xsi_scaling_carbon}.  F$_2$ falls off less with
increasing Q$^2$ for fixed $\xi$ as compared to fixed
$x$ values.  The most significant scaling violations are at low Q$^2$
values around the quasielastic peak, and decrease quickly with higher
momentum transfers.  These violations are much smaller in heavier
nuclei whose broader peak in n(k) averages over a wider range in $\xi$
in what is referred to as local duality.
\begin{figure*}[h!]
\center
\includegraphics[angle=270,width=\textwidth]{f2_const_xsi_he4_thesis_2.epsi}
\caption[Q$^2$ behavior of the F$_2$ structure function for
  $^{4}$He at fixed $\xi$.]{F$_2$ structure function for $^{4}$He as a function of
  Q$^2$ for fixed values of $\xi$.  Scaling behavior is best for low
  $\xi$, but scaling extends into the quasielastic region, with some
  violations at low Q$^2$ values.}
\label{xsi_scaling_he4}
\end{figure*}
\clearpage
\section{A-dependence of F$_2$}
\begin{figure*}[h!]
\center
\includegraphics[width=\textwidth]{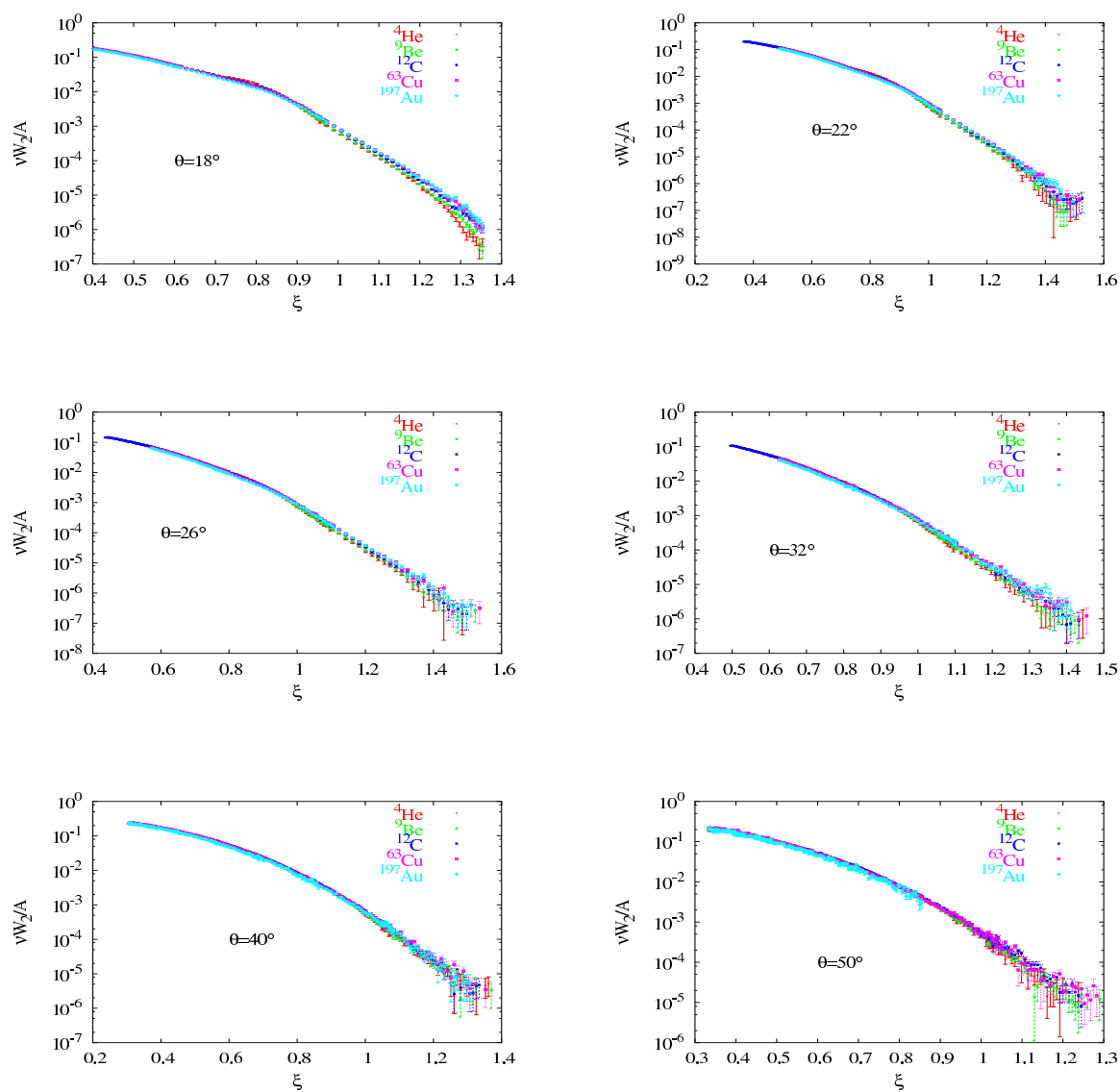}
\caption[F$_2$ structure function for A$>$3 as a function of
  $\xi$.]{F$_2$ structure function for A$>$3 as a function of $\xi$ for
  all kinematic settings.  While there's some target-dependence in the
  structure function at very high $\xi$ for the low angles (low
  Q$^2$), it is confined to light nuclei, and it goes away at larger
  angles (higher Q$^2$).}
\label{xsi_scaling_by_target}
\end{figure*}

In Sec.~\ref{inel_scaling_results}, the F$_2$ structure function was
seen to scale reasonably well for most nuclei over the entire range of
$\xi$ with the possible exception of the region around the
quasielastic peak.  From studies of the EMC effect, we know that the
structure function is modified in the nuclear medium, and the size of
the effect is similar for different nuclei.  This means, that if we
plot the structure function vs $\xi$ for a variety of nuclei at a
given Q$^2$, we expect scaling for the high energy loss side of the
quasielastic peak.  Fig.~\ref{xsi_scaling_by_target} shows that this
is indeed what happens.  Scaling in $\xi$ extends into the
quasielastic regime as well.  This is interesting because different
reaction mechanisms are at work in different kinematic ranges, but
they all conspire to yield scaling of the F$_2$ structure function in
$\xi$ for all nuclei at a common Q$^2$.

Fig.~\ref{xsi_scaling_by_target} shows that there's some
target-dependence for low Q$^2$ values around the quasielastic peak,
which is to be expected, since the light nuclei have different
momentum distributions from those of heavy nuclei.  There's also some
target-dependence in the high momentum tail, presumably from different
FSIs, but this goes away with increasing Q$^2$, supporting O.Benhar's
idea that FSIs become Q$^2$-independent, and it would seem also target-independent.

\begin{figure*}[h!]
\center
\includegraphics[width=\textwidth]{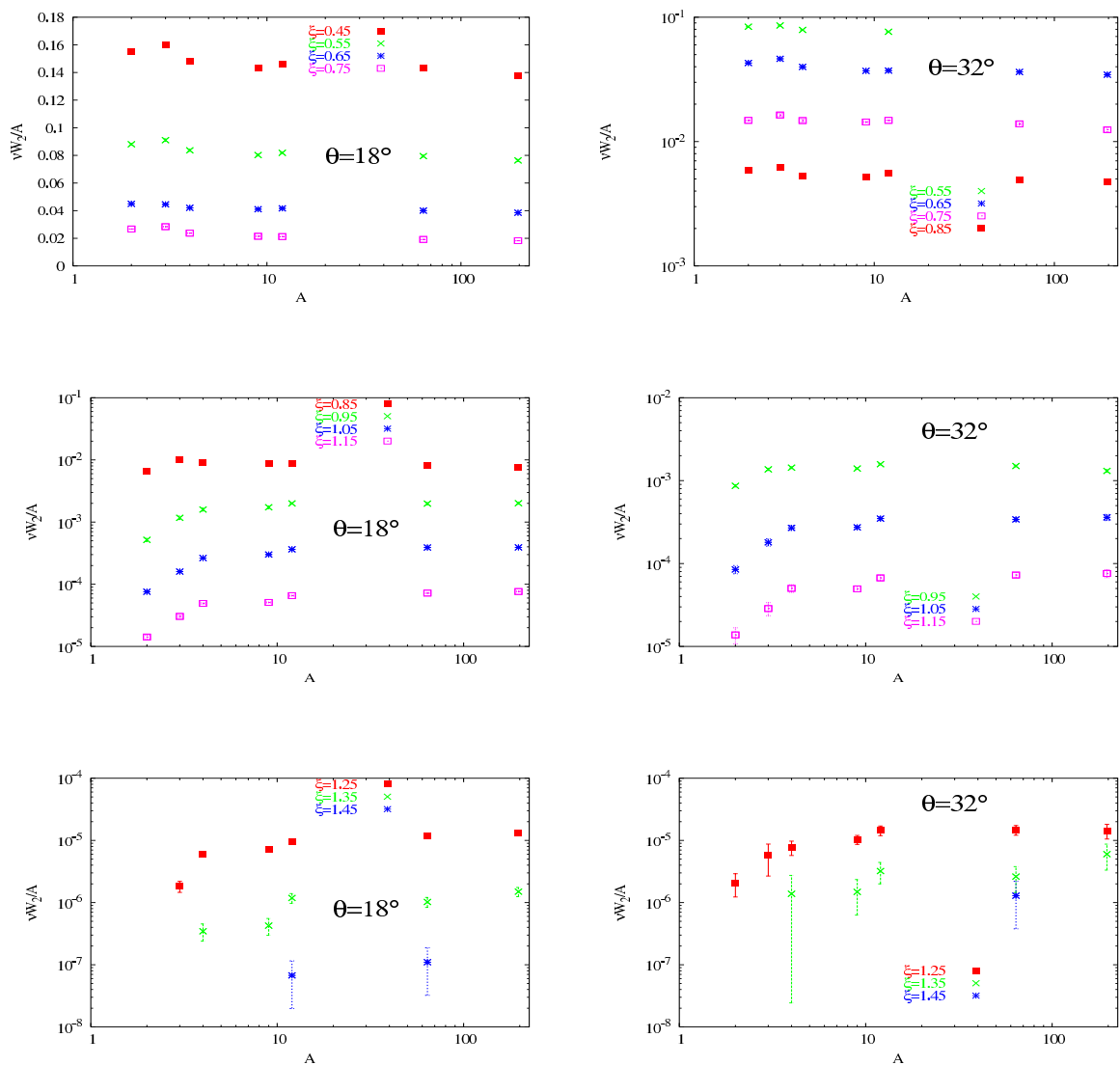}
\caption[A-dependence of the F$_2$ structure function for fixed $\xi$
  and $\theta_{HMS}$]{F$_2$ structure function as a function of A for
  fixed values of $\xi$ at 18$^{\circ}$ (left) and 32$^{\circ}$
  (right).  Note that there is little A-dependence for the lowest
  values of $\xi$ for all nuclei and the observed A-dependence at
  higher values of $\xi$ occurs for light nuclei only.}

\label{xsi_scaling_adep}
\end{figure*}

The F$_2$ structure function can also be examined for a given angle
setting and a fixed $\xi$ as a function of atomic number A, which is done in
Fig.~\ref{xsi_scaling_adep} for 18$^{\circ}$ and 22$^{\circ}$.  This
figure clearly shows the deviation of light nuclei from the scaling
curve in the quasielastic regime and excellent scaling for low values
of $\xi$, corresponding to the high energy loss side of the
quasielastic peak.
\clearpage
\section{Systematic Uncertainties\label{sys_err_section}}
The extracted cross sections were corrected for many effects, such as
the detector efficiencies, target densities, and radiative effects.
Table \ref{sys_error_table} lists the systematic uncertainties
associated with each of the corrections.  Those that quote a range in
uncertainty, such as the uncertainty due to the subtraction of the
charge symmetric background, varied with kinematics and were either
parametrized or calculated at each data point using the cross section
model.
\begin{table}[h!]
\begin{center}
\caption{Systematic Uncertainties.  They are discussed in detail in
  the sections indicated.  When a range is quoted on the scale
  uncertainties, it is to reflect the target-dependence of the
  quantity. The ($^*$) next to the uncertainties in the kinematic
  quantities denotes the maximum effect from the uncertainty at the
  most extreme kinematics, rather than the typical uncertainty.  The
  range of the total uncertainties is a result of combining the
  individual uncertainties in quadrature for best and worst case scenarios.}
\vspace*{0.25in}

\begin{tabular}{|c|c|c|c|c|}
\hline
\textbf{Source} &
\textbf{Scale} &
\textbf{Relative} &
\textbf{$\Delta \sigma / \sigma$} &
\textbf{Details} \\
\hline
Trigger Efficiency & - &  0.1$\%$ & 0.1$\%$ & Sec.~\ref{trig_eff_section} \\
Tracking Efficiency & 0.4$\%$ & 0.3$\%$ & 0.5$\%$ & Sec.~\ref{tracking_eff_section} \\
Pion Contamination & - & 0.2$\%$ &0.2$\%$ & Sec. \ref{pion_contam_section}\\
Computer Dead Time & 0.1$\%$ & -  &0.1$\%$ & Sec.~\ref{comp_dt}\\
Calorimeter Efficiency & - & -& - & Sec.~\ref{detector_eff_section}\\
Positron Subtraction  &- & 0-2$\%$ & 0-2$\%$ & Sec.~\ref{csbg_subtraction_section}\\
\v{C}erenkov Efficiency & 0.05 & 0.3$\%$ & 0.3$\%$ & Sec.~\ref{detector_eff_section}\\
\hline
Acceptance Correction & 1$\%$ &  1$\%$ & 1.4$\%$ & Sec.~\ref{accp_section}\\
Charge & 0.4$\%$ & 0.3$\%$ &0.5$\%$ & Sec.~\ref{bcmcal_sec} \\
Target Thickness & 0.5-2.4$\%$ & - &0.5-2.4$\%$ &
Secs. \ref{reg_target_section} \&\& \ref{cryo_target_section}\\
\hline
Bin-Centering  & - &0.5$\%$& 0.5$\%$ & Sec.~\ref{bc_section}\\
Radiative Correction &1$\%$ & 1$\%$ & 1.4$\%$ & Sec.~\ref{rc_section}\\
Coulomb Correction  & - & 0-2$\%$ & 0-2$\%$ & Sec.~\ref{cc_section}\\
\hline
HMS Momentum & 0.05$\%$ &  0.01$\%$ & $<$5$\%^{*}$ & Sec.~\ref{kinem_section}\\
Beam Energy &  0.05$\%$&  0.02$\%$ & $<$5$\%^{*}$ & Sec.~\ref{kinem_section}\\
HMS angle & 0.5mr & 0.2mr & $<$6$\%^{*}$ & Sec.~\ref{kinem_section}\\
\hline 
\textbf{Cryo targets only} &&&& \\
\hline
Target Boiling & 0.45$\%$ & 0.2$\%$ & 0.5$\%$ & Sec.~\ref{target_boiling_section}\\
Dummy Subtraction & $<$0.6$\%$ & 0-1.8$\%$ & 0.3-2.4$\%$ & Sec.~\ref{dummy_subtraction_section}\\
\hline\hline
Total & $\approx$1.7$\%$ &  $\approx$2.4$\%$ & 2.4-6.5$\%$& \\
\hline
\end{tabular}
\label{sys_error_table}
\end{center}
\end{table}
\subsection{Kinematic Uncertainties \label{kinem_section}}
The uncertainties in the measured kinematic quantities (beam energy,
HMS momentum, and HMS angle) are determined
based on the reproducibility of those quantities. These uncertainties do not enter into the
cross section measurement as directly as, for example, a detector
efficiency.  Instead, the effect varies with kinematics.  The point-to-point uncertainties were determined by
evaluating the cross section model at shifted kinematics.  The shifts
were 0.02$\%$, 0.01$\%$, and 0.2mr for beam, HMS momentum, and HMS
angle, respectively.  The effect of these shifts on the model cross section gives the size of the point-to-point
uncertainty in the measured cross section. 

The scale uncertainty was similarly determined.  Here, the fact that
the HMS momentum is coupled to the beam energy was taken into effect
as a correlation between the two uncertainties.  This time, both the
energy and momentum inputs into the cross section model were scaled at
the same time, by 0.05$\%$ and 0.03$\%$, respectively.  For the scale
uncertainty on the angle, a shift of 0.5mr was applied to the model.

\chapter{Conclusion and Outlook}
The results from E02-019 support previous observations and extend their
kinematic range.   Cross sections were measured for $^2$H, $^3$He,
$^4$He,$^9$Be, $^{12}$C, $^{63}$Cu, and $^{197}$Au and were compared
to calculations, whenever possible.
It appears that current theoretical calculations~\cite{Benhar:2006hh} describe the
inclusive data very well, including contributions from FSIs and SRCs.

The scaling function F($|\textbf{q}|$,y) was extracted for all
nuclei.  Scaling was observed for $y<$0, approached from above, with
violations at low Q$^2$ due to contributions from FSIs.  Momenta as
high as $-$1.2 GeV/c were observed.  It can be argued that the $y_{cw}$ scaling
variable is more appropriate for this analysis, but a relativistic
definition is needed given the wide kinematic coverage of the data.

The nuclear inelastic structure function F$_2$ was extracted for all
nuclei.  No scaling was observed in $x$, since most data settings do
not probe the kinematic regime where DIS is dominant.  However, F$_2$
shows excellent scaling in $\xi$ for most nuclei.  In addition, scaling
of F$_2$ was observed for different nuclei at the same kinematics for
the first time,
with some violations for light nuclei at low Q$^2$.  

Cross section ratios of heavy nuclei to light ones were taken in order
to study 2N- and 3N- correlations.  Further work needs to be done
before the ratios in the $x>$2.5 region can be quoted with
confidence.  The contribution from the end-caps of the cryogenic
target is very large and a careful study of possible sources of errors
is underway.  

A follow-up experiment, E12-06-105~\cite{Arrington:2006xx}, has been approved to run at
Jefferson Lab after the 12 GeV upgrade.  This measurement will extend
to the coverage to higher values of Q$^2$, into a region where the DIS
contribution to the cross sections at $x>$1 is significant.


\bibliography{allmybiblio}
\end{document}